\documentclass[journal]{IEEEtran}
\usepackage{cite}
\usepackage{amsmath,amssymb,amsfonts}
\usepackage{algorithmic}
\usepackage{graphicx}
\usepackage{subfigure}
\usepackage{amsmath}
\usepackage{textcomp}
\usepackage{xcolor}
\usepackage{array}
\usepackage{makecell}
\usepackage{multirow}
\usepackage{epstopdf}
\usepackage{threeparttable}
\usepackage{booktabs}
\usepackage{tablefootnote}
\ifCLASSINFOpdf
\else
\fi
\begin{document}
%
% paper title
% Titles are generally capitalized except for words such as a, an, and, as,
% at, but, by, for, in, nor, of, on, or, the, to and up, which are usually
% not capitalized unless they are the first or last word of the title.
% Linebreaks \\ can be used within to get better formatting as desired.
% Do not put math or special symbols in the title.
\title{An Unconditionally Stable Conformal LOD-FDTD Method for Curved PEC Objects and Its Application to EMC Problems}
%
%
% author names and IEEE memberships
% note positions of commas and nonbreaking spaces ( ~ ) LaTeX will not break
% a structure at a ~ so this keeps an author's name from being broken across
% two lines.
% use \thanks{} to gain access to the first footnote area
% a separate \thanks must be used for each paragraph as LaTeX2e's \thanks
% was not built to handle multiple paragraphs
%

\author{Hanhong~Liu,~\IEEEmembership{Graduate Student Member,~IEEE,} Xiaoying~Zhao,~Xiang-Hua~Wang, \\Shunchuan~Yang,~\IEEEmembership{Member,~IEEE,}
	and~Zhizhang~(David) Chen,~\IEEEmembership{Fellow,~IEEE}% <-this % stops a space
	
	\thanks{Manuscript received xxx; revised xxx.}
	\thanks{This work was supported in part by the National Natural Science Foundation of China through Grant 61801010 and Grant 62071125,  in part by Pre-Research Project through Grant J2019-VIII-0009-0170, and Fundamental Research Funds for the Central Universities. \textit{(Corresponding author: Shunchuan Yang)}}
	\thanks{H. H. Liu and X. Y. Zhao are with the School of Electronic and Information Engineering of Electronic and Information Engineering, Beihang University, Beijing, China (e-mail: liu759753745@buaa.edu.cn, xyz@buaa.edu.cn).}% 
	\thanks{X. H. Wang is with School of Science, Tianjin University of Technology and Education, Tianjin, China (e-mail: xhwang199@outlook.com).}% <-this % stops a space
	\thanks{S. C. Yang is with Research Institute for Frontier Science and School of Electronic and Information Engineering, Beihang University, Beijing, China (e-mail: scyang@buaa.edu.cn).}% <-this % stops a space
	\thanks{Z. D. Chen is currently with the College of Physics and Information Engineering, Fuzhou University, Fuzhou, Fujian. P. R. China, on leave from the Department of Electrical and Computer Engineering, Dalhousie University, Halifax, Nova Scotia, Canada B3H 4R2  (email: zz.chen@ieee.org).}}% <-this % stops a space

\markboth{Journal of \LaTeX\ Class Files,~Vol.~xx, No.~x, xx~xxxx}%
{Liu \MakeLowercase{\textit{et al.}}: Unconditionally Stable Conformal LOD-FDTD Method and Its EMC Applications}
% The only time the second header will appear is for the odd numbered pages
% after the title page when using the twoside option.
% 
% *** Note that you probably will NOT want to include the author's ***
% *** name in the headers of peer review papers.                   ***
% You can use \ifCLASSOPTIONpeerreview for conditional compilation here if
% you desire.

% If you want to put a publisher's ID mark on the page you can do it like
% this:
%\IEEEpubid{0000--0000/00\$00.00~\copyright~2015 IEEE}
% Remember, if you use this you must call \IEEEpubidadjcol in the second
% column for its text to clear the IEEEpubid mark.

% use for special paper notices
%\IEEEspecialpapernotice{(Invited Paper)}

% make the title area
\maketitle

% in the abstract or keywords.
% As a general rule, do not put math, special symbols or citations
\begin{abstract}
The traditional finite-difference time-domain (FDTD) method is constrained by the Courant-Friedrich-Levy (CFL) condition and suffers from the notorious staircase error in electromagnetic simulations. This paper proposes a three-dimensional conformal locally-one-dimensional FDTD (CLOD-FDTD) method to address the two issues for modeling perfectly electrical conducting (PEC) objects. By considering the partially filled cells, the proposed CLOD-FDTD method can significantly improve the accuracy compared with the traditional LOD-FDTD method and the FDTD method. At the same time, the proposed method preserves unconditional stability, which is analyzed and numerically validated using the Von-Neuman method. Significant gains in Central Processing Unit (CPU) time are achieved by using large time steps without sacrificing accuracy. Two numerical examples include a PEC cylinder and a missile are used to verify its accuracy and efficiency with different meshes and time steps. It can be found from these examples, the CLOD-FDTD method show better accuracy and can improve the efficiency compared with those of the traditional FDTD method and the traditional LOD-FDTD method.
\end{abstract}

% Note that keywords are not normally used for peerreview papers.

% For peer review papers, you can put extra information on the cover
% page as needed:
% \ifCLASSOPTIONpeerreview
% \begin{center} \bfseries EDICS Category: 3-BBND \end{center}
% \fi
%
% For peerreview papers, this IEEEtran command inserts a page break and
% creates the second title. It will be ignored for other modes.
\IEEEpeerreviewmaketitle

\section{Introduction}
\IEEEPARstart{T}{he} finite-difference time-domain (FDTD) method has been widely applied in electromagnetic simulations for its generality and simplicity \cite{FDTD,B1,TAFLOVEFDTDAD}. However, it mainly suffers from two issues: one is that time steps are restricted by the Courant-Friedrich-Levy (CFL) condition, which requires prohibitively long simulation time if geometrically fine structures are involved. The other is staircase error since orthogonal hexagonal elements are used to model irregular objects with curved surfaces. 

To address the first issue, many unconditionally stable FDTD methods, such as the alternately-direction-implicit finite-difference time-domain (ADI-FDTD) method \cite{B2}\cite{CHENADI}, the locally-one-dimensional FDTD (LOD-FDTD) method \cite{LODB3, LODChen, LODLIU}, the Crank-Nicolson FDTD (CN-FDTD) method \cite{CNB4, B4, CNB42}, the split-step FDTD (SS-FDTD) method \cite{SSB5, B5, SSB52}, the leapfrog ADI-FDTD methods \cite{LEAPFROGADICOOKE, LEAPFROGADIYANG, LEAPFROGADIYANGAP} and others, were proposed to remove this constraint, where time steps are independent of mesh sizes. Therefore, stable numerical results can be obtained no matter how large time steps are used in the simulations.

To address the staircase issue, various conformal techniques were proposed to improve the accuracy of the traditional FDTD methods. Then, those conformal techniques were extended for the unconditionally stable FDTD methods, and they can be divided into two groups: one for dielectric media and the other for perfectly electrical conducting (PEC) objects. In \cite{B6}\cite{B7}, area- and volume-weighted techniques were used to calculate the effective permittivity of partially filled cells in the traditional FDTD method. In \cite{B8}, the conformal techniques were further extended into the high-order FDTD (2,4) method to handle curved dielectric objects. All these techniques show significant accuracy and efficiency improvement compared with the traditional FDTD method. Since only averaged effective parameters are used, their stability can be guaranteed. However, for the curved PEC objects, it is quite challenging. Although these conformal techniques can indeed improve the accuracy, so-called late-time instability occurs if no special treatments are made upon small filled cells \cite{B9}\cite{B10}. Reduction of time steps is usually required to guarantee stability, which would inevitably increase the cost of Central Processing Unit (CPU) time in practical simulations. Several efforts were made to develop the conformal techniques without reduction of time steps, such as the uniformly-stable-conformal (USC) approach in \cite{B11} and the extended cell technique in \cite{B12}. However, they are usually complex to implement for practical structures. 

To reduce the staircase error in the unconditionally stable FDTD methods, one idea is to extend the conformal techniques employed in the traditional FDTD methods to the implicit FDTD methods to improve the accuracy and keep the unconditional stability. Some efforts, such as the approaches proposed in \cite{B13} \cite{B14}, have been made for this purpose. However, careful investigations show that they are only conditionally stable and suffer from the late-time instability issue. Recently, a two-dimensional conformal LOD-FDTD method was proposed in \cite{B16} to solve the scattering problems. 

In this paper, a three-dimensional conformal LOD-FDTD (CLOD-FDTD) method with preserving the unconditional stability for curved PEC objects is proposed to reduce staircase errors. By carefully incorporating partially filled cells into the time-marching formulations, curved surfaces can be accurately modeled through the proposed CLOD-FDTD method. In addition, we numerically proved its stability through the Von-Neuman method with different CFL numbers (CFLNs) and cell sizes. One obvious advantage of the proposed CLOD-FDTD method is that it can improve the accuracy compared with the traditional FDTD method and the LOD-FDTD method and preserve unconditional stability. Therefore, it is much preferred in practical engineering simulations. In \cite{CLODACES}, we have reported the primary idea about the CLOD-FDTD method. This paper significantly extends our previous work. 

The remaining paper is organized as follows. In Section II, a detailed derivation of the proposed CLOD-FDTD method is presented. In Section III, its stability is numerically proved and validated through the Von-Neuman method. For the comparison purpose, the stability of the conformal FDTD method is also included. Then, two numerical examples are carried out to verify its effectiveness in Section IV.  At last, we draw some conclusions in Section V.

\section{FORMULATIONS FOR THE CLOD-FDTD METHOD}
Without loss of generality, a PEC object hosted by linear, lossless, isotropic, and homogeneous medium with permittivity $\varepsilon$ and permeability $\mu$ is considered in our following derivation. The Maxwell's curl equations can be expressed as
\begin{subequations}
\begin{align}
\nabla \times & \boldsymbol{\rm{H}} = \varepsilon \frac{\partial \boldsymbol{\rm{E}}}{\partial t}, \label{E1} \\
\nabla \times & \boldsymbol{\rm{E}} = - \mu \frac{\partial \boldsymbol{\rm{H}}}{\partial t}. \label{E2}
\end{align}
\end{subequations}

\begin{figure}[h]
	\centering
	\includegraphics[scale=0.4]{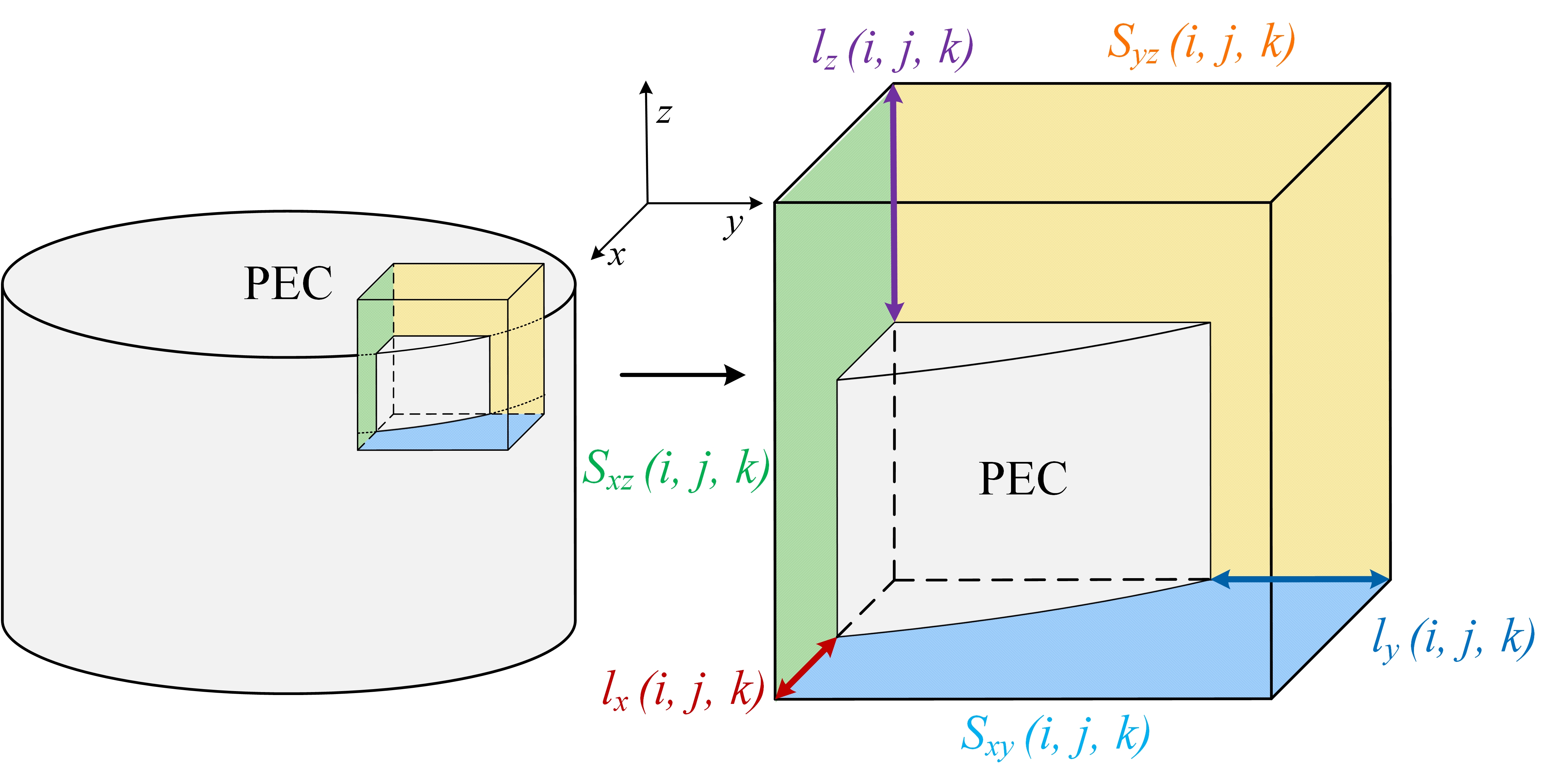}
	\caption{The view of cells with the curved PEC object projected.}
	\label{Fig.1}
\end{figure}

To numerically solve (\ref{E1}) and (\ref{E2}) with the LOD-FDTD method, they are sampled by Yee's grids in the spatial domain. Two sub-steps scheme is used in the time domain \cite{B18}. In this way, each time step is split into two sub-steps. By carefully considering the partially filled cells in Fig. \ref{Fig.1}, the time-marching formulations for the proposed CLOD-FDTD method in sub-step\#1 can be expressed as follows. 

\textbf{Sub-step\#1:}
\begin{subequations}
\begin{equation}\label{E3}
\begin{aligned}
    E_x|_{i+\frac{1}{2},j,k}^{n+\frac{1}{2}}&=E_x|_{i+\frac{1}{2},j,k}^{n}\\
    &+\frac{{\Delta}t}{2{\varepsilon}}\delta_y\left(H_z|_{i+\frac{1}{2},j+\frac{1}{2},k}^{n}+H_z|_{i+\frac{1}{2},j+\frac{1}{2},k}^{n+\frac{1}{2}}\right),
\end{aligned}
\end{equation}
\begin{equation}\label{E4}
\begin{aligned}
    E_y|_{i,j+\frac{1}{2},k}^{n+\frac{1}{2}}&=E_y|_{i,j+\frac{1}{2},k}^{n}\\
    &+\frac{{\Delta}t}{2{\varepsilon}}\delta_z\left(H_x|_{i,j+\frac{1}{2},k+\frac{1}{2}}^{n}+H_x|_{i,j+\frac{1}{2},k+\frac{1}{2}}^{n+\frac{1}{2}}\right),
\end{aligned}
\end{equation}
\begin{equation}\label{E5}
\begin{aligned}
    E_z|_{i,j,k+\frac{1}{2}}^{n+\frac{1}{2}}&=E_z|_{i,j,k+\frac{1}{2}}^{n}\\
    &+\frac{{\Delta}t}{2{\varepsilon}}\delta_x\left(H_y|_{i+\frac{1}{2},j,k+\frac{1}{2}}^{n}+H_y|_{i+\frac{1}{2},j,k+\frac{1}{2}}^{n+\frac{1}{2}}\right),
\end{aligned}
\end{equation}
\begin{equation}\label{E6}\!\!\!\!
\begin{aligned}
	H_x|_{i,j+\frac{1}{2},k+\frac{1}{2}}^{n+\frac{1}{2}}&=H_x|_{i,j+\frac{1}{2},k+\frac{1}{2}}^{n}\\
	&+\frac{{\Delta}t{\Delta}z}{2{\mu}S_{yz}|_{i,j+\frac{1}{2},k+\frac{1}{2}}}\\
	&\times\delta_z\left[l_y|_{i,j+\frac{1}{2},k}
	\right.\\&\left.\times\left(E_y|_{i,j+\frac{1}{2},k}^{n}+E_y|_{i,j+\frac{1}{2},k}^{n+\frac{1}{2}}\right)\right],
\end{aligned}
\end{equation}
\begin{equation}\label{E7}\!\!\!\!
\begin{aligned}
	H_y|_{i+\frac{1}{2},j,k+\frac{1}{2}}^{n+\frac{1}{2}}&=H_y|_{i+\frac{1}{2},j,k+\frac{1}{2}}^{n}\\
	&+\frac{{\Delta}t{\Delta}x}{2{\mu}S_{xz}|_{i+\frac{1}{2},j,k+\frac{1}{2}}}\\
	&\times\delta_x\left[l_z|_{i,j,k+\frac{1}{2}}
	\right.\\&\left.\times\left(E_z|_{i,j,k+\frac{1}{2}}^{n}+E_z|_{i,j,k+\frac{1}{2}}^{n+\frac{1}{2}}\right)\right],
\end{aligned}
\end{equation}
\begin{equation}\label{E8}\!\!\!\!
\begin{aligned}
	H_z|_{i+\frac{1}{2},j+\frac{1}{2},k}^{n+\frac{1}{2}}&=H_z|_{i+\frac{1}{2},j+\frac{1}{2},k}^{n}\\
	&+\frac{{\Delta}t{\Delta}y}{2{\mu}S_{xy}|_{i+\frac{1}{2},j+\frac{1}{2},k}}\\
	&\times\delta_y\left[l_x|_{i+\frac{1}{2},j,k}
	\right.\\&\left.\times\left(E_x|_{i+\frac{1}{2},j,k}^{n}+E_x|_{i+\frac{1}{2},j,k}^{n+\frac{1}{2}}\right)\right],
\end{aligned}
\end{equation}
\end{subequations}
where $E_x,E_y,E_z,H_x,H_y,H_z$ denote the electric and magnetic field components, and their subscripts $i$, $j$, $k$ are the $i$th, $j$th, and $k$th nodes in the \textit{x}-, \textit{y}-, \textit{z}-directions, respectively. The superscript $n$ is the $n$th time step in simulations and $\delta_x,\delta_y,\delta_z$ denote the differential operators in the ${x}$-, ${y}$-, and ${z}$-directions, respectively. As shown in Fig. \ref{Fig.1}, when the curved PEC surfaces exist in the computational domain, $l_x$, $l_y$, $l_z$ are edge lengths outside the PEC objects in each cell along the ${x}$-, ${y}$-, and ${z}$-directions, respectively. $S_{yz}$, $S_{xz}$, $S_{xy}$ denote the areas outside curved PEC surface projected onto the \textit{yoz}, \textit{xoz}, and \textit{xoy} plane, respectively. It should be noted that in the free space, $l_x$, $l_y$, $l_z$ are equal to ${\Delta}x$, ${\Delta}y$, ${\Delta}z$, and $S_{yz}$, $S_{xz}$, $S_{xy}$ are equal to areas of the corresponding Yee's cell, such as $S_{xy}={\Delta}x{\Delta}y$. Therefore, the time-marching formulations of the CLOD-FDTD method are exactly the same as those of the LOD-FDTD method in the free space. $l_x,l_y,l_z$ are zeros to make $H_x,H_y,H_z$ vanish inside the PEC objects. 

Since the conformal information of partially filled cells is embedded into the time-marching formulations, the curved PEC surface can be considered in the proposed CLOD-FDTD method. Several techniques based on geometrical kernels \cite{CONFORMALMESHYU, CONFORMALMESHGEOFF, CONFORMALMESHBERENS, CONFORMALMESHBO} were reported to accurately extract the conformal information for the FDTD method. In our simulations, we also developed a conformal FDTD meshing program based on the ray-tracing algorithm \cite{CONFORMALMESHYU} for this purpose. Therefore, the practical engineering structures can be automatically discretized, and the conformal information is accurately extracted. 

In a similar manner, the time-marching formulations of  the proposed CLOD-FDTD method in sub-step\#2 can be expressed as follows. 

\textbf{Sub-step\#2:}
\begin{subequations}
\begin{equation}\label{E9}
\begin{aligned}
    E_x|_{i+\frac{1}{2},j,k}^{n+1}&=E_x|_{i+\frac{1}{2},j,k}^{n+\frac{1}{2}}\\
    &-\frac{{\Delta}t}{2{\varepsilon}}\delta_z\left(H_y|_{i+\frac{1}{2},j,k+\frac{1}{2}}^{n+\frac{1}{2}}+H_y|_{i+\frac{1}{2},j,k+\frac{1}{2}}^{n+1}\right),
\end{aligned}
\end{equation}
\begin{equation}\label{E10}
\begin{aligned}
    E_y|_{i,j+\frac{1}{2},k}^{n+1}&=E_y|_{i,j+\frac{1}{2},k}^{n+\frac{1}{2}}\\
    &-\frac{{\Delta}t}{2{\varepsilon}}\delta_x\left(H_z|_{i+\frac{1}{2},j+\frac{1}{2},k}^{n+\frac{1}{2}}+H_z|_{i+\frac{1}{2},j+\frac{1}{2},k}^{n+1}\right),
\end{aligned}
\end{equation}
\begin{equation}\label{E11}
\begin{aligned}
    E_z|_{i,j,k+\frac{1}{2}}^{n+1}&=E_z|_{i,j,k+\frac{1}{2}}^{n+\frac{1}{2}}\\
    &-\frac{{\Delta}t}{2{\varepsilon}}\delta_y\left(H_x|_{i,j+\frac{1}{2},k+\frac{1}{2}}^{n+\frac{1}{2}}+H_x|_{i,j+\frac{1}{2},k+\frac{1}{2}}^{n+1}\right),
\end{aligned}
\end{equation}
\begin{equation}\label{E12}\!\!\!\!
\begin{aligned}
	H_x|_{i,j+\frac{1}{2},k+\frac{1}{2}}^{n+1}&=H_x|_{i,j+\frac{1}{2},k+\frac{1}{2}}^{n+\frac{1}{2}}\\
	&-\frac{{\Delta}t{\Delta}y}{2{\mu}S_{yz}|_{i,j+\frac{1}{2},k+\frac{1}{2}}}\\
	&\times\delta_y\left[l_z|_{i,j,k+\frac{1}{2}}
	\right.\\&\left.\times\left(E_z|_{i,j,k+\frac{1}{2}}^{n+\frac{1}{2}}+E_z|_{i,j,k+\frac{1}{2}}^{n+1}\right)\right],
\end{aligned}
\end{equation}
\begin{equation}\label{E13}\!\!\!\!
\begin{aligned}
	H_y|_{i+\frac{1}{2},j,k+\frac{1}{2}}^{n+1}&=H_y|_{i+\frac{1}{2},j,k+\frac{1}{2}}^{n+\frac{1}{2}}\\
	&-\frac{{\Delta}t{\Delta}z}{2{\mu}S_{xz}|_{i+\frac{1}{2},j,k+\frac{1}{2}}}\\
	&\times\delta_z\left[l_x|_{i+\frac{1}{2},j,k}
	\right.\\&\left.\times\left(E_x|_{i+\frac{1}{2},j,k}^{n+\frac{1}{2}}+E_x|_{i+\frac{1}{2},j,k}^{n+1}\right)\right],
\end{aligned}
\end{equation}
\begin{equation}\label{E14}\!\!\!\!\!
\begin{aligned}
	H_z|_{i+\frac{1}{2},j+\frac{1}{2},k}^{n+1}&=H_z|_{i+\frac{1}{2},j+\frac{1}{2},k}^{n+\frac{1}{2}}\\
	&-\frac{{\Delta}t{\Delta}x}{2{\mu}S_{xy}|_{i+\frac{1}{2},j+\frac{1}{2},k}}\\
	&\times\delta_x\left[l_y|_{i,j+\frac{1}{2},k}
	\right.\\&\left.\times\left(E_y|_{i,j+\frac{1}{2},k}^{n+\frac{1}{2}}+E_y|_{i,j+\frac{1}{2},k}^{n+1}\right)\right].
\end{aligned}
\end{equation}
\end{subequations}
Unlike the traditional FDTD method, in which the explicit time-marching formulations are used to update electric and magnetic fields in a leapfrog manner \cite{B1}, electric fields in the proposed CLOD-FDTD method cannot be explicitly calculated. It requires combining the magnetic and electric field formulations as the LOD-FDTD method. For instance, by substituting (\ref{E8}) into (\ref{E3}), $E_x$ in {{sub-step\#1}}  can be written as
\begin{equation}\label{E15}
\begin{aligned}
    &C_1|_{i+\frac{1}{2},j-1,k}E_x|_{i+\frac{1}{2},j-1,k}^{n+\frac{1}{2}}+ C_2|_{i+\frac{1}{2},j,k}E_x|_{i+\frac{1}{2},j,k}^{n+\frac{1}{2}}\\
	&+ C_3|_{i+\frac{1}{2},j+1,k}E_x|_{i+\frac{1}{2},j+1,k}^{n+\frac{1}{2}}\\
	&=\frac{{\Delta}t}{{\varepsilon\Delta}y}\left(H_z|_{i+\frac{1}{2},j+\frac{1}{2},k}^{n}-H_z|_{i+\frac{1}{2},j-\frac{1}{2},k}^{n}\right)\\
	&-C_1|_{i+\frac{1}{2},j-1,k}E_x|_{i+\frac{1}{2},j-1,k}^{n}+ C_4|_{i+\frac{1}{2},j,k}E_x|_{i+\frac{1}{2},j,k}^{n}\\
	&-C_3|_{i+\frac{1}{2},j+1,k}E_x|_{i+\frac{1}{2},j+1,k}^{n},
\end{aligned}
\end{equation}
where
\begin{flalign}
& C_1|_{i+\frac{1}{2},j-1,k}=-\frac{{\Delta}t^2l_x|_{i+\frac{1}{2},j-1,k}}{4\mu\varepsilon{\Delta}yS_{xy}|_{i+\frac{1}{2},j-\frac{1}{2},k}},&
\nonumber
\end{flalign}
\begin{flalign}
& C_2|_{i+\frac{1}{2},j,k}=1+\frac{{\Delta}t^2}{4\mu\varepsilon{\Delta}y}\left(\frac{l_x|_{i+\frac{1}{2},j,k}}{S_{xy}|_{i+\frac{1}{2},j+\frac{1}{2},k}}+\frac{l_x|_{i+\frac{1}{2},j,k}}{S_{xy}|_{i+\frac{1}{2},j-\frac{1}{2},k}}\right),&
\nonumber
\end{flalign}
\begin{flalign}
& C_3|_{i+\frac{1}{2},j+1,k}=-\frac{{\Delta}t^2l_x|_{i+\frac{1}{2},j+1,k}}{4\mu\varepsilon{\Delta}yS_{xy}|_{i+\frac{1}{2},j+\frac{1}{2},k}}, &
\nonumber
\end{flalign}
\begin{flalign}
& C_4|_{i+\frac{1}{2},j,k}=1-\frac{{\Delta}t^2}{4\mu\varepsilon{\Delta}y}\left(\frac{l_x|_{i+\frac{1}{2},j,k}}{S_{xy}|_{i+\frac{1}{2},j+\frac{1}{2},k}}+\frac{l_x|_{i+\frac{1}{2},j,k}}{S_{xy}|_{i+\frac{1}{2},j-\frac{1}{2},k}}\right).&
\nonumber
\end{flalign}

Similarly, by substituting (\ref{E14}) into (\ref{E9}), $E_x$ in {sub-step\#2} can be expressed as
\begin{equation}\label{E16}
\begin{aligned}
    &C_5|_{i+\frac{1}{2},j,k-1}E_x|_{i+\frac{1}{2},j,k-1}^{n+1}+ C_6|_{i+\frac{1}{2},j,k}E_x|_{i+\frac{1}{2},j,k}^{n+1}\\
&+ C_7|_{i+\frac{1}{2},j,k+1}E_x|_{i+\frac{1}{2},j,k+1}^{n+1}\\
&=-\frac{{\Delta}t}{{\varepsilon\Delta}z}\left(H_y|_{i+\frac{1}{2},j,k+\frac{1}{2}}^{n+\frac{1}{2}}-H_y|_{i+\frac{1}{2},j,k-\frac{1}{2}}^{n+\frac{1}{2}}\right)\\
&-C_5|_{i+\frac{1}{2},j,k-1}E_x|_{i+\frac{1}{2},j,k-1}^{n+\frac{1}{2}}+ C_8|_{i+\frac{1}{2},j,k}E_x|_{i+\frac{1}{2},j,k}^{n+\frac{1}{2}}\\
&-C_7|_{i+\frac{1}{2},j,k+1}E_x|_{i+\frac{1}{2},j,k+1}^{n+\frac{1}{2}},
\end{aligned}
\end{equation}
where
\begin{flalign}
& C_5|_{i+\frac{1}{2},j,k-1}=-\frac{{\Delta}t^2l_x|_{i+\frac{1}{2},j,k-1}}{4\mu\varepsilon{\Delta}zS_{xz}|_{i+\frac{1}{2},j,k-\frac{1}{2}}},&
\nonumber
\end{flalign}
\begin{flalign}
& C_6|_{i+\frac{1}{2},j,k}=1+\frac{{\Delta}t^2}{4\mu\varepsilon{\Delta}z}\left(\frac{l_x|_{i+\frac{1}{2},j,k}}{S_{xz}|_{i+\frac{1}{2},j,k+\frac{1}{2}}}+\frac{l_x|_{i+\frac{1}{2},j,k}}{S_{xz}|_{i+\frac{1}{2},j,k-\frac{1}{2}}}\right),&
\nonumber
\end{flalign}
\begin{flalign}
& C_7|_{i+\frac{1}{2},j,k+1}=-\frac{{\Delta}t^2l_x|_{i+\frac{1}{2},j,k+1}}{4\mu\varepsilon{\Delta}zS_{xz}|_{i+\frac{1}{2},j,k+\frac{1}{2}}},&
\nonumber
\end{flalign}
\begin{flalign}
& C_8|_{i+\frac{1}{2},j,k}=1-\frac{{\Delta}t^2}{4\mu\varepsilon{\Delta}z}\left(\frac{l_x|_{i+\frac{1}{2},j,k}}{S_{xz}|_{i+\frac{1}{2},j,k+\frac{1}{2}}}+\frac{l_x|_{i+\frac{1}{2},j,k}}{S_{xz}|_{i+\frac{1}{2},j,k-\frac{1}{2}}}\right).&
\nonumber
\end{flalign}

It can be found that (\ref{E15}) and (\ref{E16}) are similar to the traditional LOD-FDTD time-marching formulations while $C_1-C_8$ are modified by $S_{yz},S_{xz},S_{xy}$ and $l_x,l_y,l_z$ through carefully considering partially filled cells near the curved PEC surfaces. It is necessary to collect all the electric field components into the compact matrix equation form for solving their field values. By rewriting the time-marching formulation (\ref{E15}) into the matrix form, we obtain
\begin{equation}\label{E17}
	{{\bf{A}}}
	{{\bf{E}}_x^{n+\frac{1}{2}}}
	={\bf{r}}^n,
\end{equation}
where the square coefficient matrix $\boldsymbol{\rm{A}}$ with the dimension of $N \times N$, where $N$ is the total number of $E_x$ nodes along the $y$- direction, is tridiagonal, and the column vector ${\bf{r}}^n$ contains all values on the right-hand side of (\ref{E15}) in a column-wise fashion. (\ref{E17}) can be efficiently solved by the Thomas Algorithm \cite{THOMAS}. $E_y$ and $E_z$ can be also easily calculated in a similar manner. Once all the electric fields are calculated, the magnetic fields can be explicitly updated through (\ref{E6})-(\ref{E8}) and (\ref{E12})-(\ref{E14}).

\section{STABILITY ANALYSIS OF THE PROPOSED CLOD-FDTD METHOD AND THE CONFORMAL FDTD METHOD}
In this section, the Von-Neuman method \cite{B17} is used to prove the stability of the proposed CLOD-FDTD method and the conformal FDTD (CFDTD) method, which transforms electromagnetic field components in the time domain into time-harmonic counterparts, and then analyze the eigenvalues of the coefficient matrix. If the modulus of all eigenvalues are not larger than one, the proposed CLOD method is unconditionally stable. 

For the time-harmonic fields, we have the following relationship
	\begin{equation}\label{E18}
		{\bf{U}}^{n+1}= e^{jw{\Delta}t}{\bf{U}}^n,
	\end{equation}
where ${\bf{U}}^n$ and ${\bf{U}}^{n+1}$ denote column vectors of electromagnetic fields at the $n$th and  $n+1$th time step. For the Von-Neuman method, we have ${\bf{U}}^n=\left[E_x^n,\,E_y^n,\,E_z^n,\,H_x^n,\,H_y^n,\,H_z^n\right]^T$, where $E_x^n,\,E_y^n,\,E_z^n,\,H_x^n,\,H_y^n,$, and $H_z^n$ are field values in the spectral domain.

\subsection{Stability Analysis of the Proposed CLOD-FDTD Method}
To make our derivation clear, we define a column vector ${\bf U} = \left[{\bf{E}}_x^n,\,{\bf{E}}_y^n,\,{\bf{E}}_z^n,\,{\bf{H}}_x^n,\,{\bf{H}}_y^n,\,{\bf{H}}_z^n\right]^T$, where ${\bf{E}}_x^{n}$, ${\bf{E}}_y^{n}$, ${\bf{E}}_z^{n}$, ${\bf{H}}_x^{n}$, ${\bf{H}}_y^{n}$, and ${\bf{H}}_z^{n}$ include all the corresponding field nodes on Yee's grid in the $x$-, $y$-, $z$-directions, respectively. Take ${\bf{E}}_x^n$ as an example, ${{\bf{E}}_x^n} = \left[{ E}_x|^n_{1,1,1}, \,{\ E}_x|_{2,1,1}^{n},\,\dots,\,{ E}_x|_{m,p+1,q+1}^{n}\right]^T$, where the subscripts $m$, $p$, and $q$ denote field components' indices in $x$-, $y$- and $z$-directions, respectively.

According to (\ref{E15}), the amplification matrix ${\bf{\Lambda}_1}$ for {sub-step{\#}1} is given by
\begin{equation}\label{E19}
{\bf{U}}^{n+\frac{1}{2}} = {\bf{\Lambda}_1}  {\bf{U}}^{n}.
\end{equation}

Similarly, the amplification matrix ${\bf{\Lambda}_2}$ for {sub-step{\#}2} can be obtained as
\begin{equation}\label{E20}
{\bf{U}}^{n+1} = {\bf{\Lambda}_2}  {\bf{U}}^{n+\frac{1}{2}}.
\end{equation}

By substituting (\ref{E19}) into (\ref{E20}), we obtain the following matrix equation
\begin{equation}\label{E21}
{\bf{U}}^{n+1} = {\bf{\Lambda}}  {\bf{U}}^{n}.
\end{equation}
 If the modulus of all eigenvalues in $\bf \Lambda$ are less than or equal to one, the proposed CLOD-FDTD method is stable. When the homogeneous medium and uniform cells are used, the dimension of $\bf \Lambda$ can be reduced to 6$\times$6. Its 6 eigenvalues can be analytically calculated, and then compared with one. However, the scenario for the proposed CLOD-FDTD methods is different since $S_{yz},S_{xz},S_{xy}$ and $l_x,l_y,l_z$ are location-dependent. It cannot be dealt with the mentioned approaches in \cite{CHENADI}\cite{LEAPFROGADIYANGAP}. To address this issue, we use a specific example to demonstrate the stability of the proposed CLOD-FDTD method.  

\begin{figure}[htbp]
\centering
\subfigure[ ]{
\includegraphics[scale=0.35]{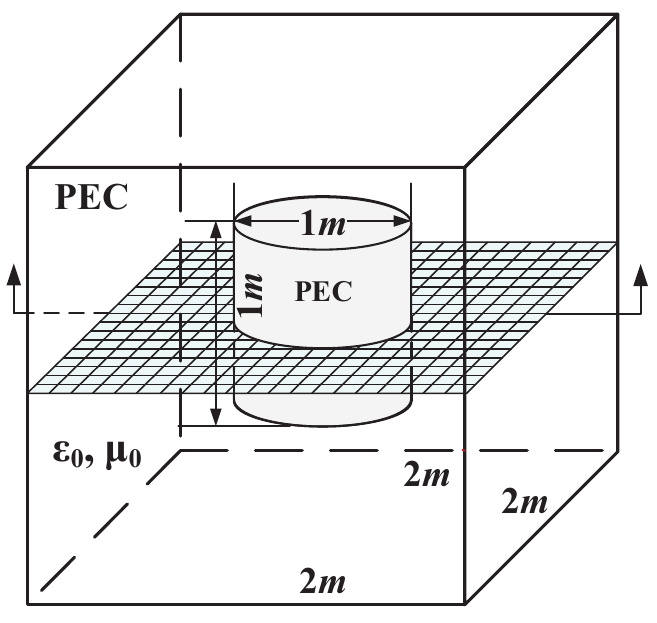}
%\caption{fig1}
}
\subfigure[ ]{
\includegraphics[scale=0.42]{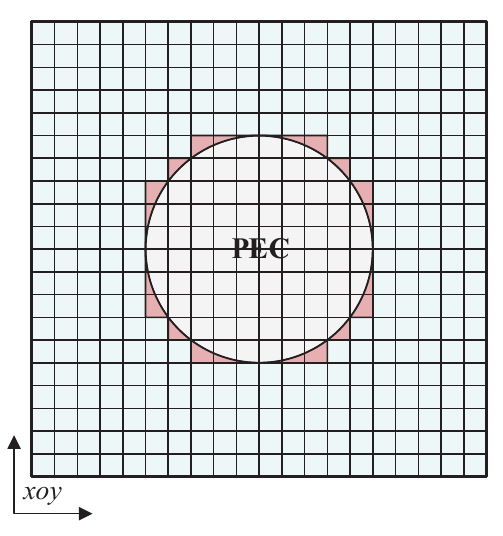}
}
\caption{(a) The geometrical configuration of the PEC cylinder and the PEC cavity, (b) the Yee's grid with cell size 0.1$m$ at $z=1m$.}
\label{Fig.2}
\end{figure}

Fig. \ref{Fig.2} shows the geometrical configuration to validate the numerical stability. A cubic cavity filled with air is considered, and its boundaries are the perfect electric conductor. The size of the cavity is $2m\times2m\times2m$. A PEC cylinder is placed at the middle of the cavity, and its diameter and height are both 1$m$. 

Three grids with cell sizes 0.25$m$, 0.125$m$, and 0.1$m$ are used to discrete the structures, and a horizontal view of Yee's grid with cell size 0.1$m$ at $z = 1m$ is shown in Fig. \ref{Fig.2}(b). It is easy to find that many partially filled cells, which are marked in light red, exist. Once the conformal meshes are generated, the coefficients in (\ref{E15})-(\ref{E16}) can be calculated, and then all the eigenvalues of $\bf  \Lambda$ can be numerically computed through the Matlab command ``eig($\cdot$)''. 

\begin{figure}[h]
\centering
\subfigure[CFLN=1]{
\includegraphics[scale=0.275]{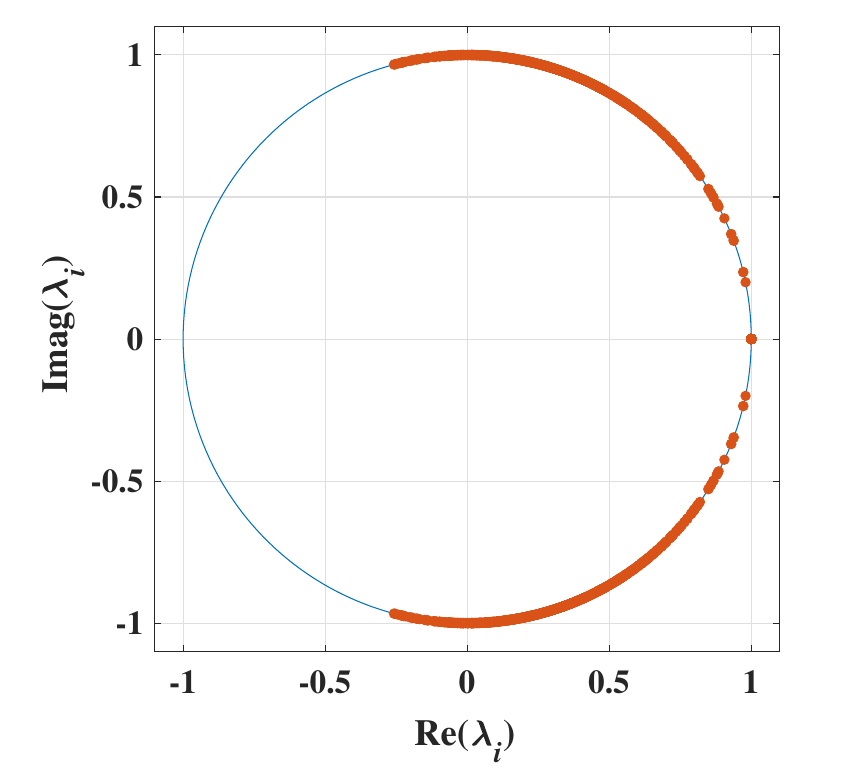}
%\caption{fig1}
}
\subfigure[CFLN=4]{
\includegraphics[scale=0.275]{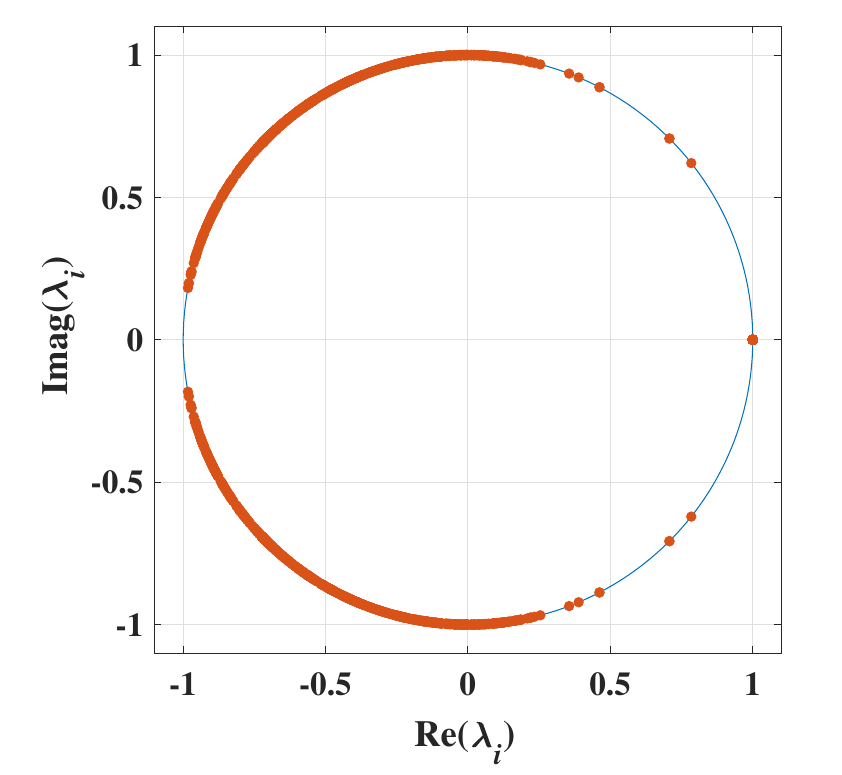}
}
\\ 
\centering
\subfigure[CFLN=8]{
\includegraphics[scale=0.275]{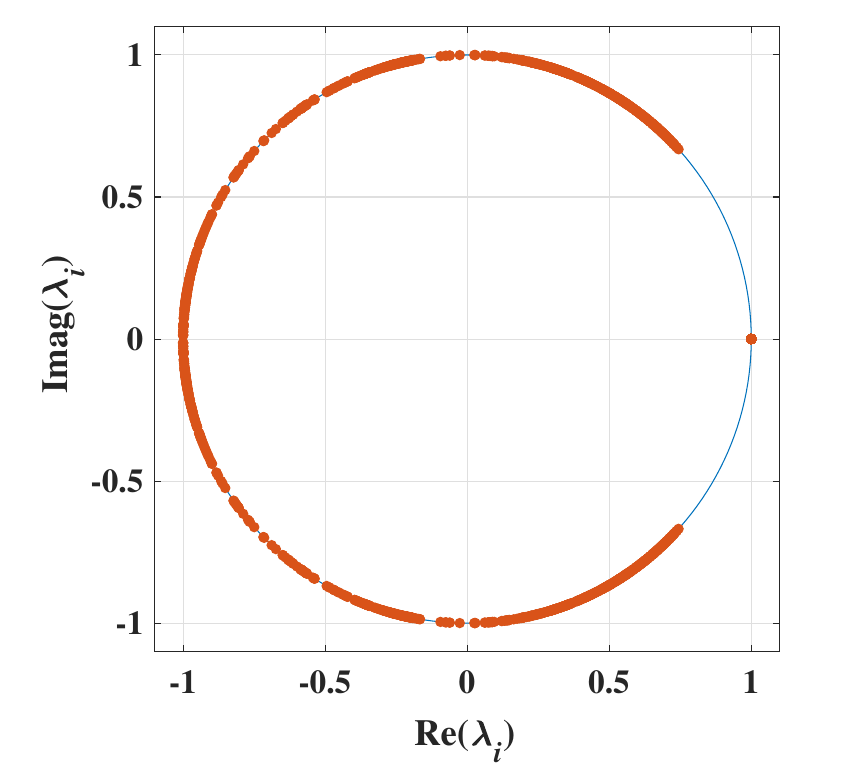}
}
\subfigure[CFLN=64]{
\includegraphics[scale=0.275]{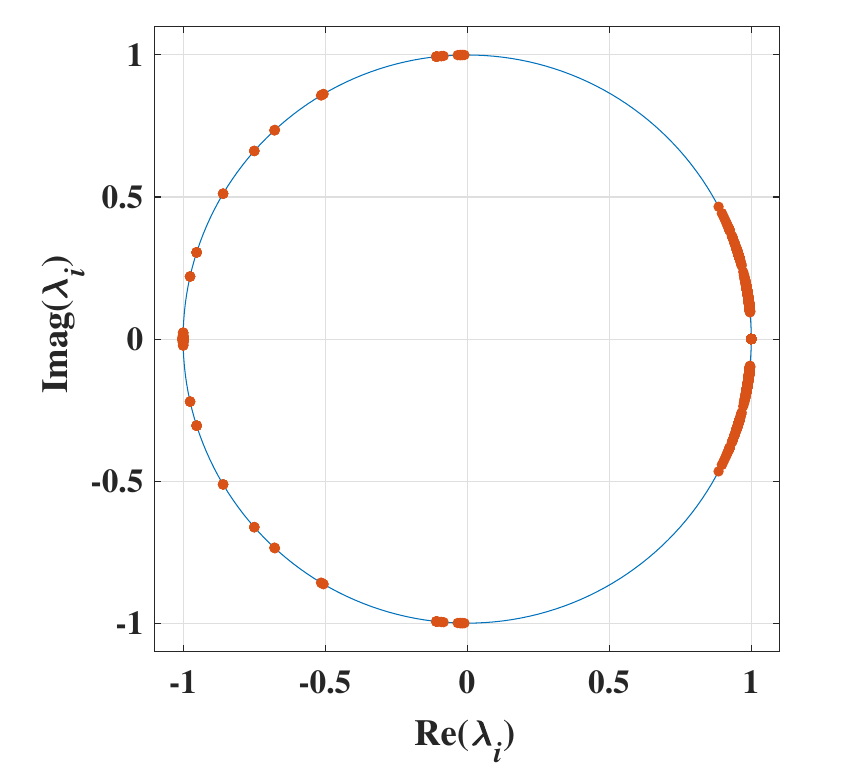}
}
\caption{The eigenvalues of $\bf  \Lambda$ for the proposed CLOD-FDTD method with mesh size 0.25$m$ for CFLN = 1, 4, 8, and 64, respectively.}
\label{Fig.3}
\end{figure}

\begin{figure}[h]\!\!\!\!\!\!
\centering
\subfigure[CFLN=1]{
\includegraphics[scale=0.25]{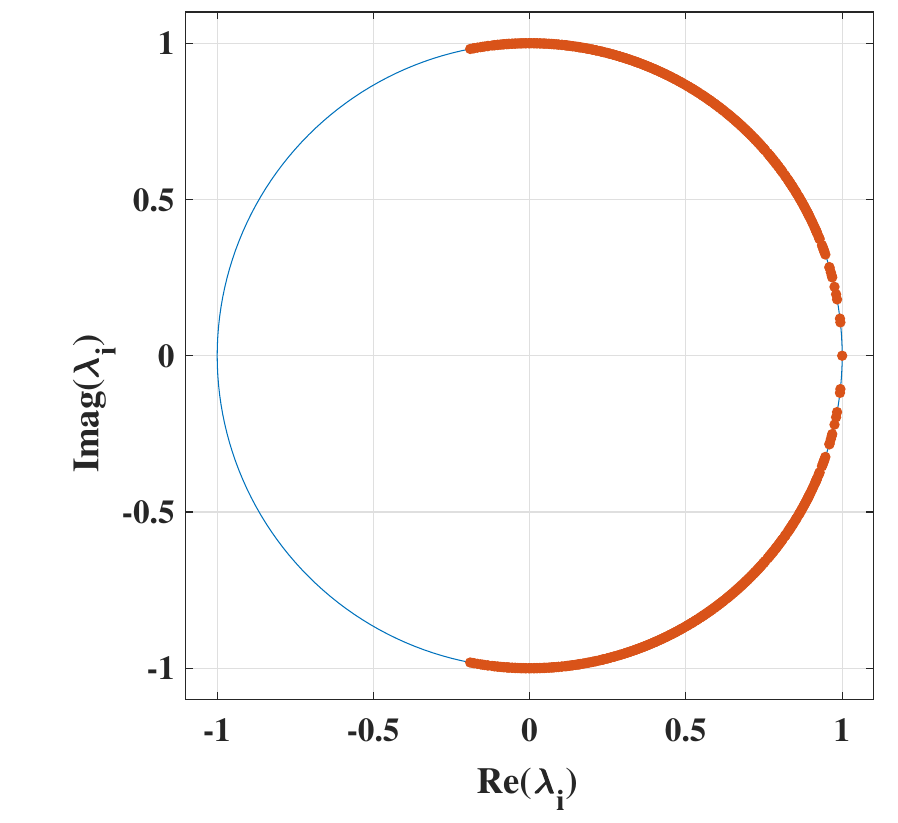}
}
\subfigure[CFLN=4]{
\includegraphics[scale=0.25]{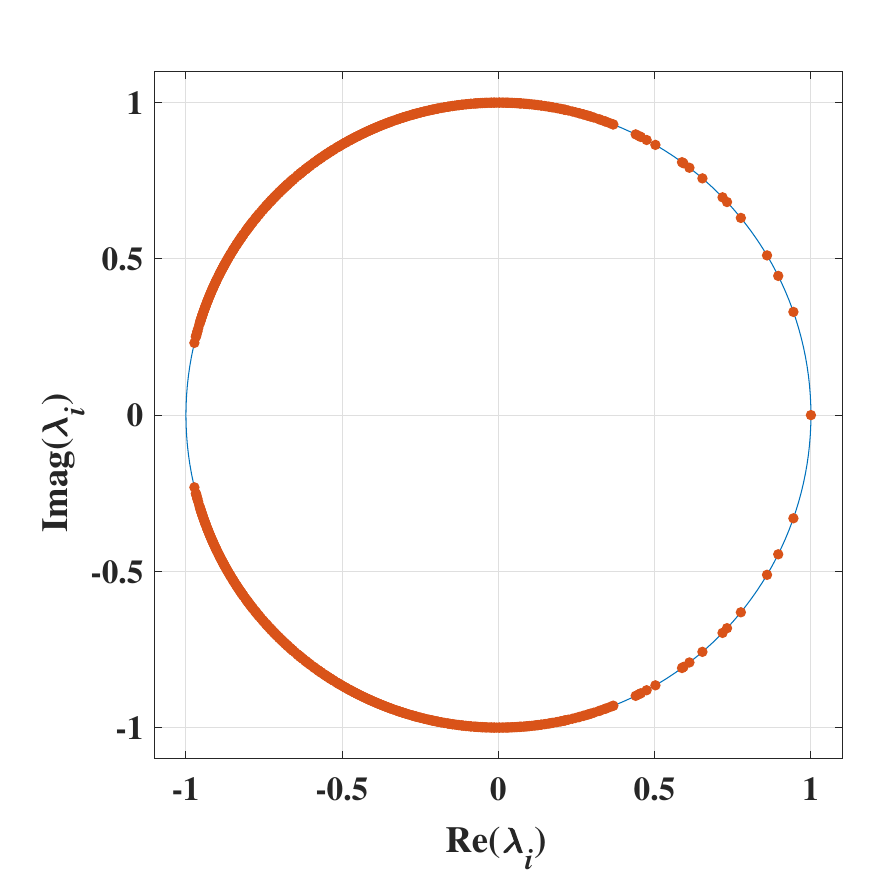}
}
\\
\centering\!\!\!\!\!\!
\subfigure[CFLN=8]{
\includegraphics[scale=0.25]{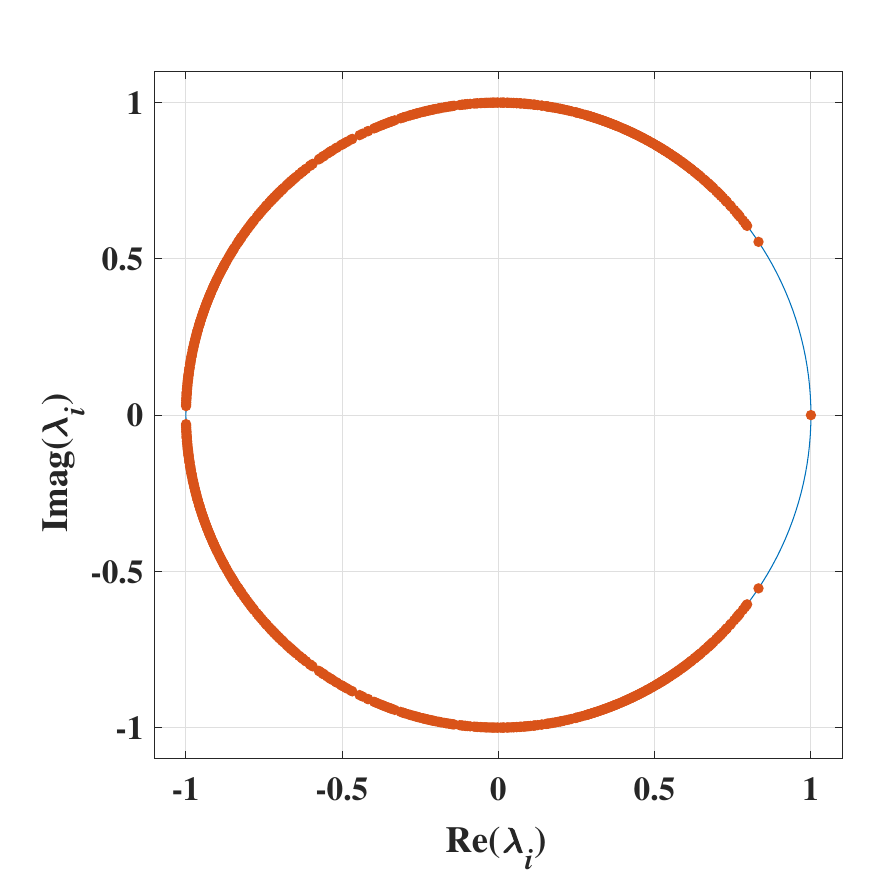}
}
\subfigure[CFLN=64]{
\includegraphics[scale=0.25]{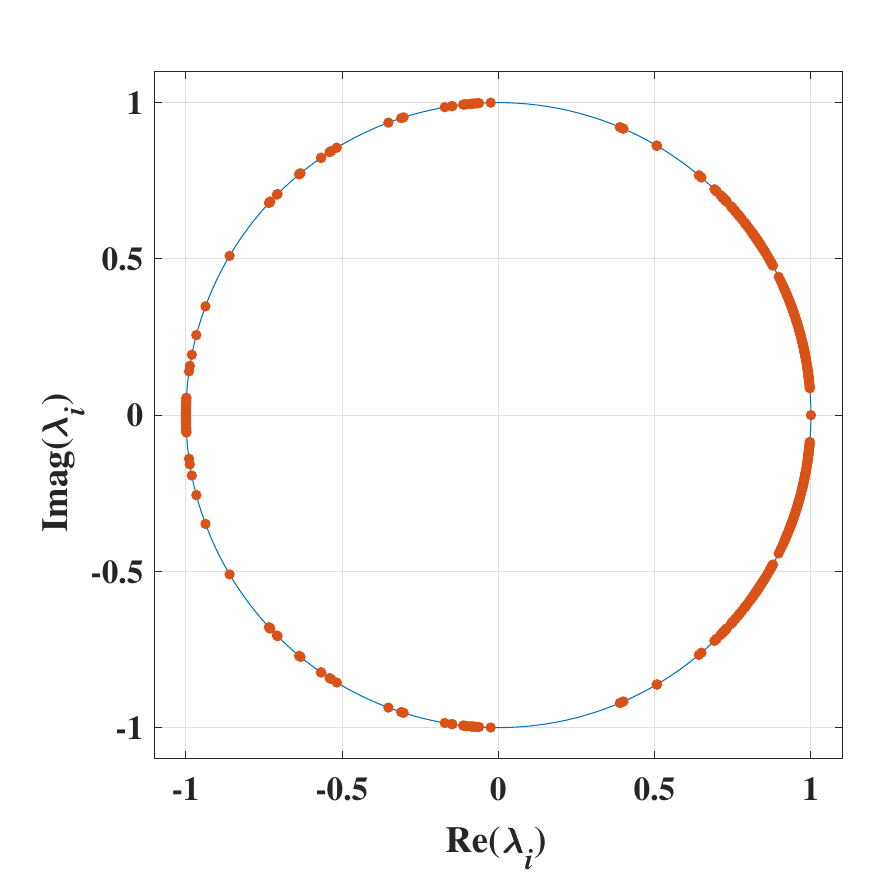}
}
\caption{The eigenvalues of $\bf  \Lambda$ for the proposed CLOD-FDTD method with mesh size 0.125$m$ for CFLN = 1, 4, 8, and 64, respectively.}
\label{Fig.4}
\end{figure}

\begin{figure}[h]
\centering
\subfigure[CFLN=1]{
\includegraphics[scale=0.275]{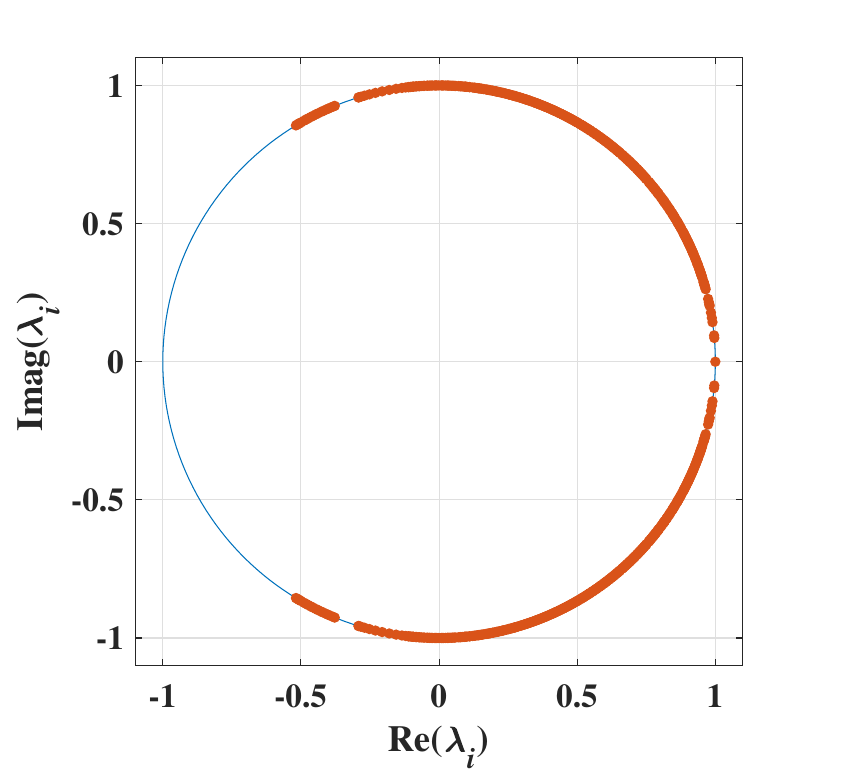}
}
\subfigure[CFLN=4]{
\includegraphics[scale=0.275]{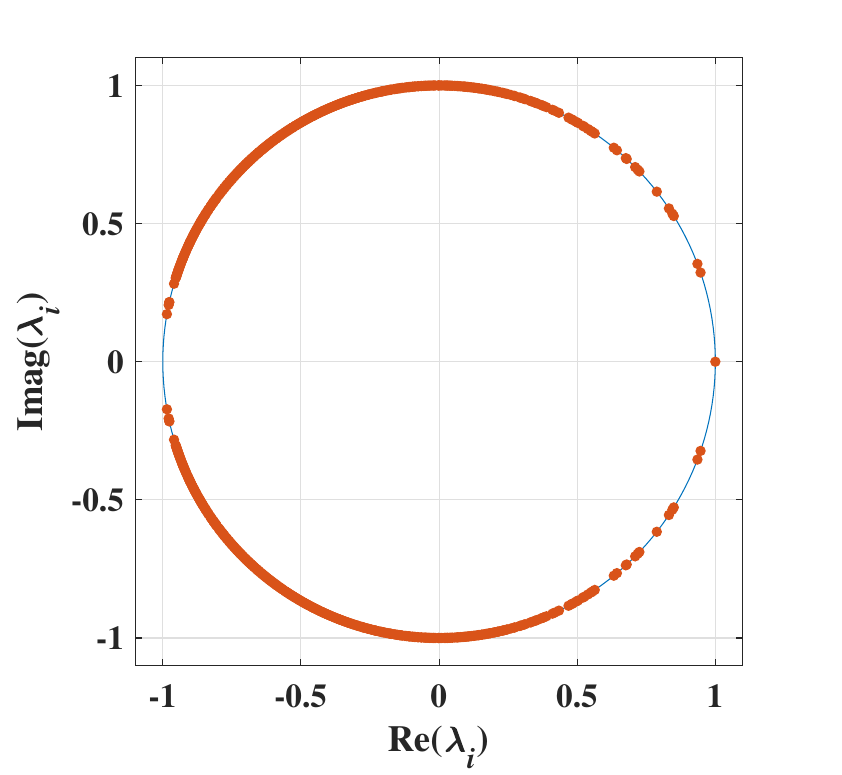}
}
\\
\centering
\subfigure[CFLN=8]{
\includegraphics[scale=0.275]{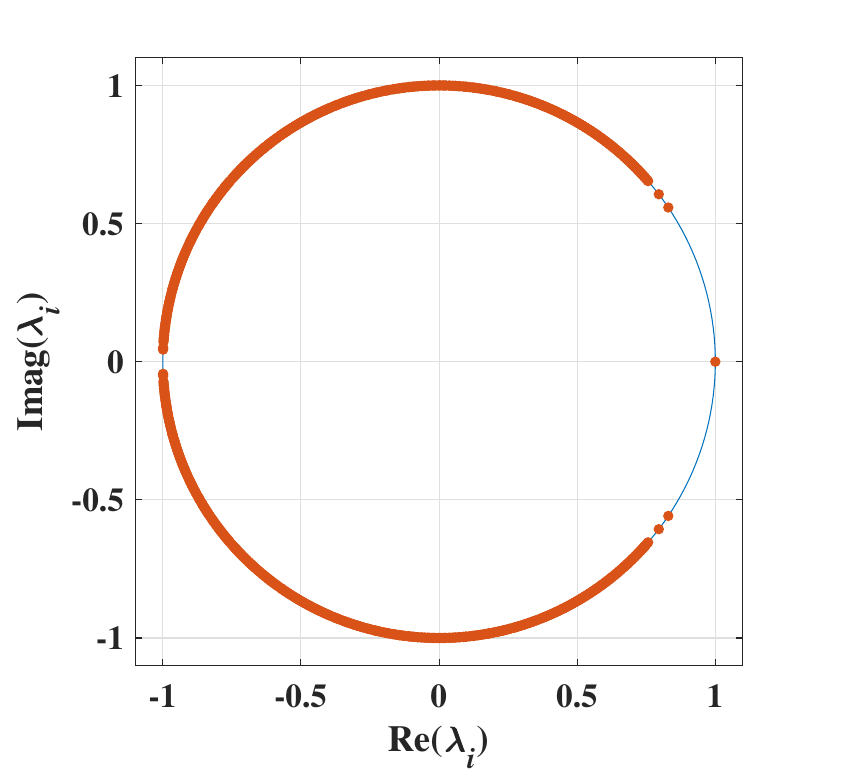}
}
\subfigure[CFLN=64]{
\includegraphics[scale=0.275]{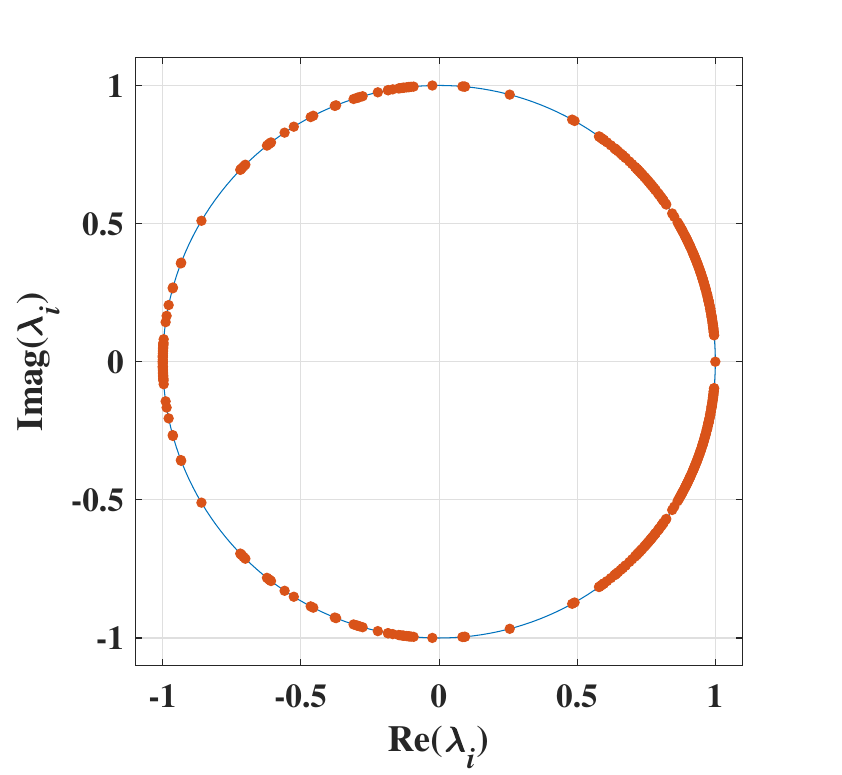}
}
\caption{The eigenvalues of $\bf  \Lambda$ for the proposed CLOD-FDTD method with mesh size 0.1$m$ for CFLN = 1, 4, 8, and 64, respectively.}
\label{Fig.5}
\end{figure}

Figs. \ref{Fig.3}-\ref{Fig.5} show the eigenvalues of $\bf  \Lambda$ for the proposed CLOD-FDTD method with different CFLNs. CFLN is the ratio of time step $\Delta t$ used in the simulation to the maximum time step $\Delta t_{\text{max}}$ defined by the CFL condition. $\Delta t_{\text{max}}$ is given by
\begin{equation}\label{E22}
\Delta t_{\text{max}} = \frac{\text{1}}{{c \sqrt {{{{{\left( {\Delta x} \right)}^{-2}}}} + {{{{\left( {\Delta y} \right)}^{-2}}}} + {{{{\left( {\Delta z} \right)}^{-2}}}}} }},
\end{equation}
where $c$ denotes the electromagnetic wave velocity in the air. In each grid size, four scenarios with CFLN = 1, 4, 8 and 64 are shown in this paper. It can be found that all the eigenvalues fall on the unit circle in Figs. \ref{Fig.3}-\ref{Fig.5}. Although only twelve specific scenarios are presented in this subsection, we have done a number of numerical verification to cover the general situations in the practical simulations. Without exception, all the eigenvalues of $\bf  \Lambda$ fall on the unit circle. Therefore, we can safely draw the conclusion that the proposed conformal LOD-FDTD method is unconditionally stable.  

\begin{figure}
	\centering
	\subfigure[CFLN=0.25]{
		\includegraphics[scale=0.22]{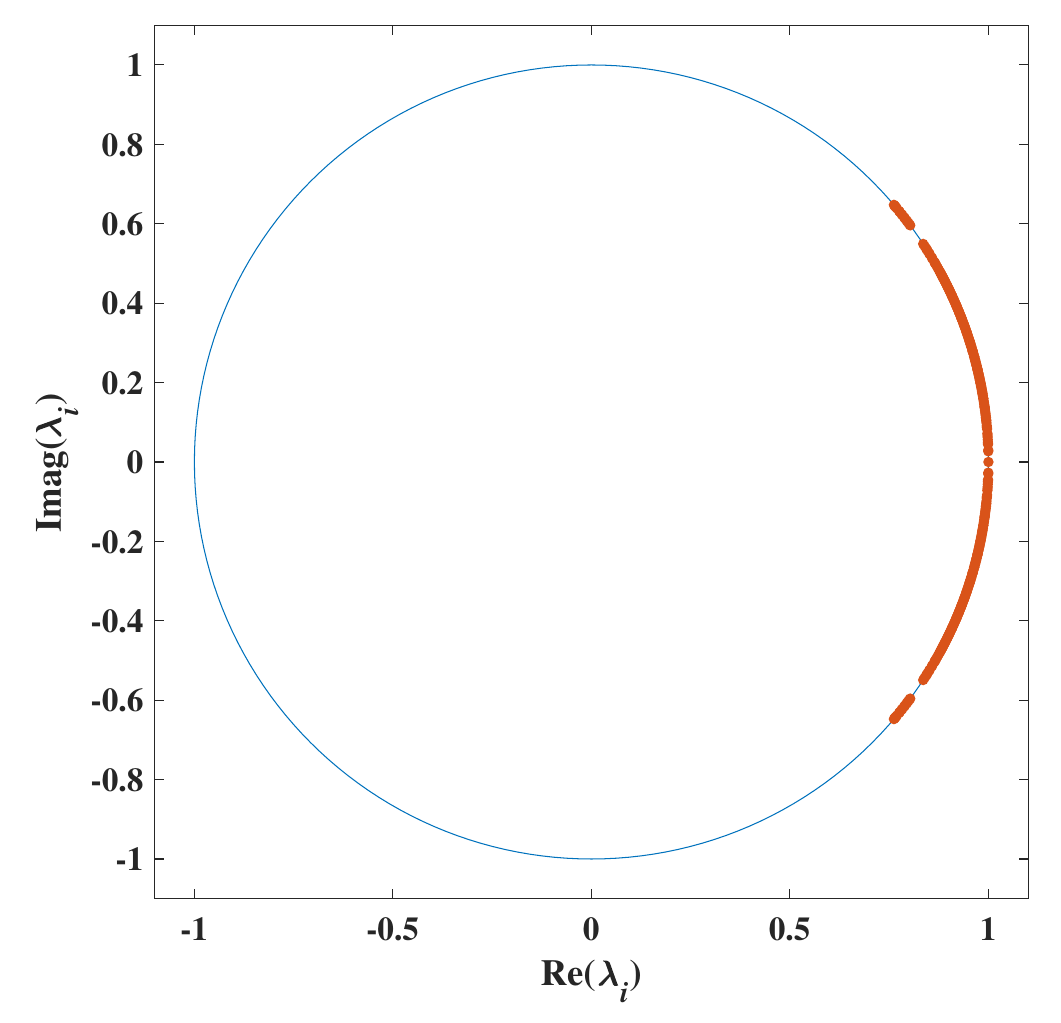}
	}
	\subfigure[CFLN=0.5]{
		\includegraphics[scale=0.22]{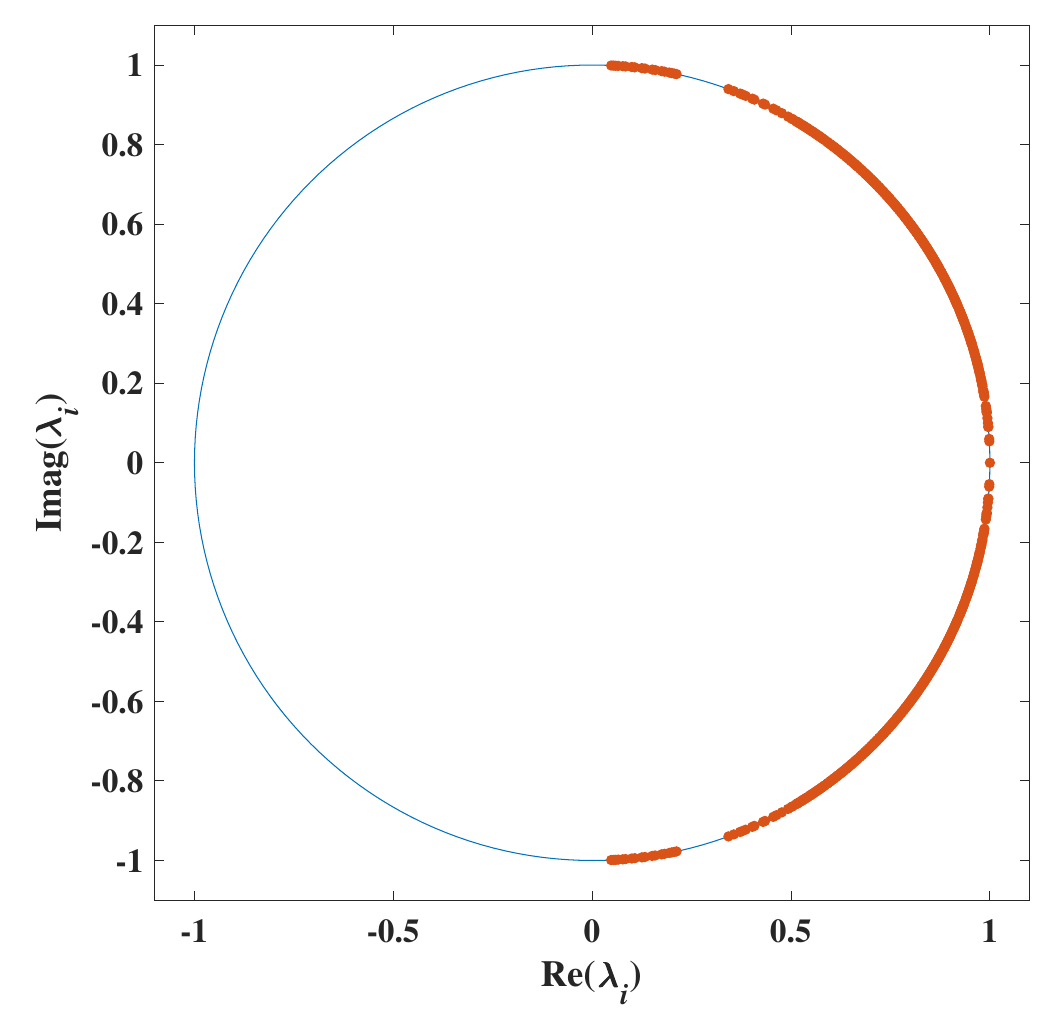}
	}
	\\
	\subfigure[CFLN=0.75]{
		\includegraphics[scale=0.18]{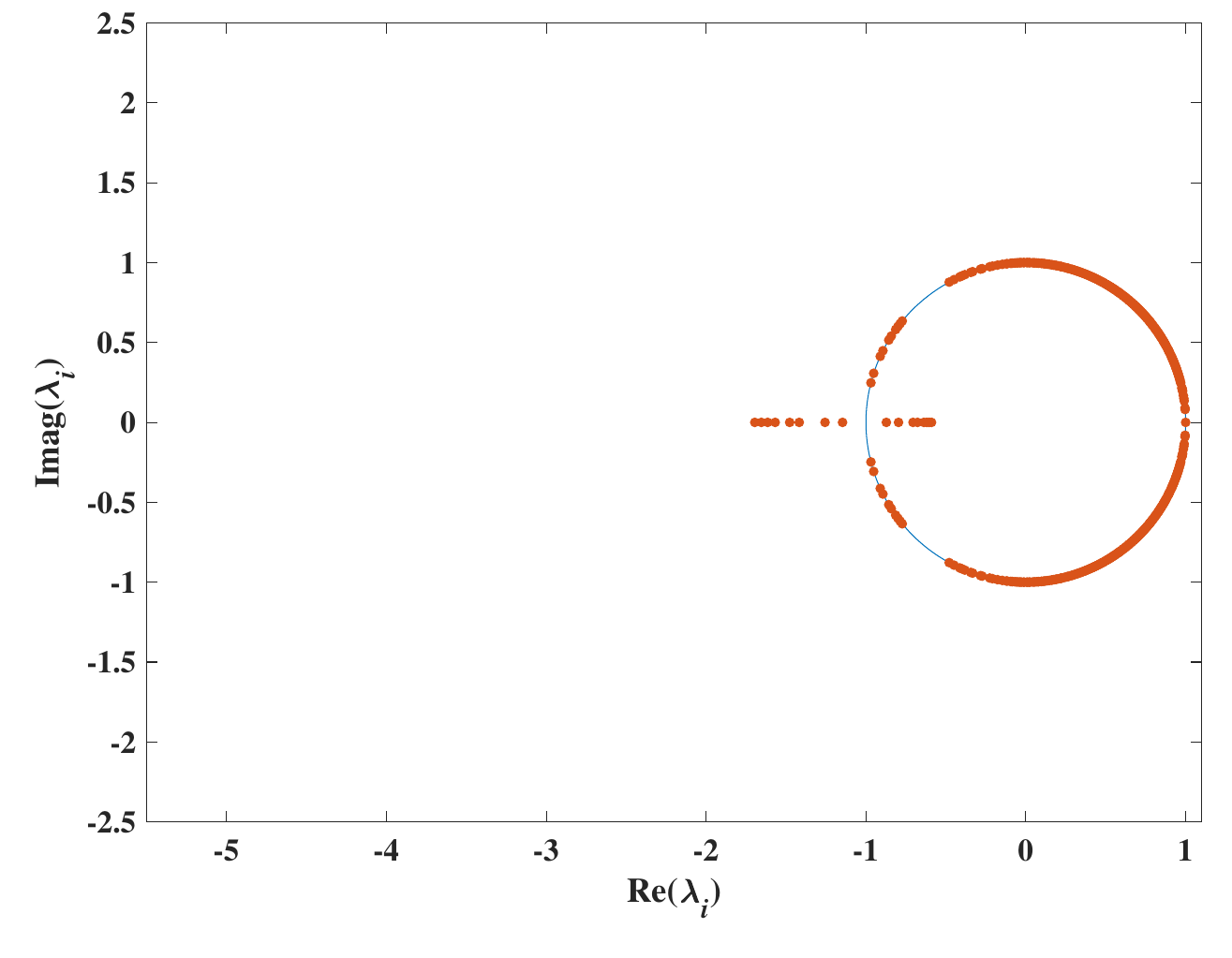}
	}
	\subfigure[CFLN=1.0]{
		\includegraphics[scale=0.18]{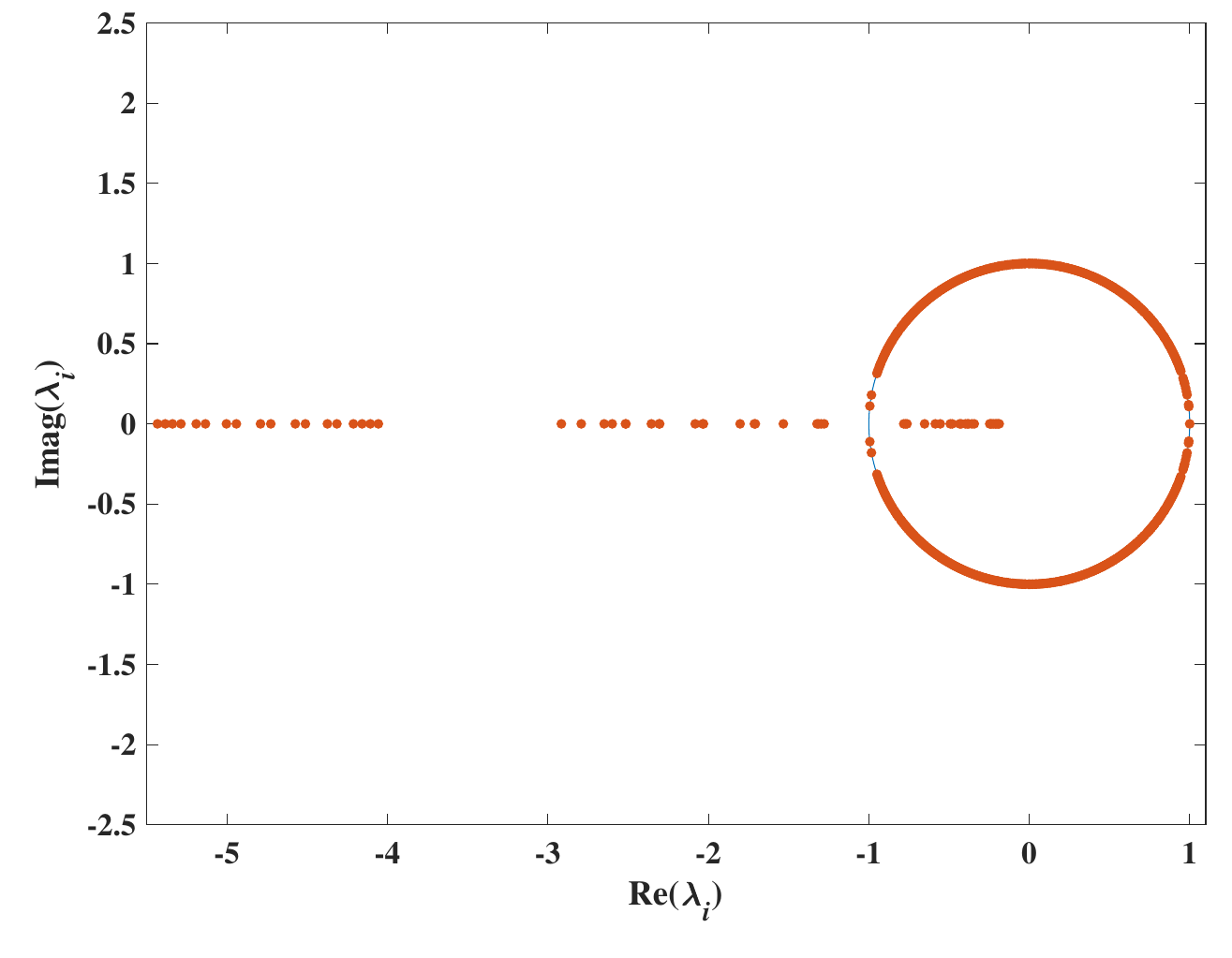}
	}
	\caption{The eigenvalues of $\bf  \Lambda$ in the CFDTD method with CFLN = 0.25, 0.5, 0.75 and 1.0, respectively.}
	\label{Fig.6}
\end{figure}

\subsection{Stability Analysis of the CFDTD Method}
For comparison purposes, we further analyze the stability of the traditional CFDTD method based on \cite{CFDTD} in a similar manner. By taking $E_x,H_z$ as examples, the time-marching formulations of the CFDTD method can be expressed as 
\begin{subequations}
\begin{equation}\label{E26}
\begin{aligned}
	E_x|_{i+\frac{1}{2},j,k}^{n+1}&=E_x|_{i+\frac{1}{2},j,k}^{n}\\
	&+\frac{{\Delta}t}{{\varepsilon}{\Delta}y}\left(H_z|_{i+\frac{1}{2},j+\frac{1}{2},k}^{n-\frac{1}{2}}-H_z|_{i+\frac{1}{2},j-\frac{1}{2},k}^{n-\frac{1}{2}}\right)\\
	&-\frac{{\Delta}t}{{\varepsilon}{\Delta}z}\left(H_y|_{i+\frac{1}{2},j,k+\frac{1}{2}}^{n-\frac{1}{2}}-H_y|_{i+\frac{1}{2},j,k-\frac{1}{2}}^{n-\frac{1}{2}}\right),
\end{aligned}
\end{equation}
\begin{align}
	H_z|_{i+\frac{1}{2},j+\frac{1}{2},k}^{n+\frac{1}{2}}&=H_z|_{i+\frac{1}{2},j+\frac{1}{2},k}^{n-\frac{1}{2}}  \\\notag
	&+\frac{{\Delta}t{\Delta}y}{{\mu}S_{xy}|_{i+\frac{1}{2},j+\frac{1}{2},k}}{\delta}y\left(l_x|_{i+\frac{1}{2},j,k}E_x|_{i+\frac{1}{2},j,k}^{n}\right)\\ 
	&-\frac{{\Delta}t{\Delta}x}{{\mu}S_{xy}|_{i+\frac{1}{2},j+\frac{1}{2},k}}{\delta}x\left(l_y|_{i,j+\frac{1}{2},k}E_y|_{i,j+\frac{1}{2},k}^{n}\right), \notag
\end{align}
\end{subequations}
where $\delta_x,\delta_y$ denote the differential operators in the $x$- and $y$-directions, respectively, and $l_x,l_y$, $S_{xy}$ are conformal coefficients, which are exactly the same as those in the proposed CLOD-FDTD method. It should be noted that the electric and magnetic components in the CFDTD method are assigned at different time instances. By using (\ref{E18}), ${\bf{U}}^{n}$ for the CFDTD method is slightly modified as
\begin{equation}\label{E28}
\begin{aligned}
{\bf{U}}^{n}&=\left[{\bf{E}}_x^n,	{\bf{E}}_y^n, {\bf{E}}_z^n, {\bf{H}}_x^{n-\frac{1}{2}},{\bf{H}}_y^{n-\frac{1}{2}}, {\bf{H}}_z^{n-\frac{1}{2}}\right]^T.
\end{aligned}
\end{equation}
The amplification matrix ${\bf{\Lambda}}_{ {\text{CFDTD}} }$ of the CFDTD method can be obtained as 
\begin{equation}\label{E29}
\begin{aligned}
{\bf{U}}^{n+1}
=
{\bf{\Lambda}}_{ {\text{CFDTD}} } {\bf{U}}^{n}.
\end{aligned}
\end{equation}

The structure in Fig. \ref{Fig.2} is still used to illustrate the stability. Yee's grids with the mesh size of 0.125$m$ are used to discrete the structure. All eigenvalues with CFLN = 0.25, 0.5, 0.75, and 1.0 are shown in Fig. \ref{Fig.6}(a)-(d). 

It can be found that from Figs. \ref{Fig.6}(a) and (b), the eigenvalues fall on the unit circle. Therefore, the CFDTD is stable when CFLN is less than 0.5 with the mesh size being 0.125$m$. However, some eigenvalues fall outside the unit circle in Figs. \ref{Fig.6}(c) and (d), which implies that the CFDTD method is unstable if the time step increases to 0.75 times of $\Delta t_{\text{max}}$. The stability of the CFDTD method strongly depends on $S_{yz},S_{xz},S_{xy}$ and $l_x,l_y,l_z$. The CFLNs must be decreased as the ratio $l_{max}/S_{min}$ decrease. Therefore, when the finer meshes are used, the CFLNs should be smaller. To obtain stable numerical results, quite small time steps may be required, which is undesirable in practical simulations. In addition, if the CFDTD is unstable, the maximum modulus value of the eigenvalues also increases as CFLN increases, which makes the CFDTD solution divergent faster as time steps increase.

\subsection{Numerical Verification}
\begin{figure}[h]
	\centerline{{\includegraphics[scale=0.35]{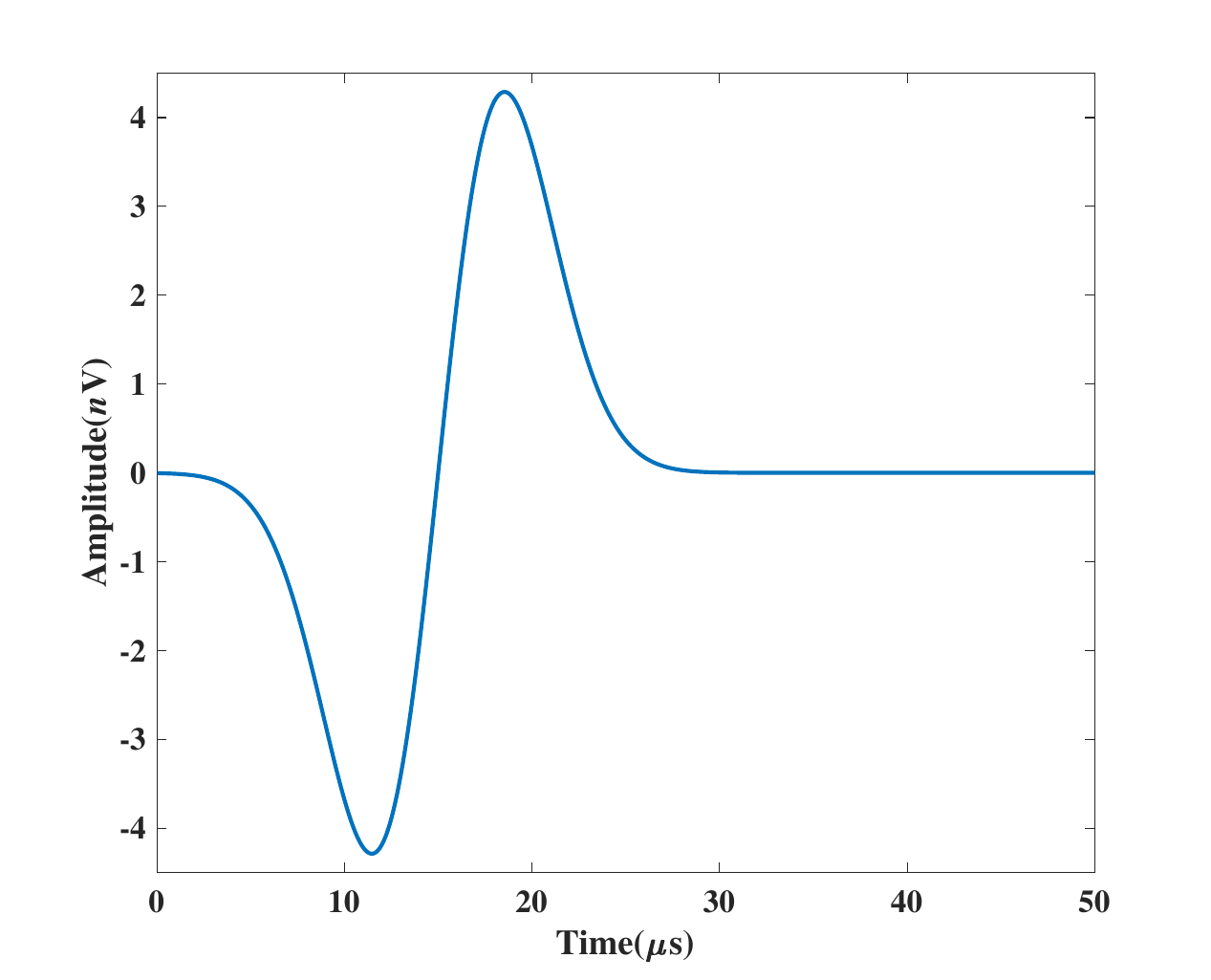}}}
	\caption{The differentiated Gaussian pulse used as the excitation.}
	\label{Fig.Gaussian}
\end{figure}

To further illustrate the stability of the proposed CLOD-FDTD method and the CFDTD method, a series of numerical simulations are performed with different CFLNs. In these simulations, the grids with cell size of 0.05$m$ are used to discretize a cubic cavity with a PEC cylinder, as shown in Fig. \ref{Fig.2}. The overall simulation time is 36$\mu s$ when CFLN = 1, 4, 8, and 64. A line differentiated Gaussian pulse current source along the z-direction, as shown in Fig. \ref{Fig.Gaussian}, is located at (0.5, 0.5, 1.0)[$m$], and an observation point is located at (0.5, 1.5, 1.0)[$m$].

\begin{figure}[h]
\centering
\subfigure[CFLN=1]{
\includegraphics[scale=0.18]{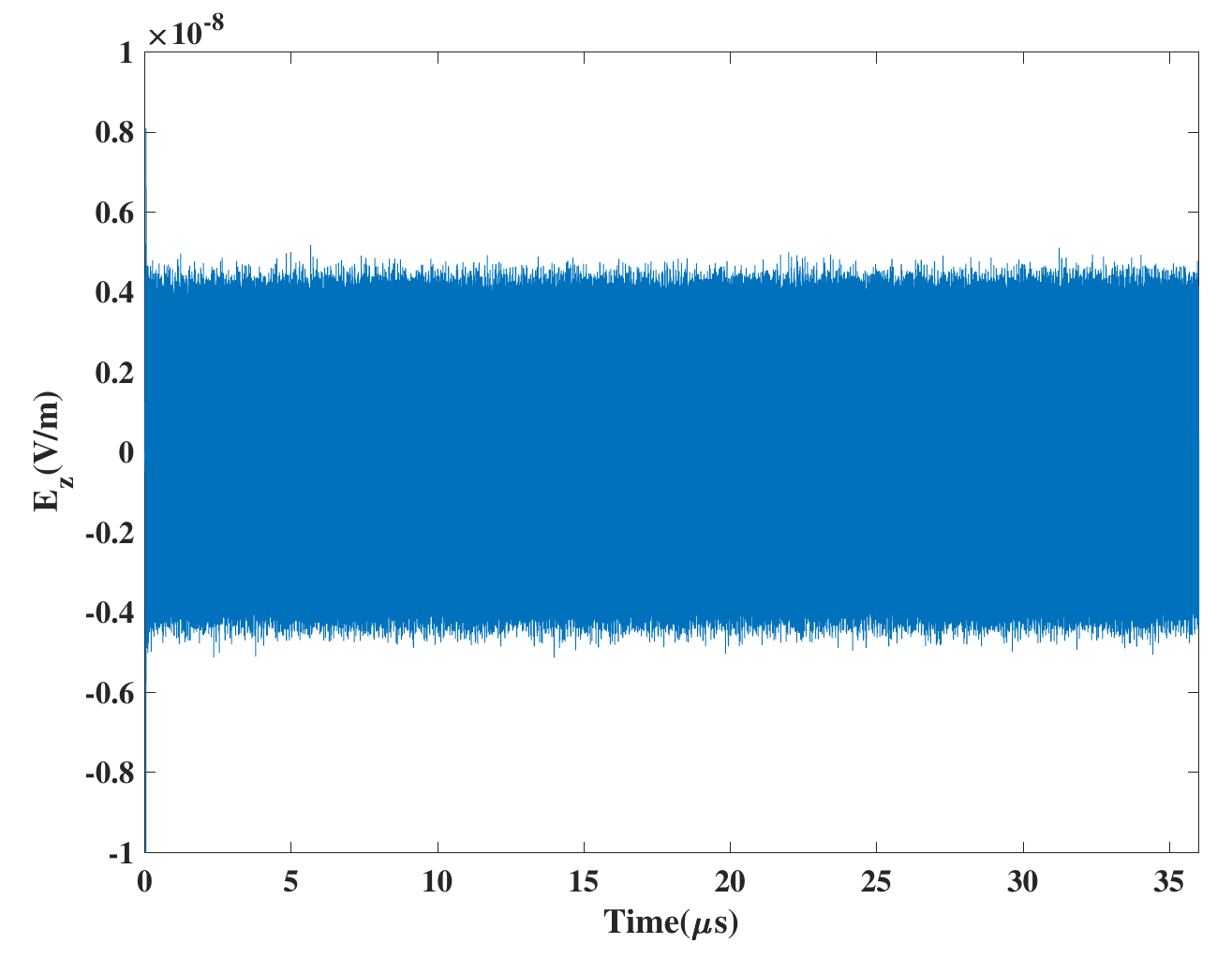}
}
\subfigure[CFLN=4]{
\includegraphics[scale=0.18]{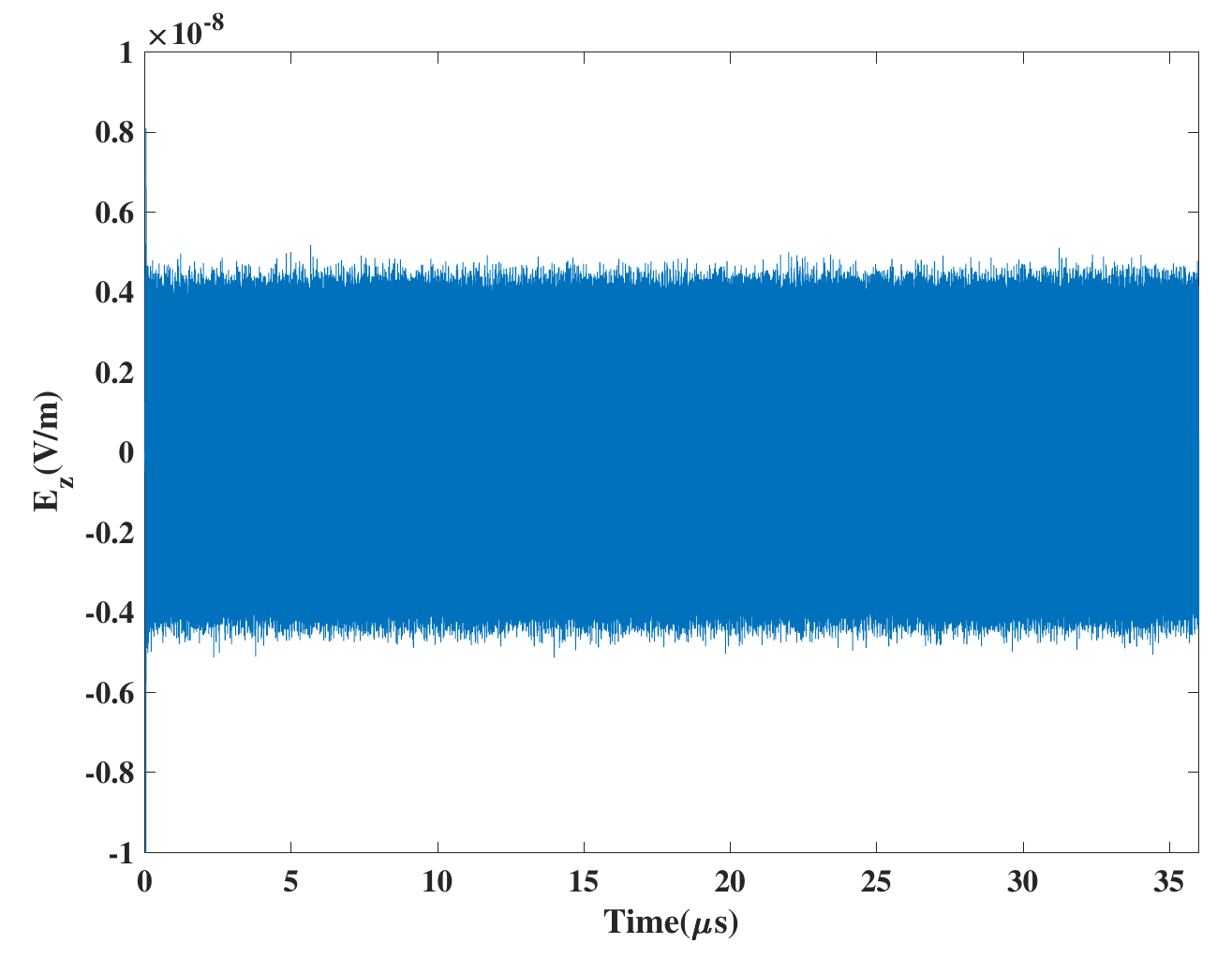}
}
\\
\subfigure[CFLN=8]{
\includegraphics[scale=0.18]{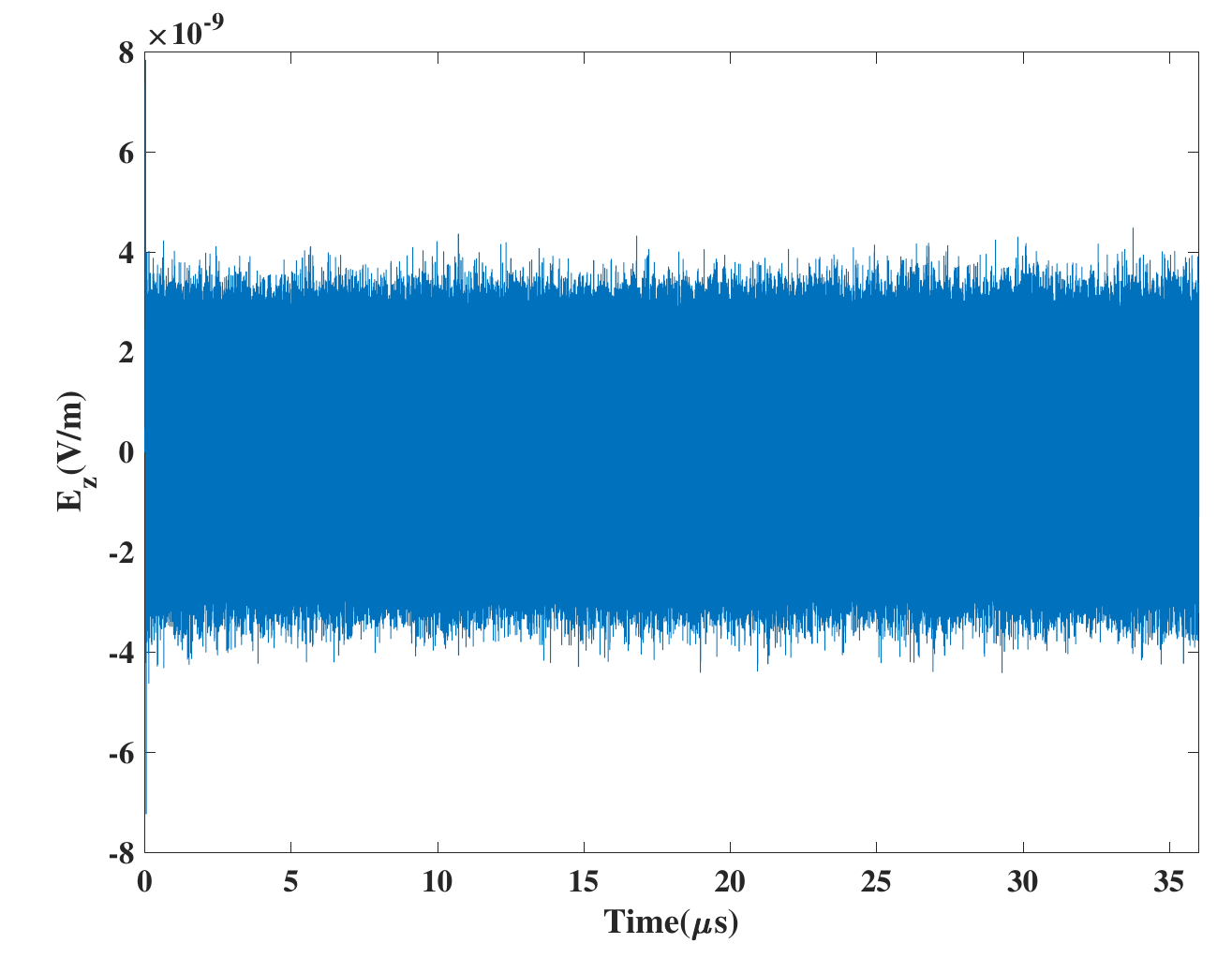}
}
\subfigure[CFLN=64]{
\includegraphics[scale=0.18]{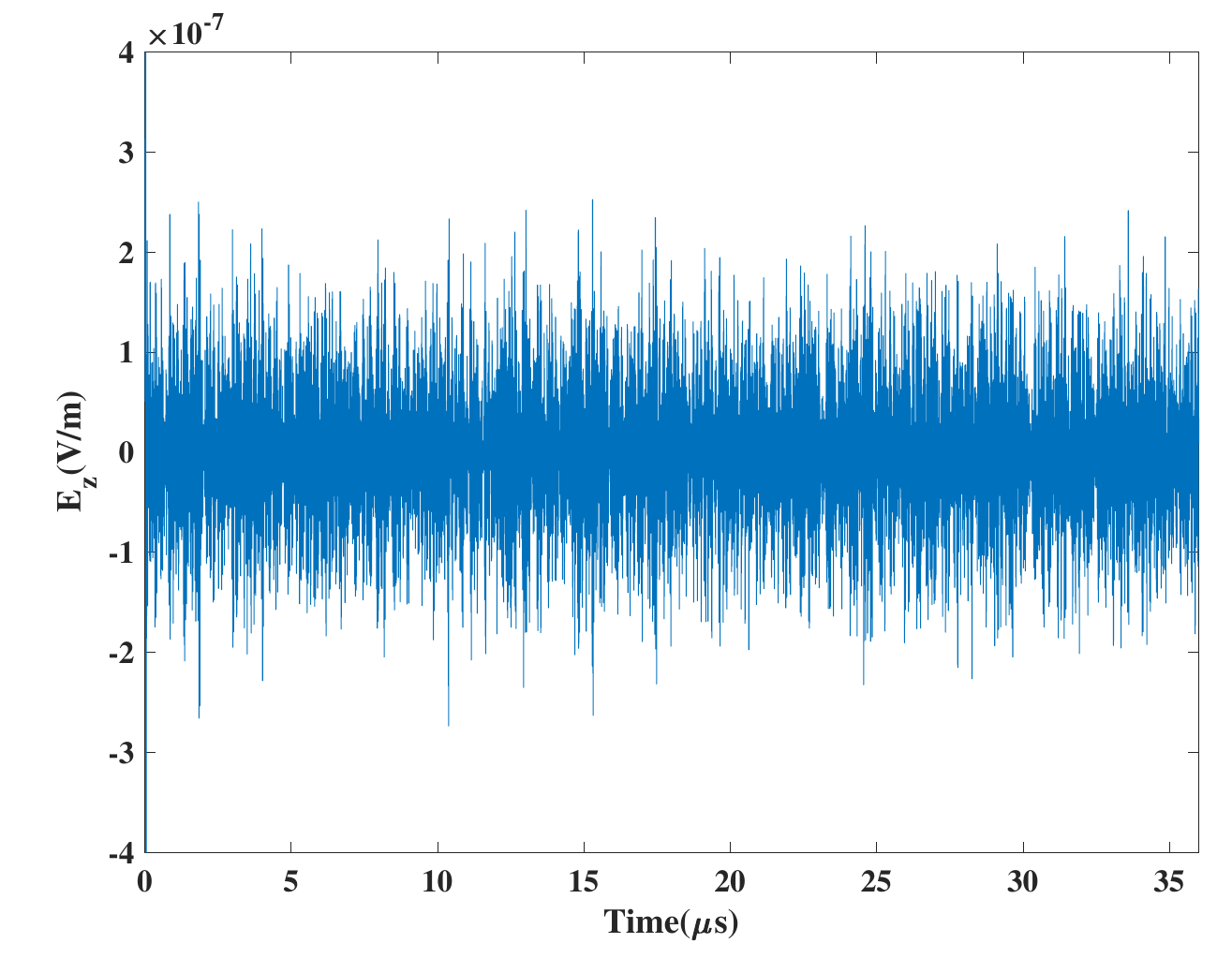}
}
\caption{The $E_z$ field values verse time steps with CFLN = 1, 4, 8 and 64, respectively.}
\label{Fig.7}
\end{figure}

Fig. \ref{Fig.7}(a)-(d) show $E_z$ verse time with CFLN = 1, 4, 8, and 64, respectively. The overall time steps are 373,864, 93,466, 46,733, and 5,841 when CFLN = 1, 4, 8 and 64, respectively. It can be found that all the simulations are stable even if quite large time steps are used, which confirms our conclusion through the eigenvalue analysis. 

In addition, it can be found that the results obtained by the CLOD-FDTD method show some difference when different CFLNs are used. The reason is that the dispersion error is diverse for different CFLNs \cite{dispersion-ADI1,dispersion-ADI2,dispersion-FDTD}. According to \cite{dispersion-LOD}, as CFLNs increase, the dispersion error in the LOD-FDTD method also increases. Therefore, results obtained with CFLN = 64 have large error comparing with other situations. Since the long time step with CFLN = 64 will cause serious impact on stability, we choose this CFLN to prove the strong stability of the proposed CLOD-FDTD method.

\begin{figure}[h]
\centerline{{\includegraphics[scale=0.4]{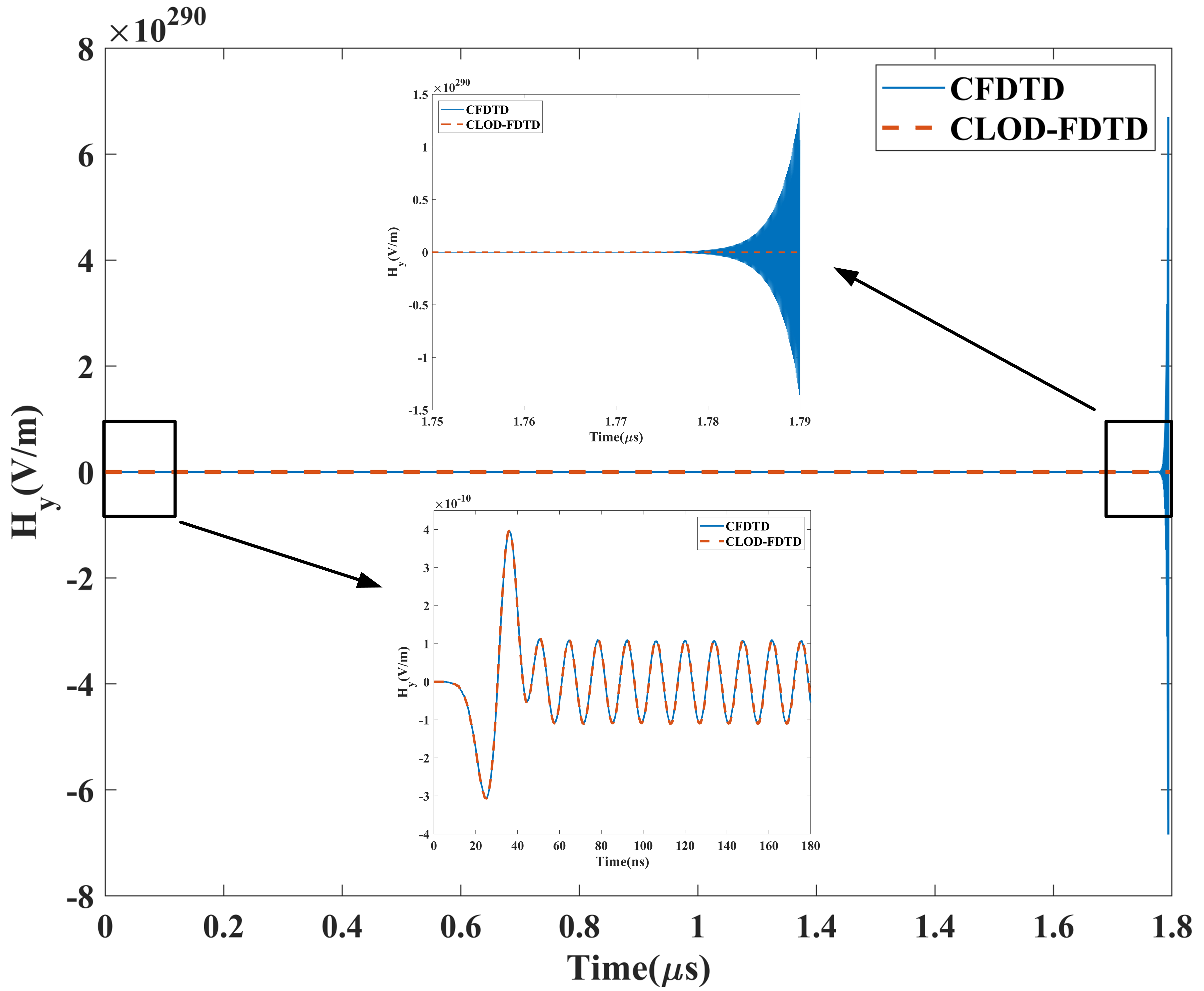}}}
\caption{The $E_z$ field values verse time step obtained from the proposed LOD-FDTD method with CFLN = 1 and the CFDTD method with CFLN = 0.51.}
\label{Fig.8}
\end{figure}

We performed another numerical simulation for comparison, where the CFDTD method with CFLN = 0.51 and the CLOD-FDTD method with CFLN = 1 are used. The overall time is 1.8$\mu s$, which requires 36,975 time steps to complete this simulation in the CFDTD method and 18,693 time steps in the proposed CLOD-FDTD method. To make the comparison fair for the CLOD-FDTD method and the CFDTD method, the same grids with cell size 0.05$m$ are used. Fig. \ref{Fig.8} shows results obtained from the CFDTD method and the proposed CLOD-FDTD method. It is easy to find that from 0$\mu s$ to 0.18$\mu s$, results obtained from the two methods agree well with each other. However, results obtained from the CFDTD method gradually divergent, and the simulation becomes unstable in the end. For the proposed CLOD-FDTD method, the simulation is always stable with no sign of instability, which confirms our previous analysis. 

Here we only present some numerical results for a specific example. In addition, we carried out a number of numerical simulations and eigenvalue analyses, which confirm our analysis. In summary, all the analyses mentioned above show that the conformal LOD-FDTD method is unconditionally stable for curved PEC objects.

\section{NUMERICAL RESULTS AND DISCUSSION}
In this section, two numerical examples are carried out to validate the accuracy and efficiency of the proposed CLOD-FDTD method. Results computed by the FDTD method and the LOD-FDTD method are also provided for comparison. All the in-house codes are written using C++ and linked in the release mode with exactly the same configurations. They are run on a workstation including an Intel Gold 6143 CPU with the operating frequency of 3.2 GHz and 256 GB memory.  For a fair comparison, all codes are run in a single thread.

\subsection{A PEC Cylinder Object in A Cavity}
\begin{figure}[h]
\centering
\subfigure[]{
\includegraphics[scale=0.46]{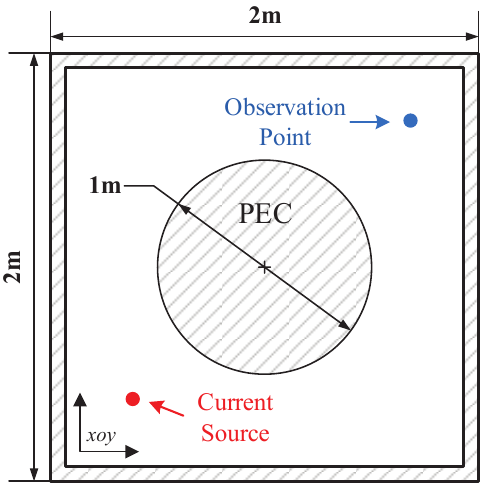}
}
\subfigure[]{
\includegraphics[scale=0.47]{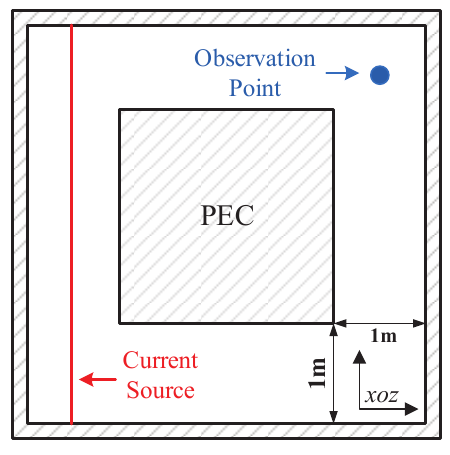}
}
\caption{The top-down and front-back views of the structure: (a) the top-down view, (b) the front-back view.}
\label{Fig.9}
\end{figure}

In this subsection, a PEC cylinder is considered to validate the accuracy of the proposed CLOD-FDTD method, as shown in the previous section. Its diameter and height are 1$m$. It is placed at the center of a PEC cubic cavity. The PEC cavity is filled with air, and its dimension is $2m\times2m\times2m$. The top-down and front-back views of this structure are shown in Figs. \ref{Fig.9}(a) and (b), respectively.
 
Two uniform meshes with cell sizes 0.05$m$ and 0.01$m$ are used to discretize the structure. Coarse meshes with cell size 0.05$m$ are used in the proposed CLOD-FDTD method, and both the coarse and fine meshes are used in the LOD-FDTD method and the FDTD method. 

A line differential Gaussian pulse current source $f=(t-t_0)e^{[-{(t-t_0)^2}/{\tau^2 }]}$ with $t_0=3\tau, \tau=10ns$ is used as an excitation function. Its location is (0.5,0.5,1.0)[$m$] in both coarse and fine meshes. The observation point is located at (1.5,1.5,1.0)[$m$]. The overall physical time is 2.4${\mu}s$.

\begin{figure}[h]
	\centerline{{\includegraphics[scale=0.4]{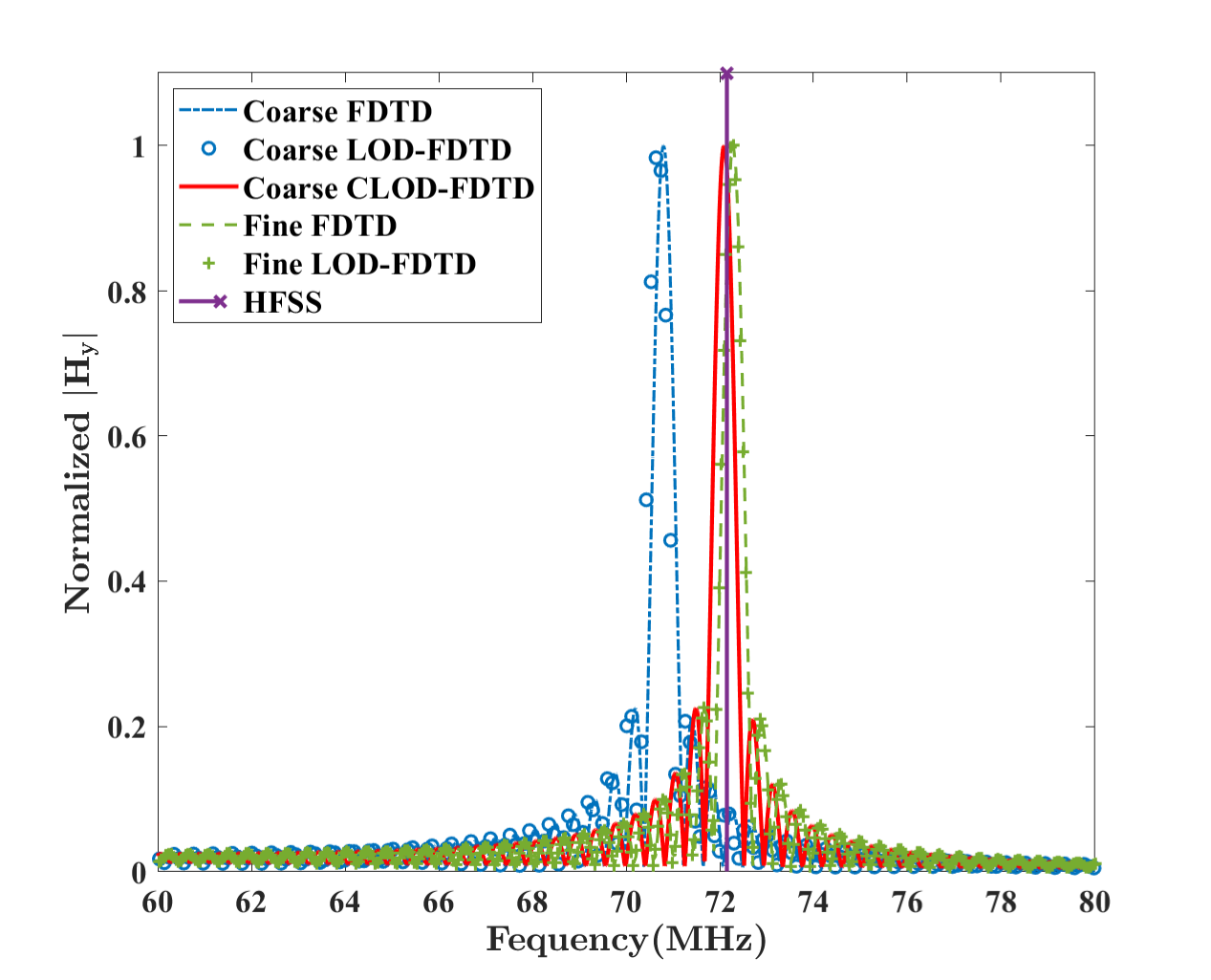}}}
	\caption{The normalized frequency responses of $H_y$ obtained from the FDTD method, the LOD-FDTD method, the proposed CLOD-FDTD method with CFLN = 1 and different mesh sizes, and the HFSS for the PEC cavity}
	\label{Fig.10}
\end{figure}

The normalized frequency responses of $H_y$ are shown in Fig. \ref{Fig.10} with CFLN = 1 for three methods. It can be found that results obtained from the LOD-FDTD method and the FDTD method show good agreement with both coarse and fine meshes. Therefore, when time steps and meshes used in the simulations are the same, the LOD-FDTD method and the FDTD method can achieve a similar level of accuracy. However, when coarse meshes with cell size 0.05$m$ are used in the proposed CLOD-FDTD method, its frequency response shows excellent agreement with those from the LOD-FDTD method and FDTD method with fine meshes of cell size 0.01$m$. Therefore, significant accuracy improvement can be achieved for the proposed CLOD-FDTD method. It should be noted that the time step used in the proposed CLOD-FDTD method is five times those for the other two methods since the cell size of coarse meshes (0.05$m$) is five times as that of fine meshes (0.01$m$). In order to compare the accuracy of various FDTD methods, the resonant frequency obtained by the HFSS based on the FEM is included, as shown in Fig. \ref{Fig.10}.

\begin{figure}[h]
\centerline{{\includegraphics[scale=0.3]{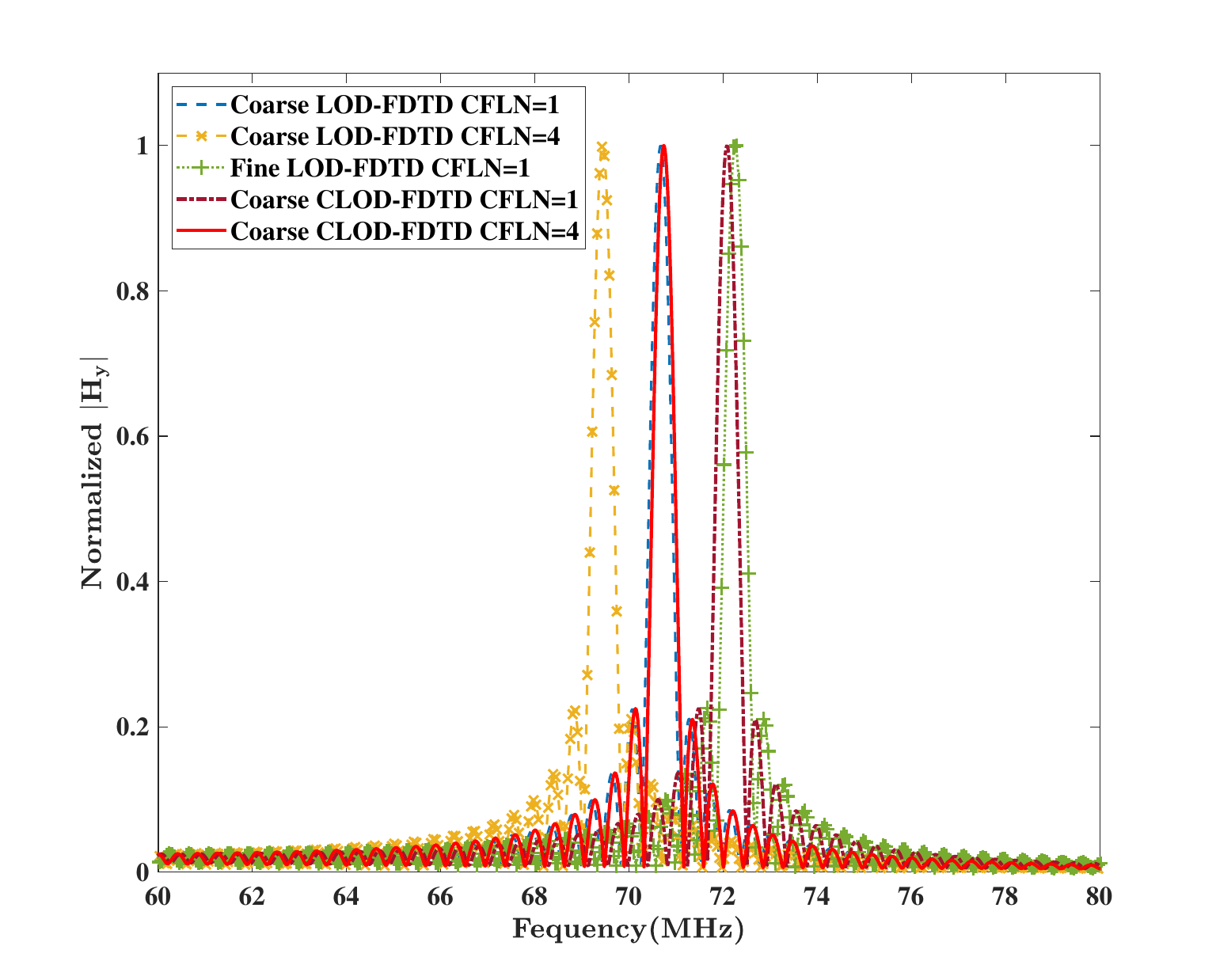}}}
\caption{The normalized frequency responses of $H_y$ obtained from the LOD-FDTD method and the CLOD-FDTD method with CFLN = 1 and 4.}
\label{Fig.11}
\end{figure}

Since the proposed CLOD-FDTD method is unconditionally stable, CPU time can be further reduced as CFLN increases. Fig. \ref{Fig.11} shows the normalized frequency responses obtained from the CLOD-FDTD method with CFLN = 1 and 4. Mesh size 0.05$m$ is used in the proposed CLOD-FDTD method. The reference results are obtained from the LOD-FDTD method with coarse meshes when CFLN = 1, 4, and the LOD-FDTD method with fine meshes when CFLN = 1. 

As shown in Fig. \ref{Fig.11}, results obtained from the CLOD-FDTD method with CFLN = 1 agree well with those obtained from the LOD-FDTD method with fine meshes of cell size 0.01$m$ and CFLN = 1. When CFLN = 1 and coarse meshes are used, the resonant frequency obtained from the LOD-FDTD method shows large discrepancy with the reference solutions. Furthermore, when CFLN = 4 and coarse meshes are used, results obtained from the LOD-FDTD method show even larger errors. Since the dispersion error in the LOD-FDTD method with CFLN = 4 is larger than that of the LOD-FDTD method with CFLN = 1, results in Fig. 11 are different from each other. However, when the proposed CLOD-FDTD method with CFLN = 4 is used to complete this simulation, and coarse meshes are used, its results show good agreement with those obtained from the LOD-FDTD method with CFLN = 1. Therefore, the efficiency of the proposed CLOD-FDTD method can be further improved compared with the conventional LOD-FDTD method. 

\begin{figure}[h]
	\centerline{{\includegraphics[scale=0.4]{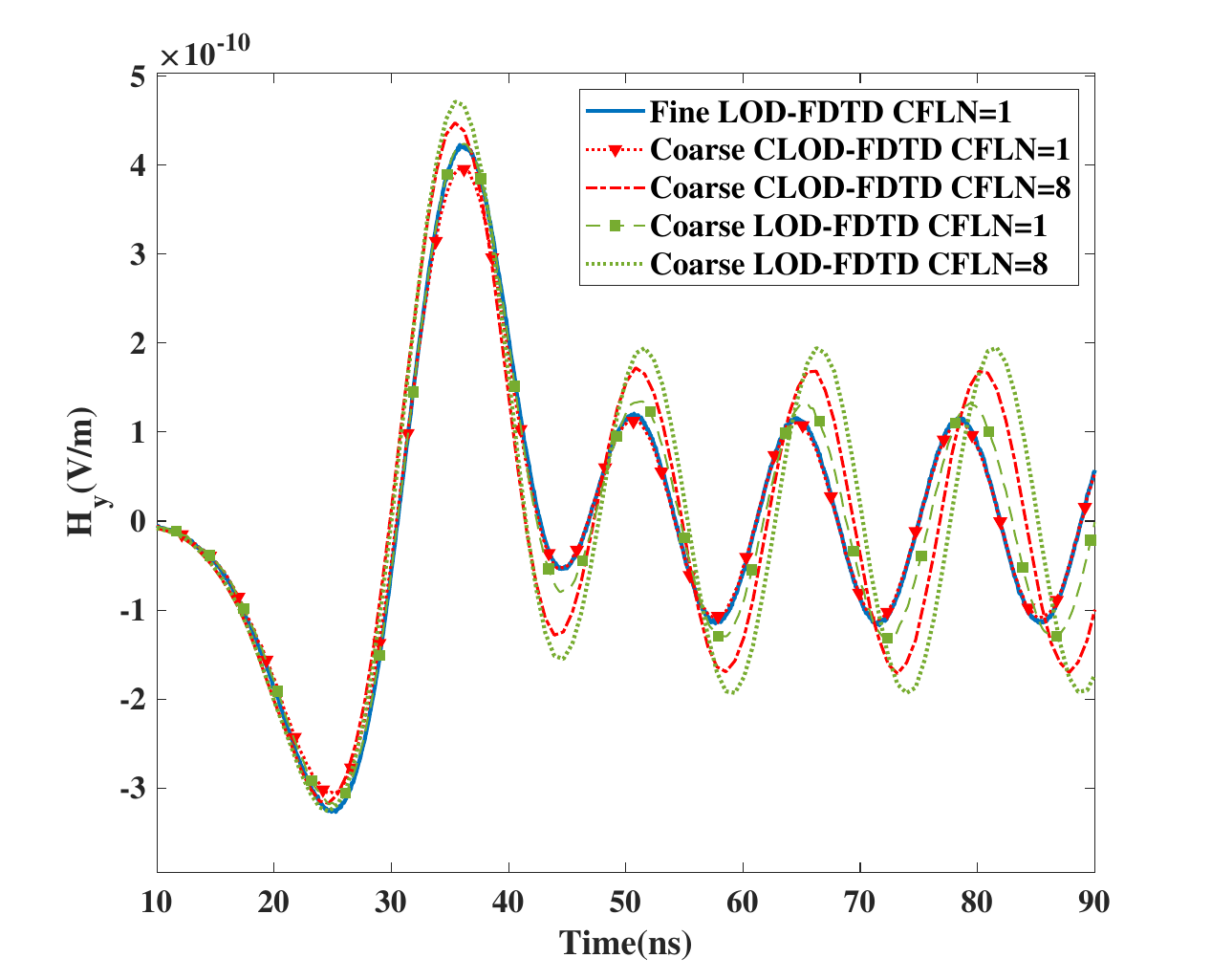}}}
	\caption{The responses of $H_y$ obtained from the LOD-FDTD method and the CLOD-FDTD method with CFLN = 1 and 8.}
	\label{Fig.CFLN8}
\end{figure}

Moreover, some comparisons between the LOD-FDTD method and the CLOD-FDTD method in the time domain, as shown in Fig. \ref{Fig.CFLN8}. The sizes of fine and coarse meshes are 0.01$m$, 0.05$m$, respectively. From Fig. \ref{Fig.CFLN8}, it can be found that the amplitudes of waveforms obtained with CFLN = 8 are larger than those with CFLN = 1. Furthermore, differences between results with CFLN = 8 and CFLN = 1 become larger as the simulation goes by. However, results obtained by the CLOD-FDTD method still have higher accuracy than that of the LOD-FDTD method.

The total time cost to finish the simulations is listed in Table \ref{T2}. The overall time is 180$ns$. Mesh sizes are 0.05$m$ and 0.01$m$ for coarse and fine meshes, respectively. Since the CFDTD method is unstable with CFLN = 0.25 in fine meshes, CFLN is set as 0.1 for fine meshes to guarantee the stability. It can be found that from Table \ref{T2}, the CFDTD method with CFLN = 0.5 is faster than the CLOD-FDTD method with CFLN = 1 when coarse meshes are used. However, as CFLN increases, the CLOD-FDTD method gradually outperforms the CFDTD method for two reasons: one is that the CFDTD method has to reduce the CFLN to guarantee the stability, and the other is that the overall count of cells significantly increases in fine meshes. 

\begin{table}[h]
	\renewcommand\arraystretch{1.5}
	\centering
	\caption{The computational consumption of the LOD-FDTD method the CLOD-FDTD method and the CFDTD method with CFLN = 1,4 and different meshes}
	\label{T2}
	\resizebox{9cm}{!}{
		\begin{threeparttable}[b]
			\begin{tabular}{ c| c| c c c c c }
				\hline
				\hline
				\multicolumn{2}{c }{\textbf{Method}}		 &\textbf{CFLN}		&${\bf{\Delta}t}$ [ns]	 &\textbf{No. of Cells}	&\textbf{Time Cost} [s]		&\textbf{Ratio$^*$}\cr
				\hline
				\hline
				\multirow{4}*{\textbf{LOD-FDTD}} 
				&\multirow{2}*{Coarse}	
					&1	&0.092		&64,000		&14.2		&1107.3\\
				&	&4	&0.385		&64,000		&3.6		&4367.7\\
				\cline{2-7}
				&\multirow{2}*{Fine}	
				    &1	&0.019		&8,000,000		&15723.7	&1.0\\
				&	&4	&0.073		&8,000,000		&4253.5		&3.7\\
				\hline
				\multirow{2}*{\textbf{CLOD-FDTD}} 
				&\multirow{2}*{Coarse}	
				    &1	&0.092		&64,000		&22.6		&695.7\\
				&   &4	&0.385		&64,000		&5.5		&2858.9\\
				% \cline{2-7}
				%&\multirow{2}*{Fine}	
				%    &1	&0.019		&8,000,000		&25078.2	&-\\
				%    & &4	&0.073		&8,000,000		&7648.3	&-\\
				\hline
				\multirow{2}*{\textbf{CFDTD}} 
				&Coarse 	&0.5	&0.048		&64,000		&7.2		&2183.8\\
				\cline{2-7}
				&Fine 	&0.1	&0.0019		&8,000,000		&21978.9		&0.71\\
				\hline
				\hline
			\end{tabular}
		\begin{tablenotes}
			\footnotesize
			\item[*]Ratio is defined as the ratio of time cost used in the LOD-FDTD method with fine meshes to that of the corresponding method.
		\end{tablenotes}
\end{threeparttable}
}
\end{table}

In summary, the proposed CLOD-FDTD method shows better accuracy compared with the FDTD method and the LOD-FDTD method, when the same meshes and time step are used. Significant accuracy improvement of the proposed CLOD-FDTD method is obtained compared with the LOD-FDTD method, when the same meshes are used. Therefore, we can obtain accurate results with relative coarse meshes and large time steps without compromising the accuracy. Consequently, CPU time can be saved in simulations. 

\subsection{The Electromagnetic Scattering from A Missile}
\begin{figure}[htbp]
\centerline{{\includegraphics[scale=0.30]{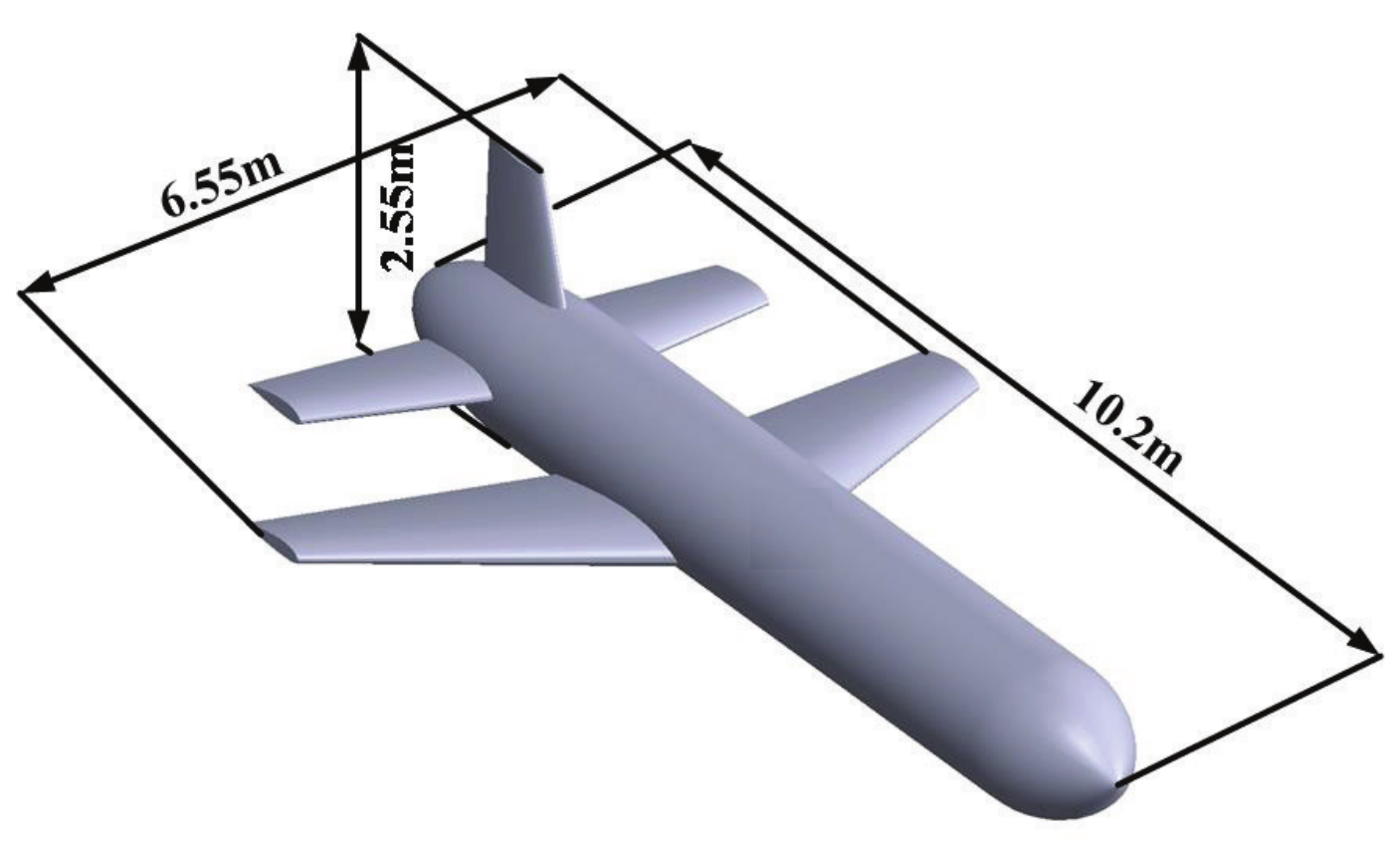}}}
\caption{The geometrical model of the missile.}
\label{Fig.12}
\end{figure}

In this section, a PEC missile, which is 10.2$m$ long, 6.55$m$ wide, and 2.55$m$ high, is considered. The convolution perfectly matched layer (CPML) based on \cite{FDTDCPML} \cite{LODCPML} is used to truncate the computational domain. The geometrical model is shown in Fig. \ref{Fig.12}. Its surrounding medium is the vacuum. 

The grids with mesh sizes 0.1$m$ and 0.025$m$ are used to discrete the missile. Coarse meshes are used in the CLOD-FDTD method, and both the coarse and fine meshes are used in the conventional LOD-FDTD method and the FDTD method for the references. 

\begin{figure}[h]
\centering
\subfigure[]{
\includegraphics[scale=0.25]{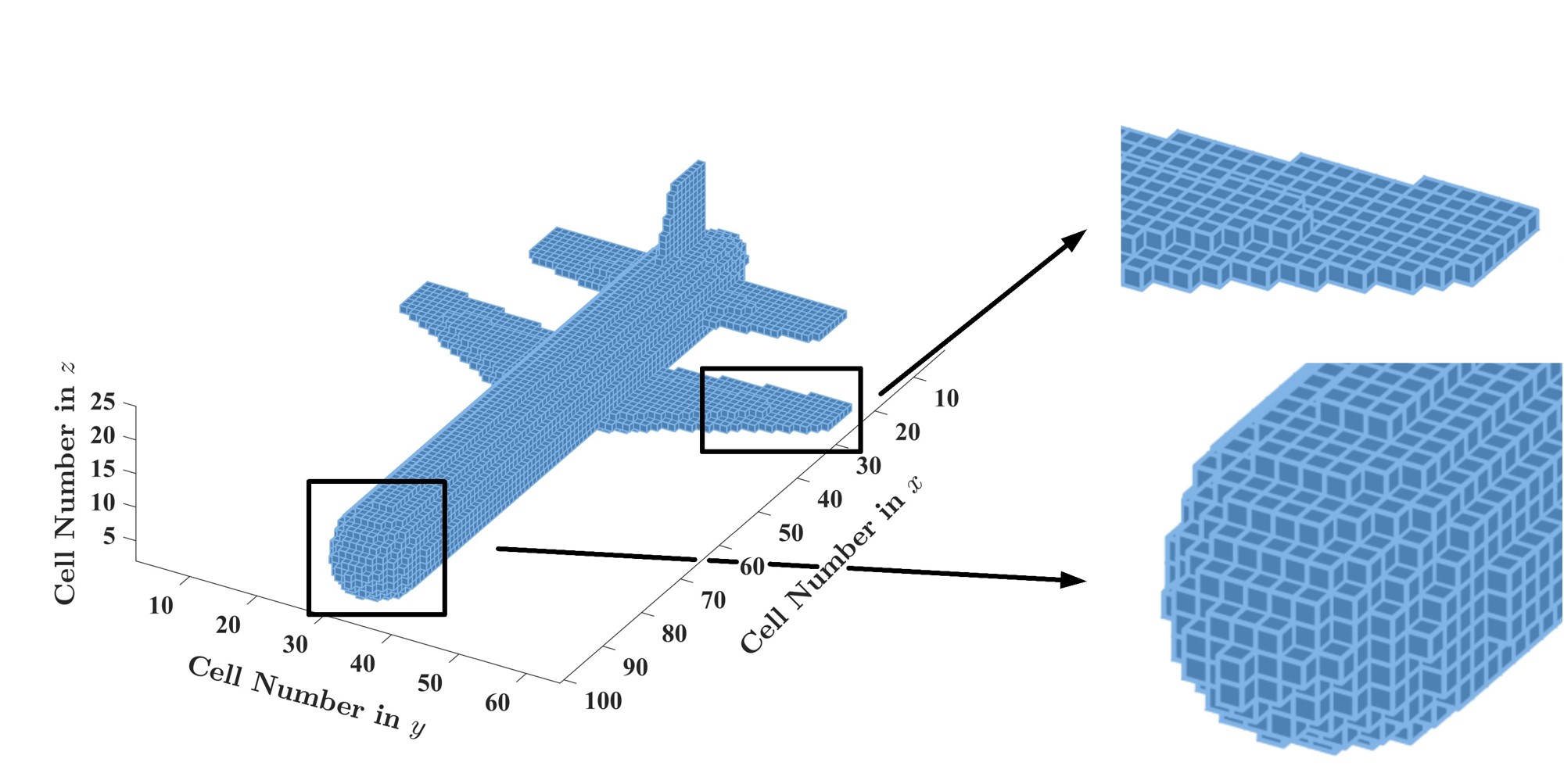}
%\caption{fig1}
}
\subfigure[]{
\includegraphics[scale=0.25]{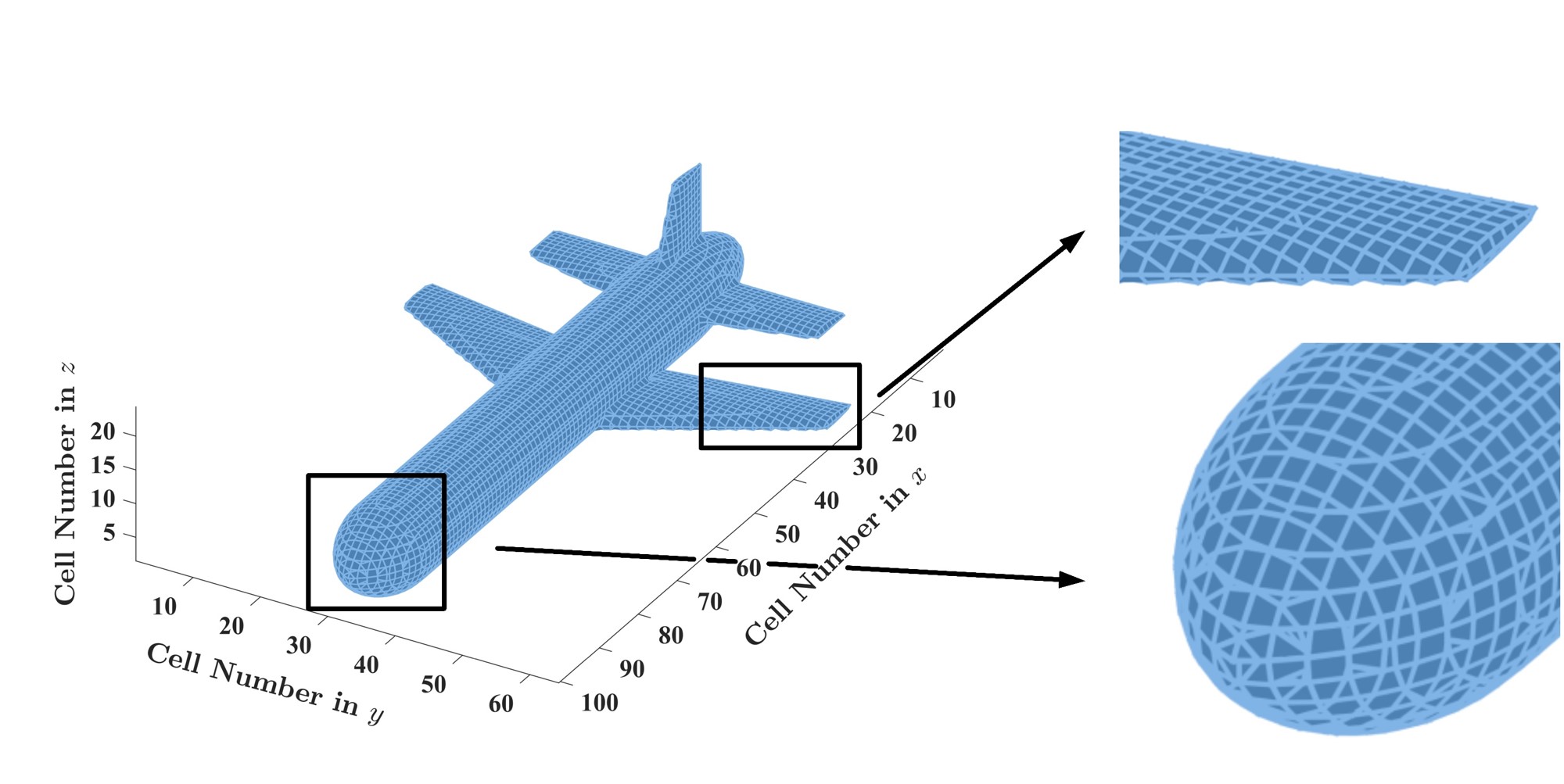}
}
\caption{The meshes of an ideal conductor missile with mesh size 0.1 $m$: (a) Yee's meshes, (b) the conformal meshes.}
\label{Fig.13}
\end{figure}

Fig. \ref{Fig.13} shows Yee's grids and the conformal grids when mesh size is 0.1$m$. The normal Yee's grids are used in the traditional FDTD method and the LOD-FDTD method, while the conformal grids are used in the CLOD-FDTD method. In Yee's grids, large geometrical modeling error occurs due to the rectangular cells. However, the conformal grids fit well with the original model. For example, the curved structures of the missile, such as wings and the missile head, are accurately modeled through the conformal grids, as shown in Fig. \ref{Fig.13} (b).

A line current source with the differential Gaussian pulse $f=(t-t_0)e^{[-{(t-t_0)^2}/{\tau^2 }]}$ $t_0=3\tau, \tau=10ns$ in the $x$ direction is used as the excitation function. The current source is located at (9, 2.5, 1.4)[$m$] in both coarse mesh and fine mesh. 

\begin{figure*}
 \begin{minipage}[h]{0.23\linewidth}
  \centerline{\includegraphics[scale=0.26]{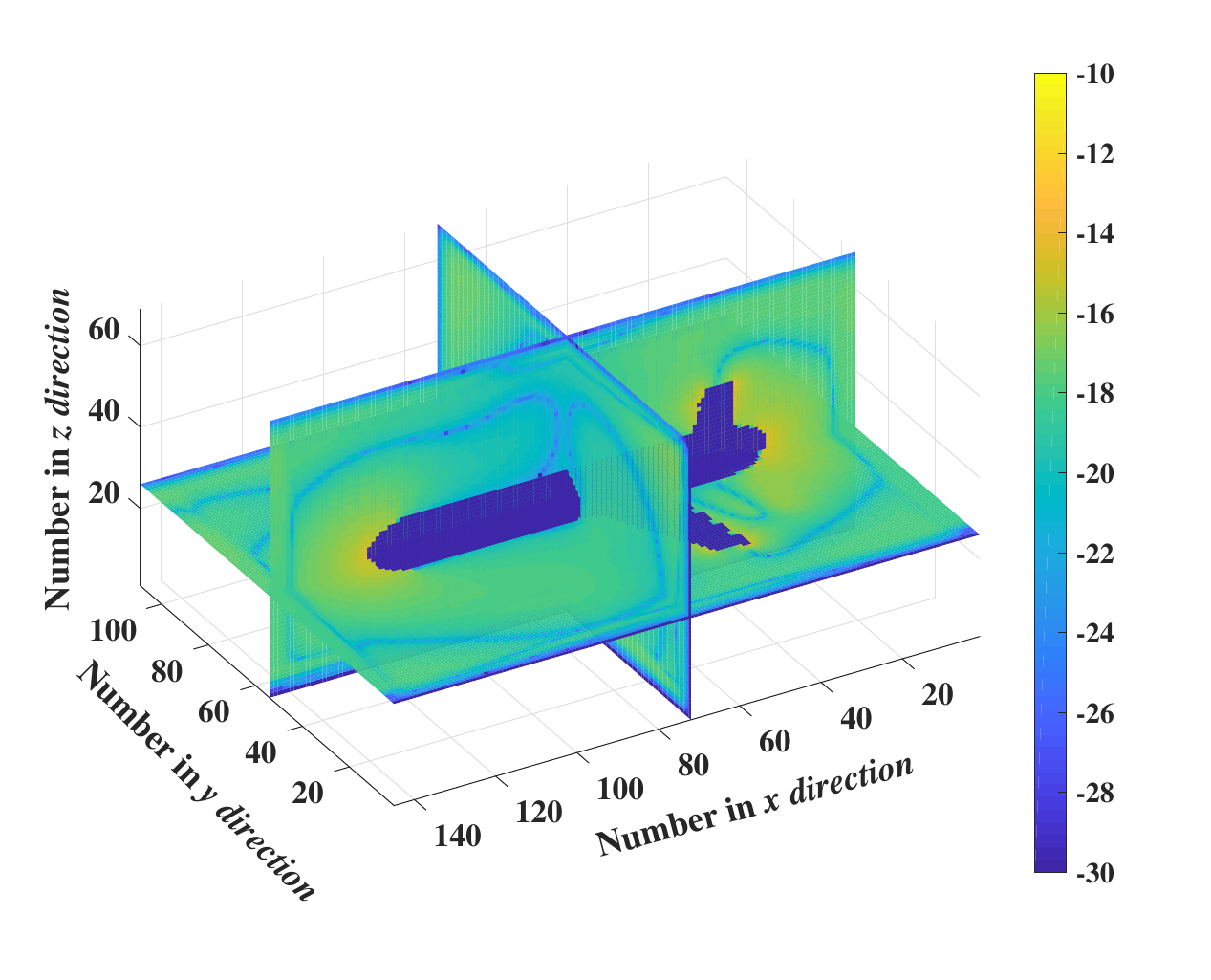}}
  \centerline{(a)}
 \end{minipage}
 \hfill
 \begin{minipage}[h]{0.23\linewidth}
  \centerline{\includegraphics[scale=0.26]{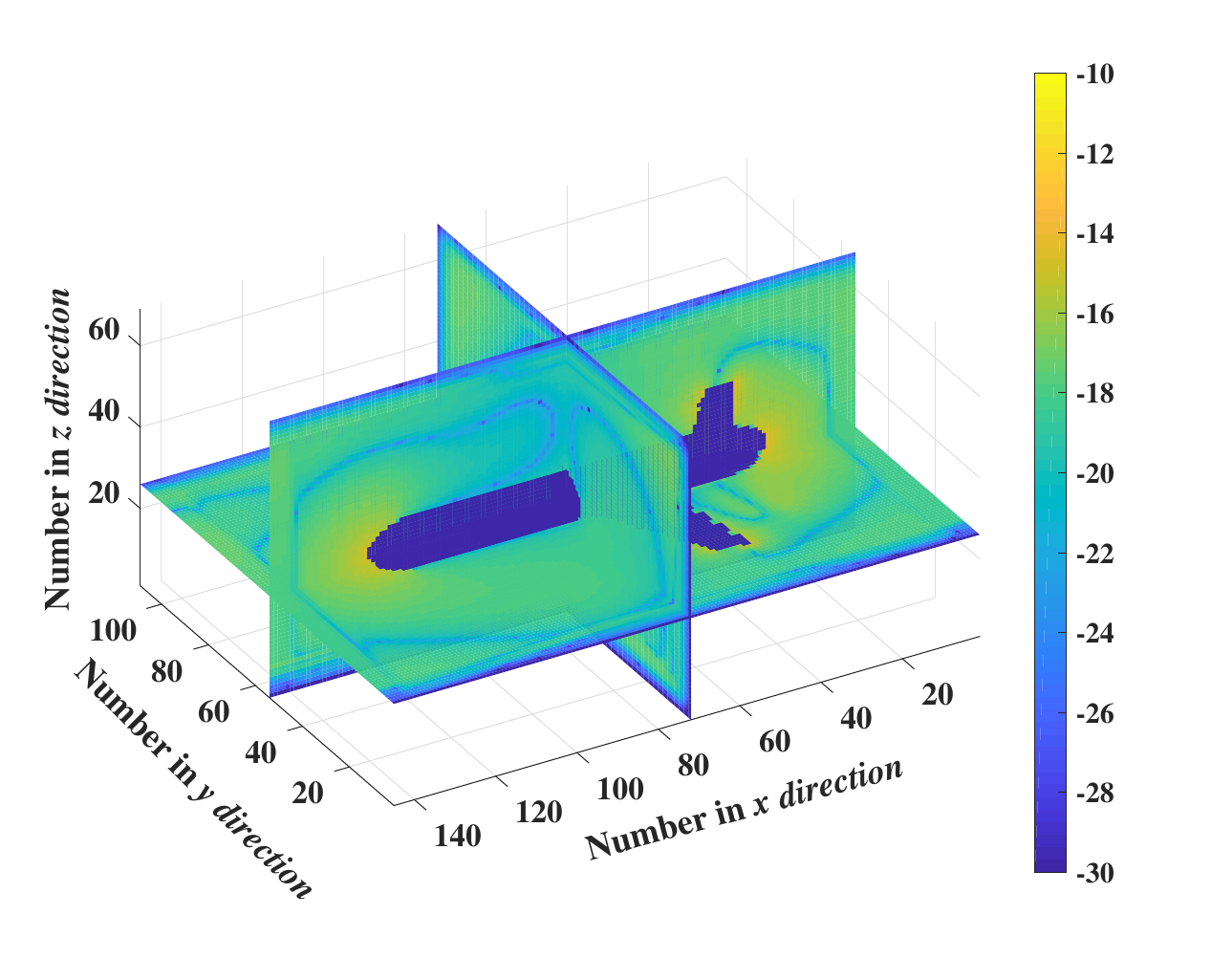}}
  \centerline{(b)}
 \end{minipage}
 \hfill
 \begin{minipage}[h]{0.23\linewidth}
  \centerline{\includegraphics[scale=0.26]{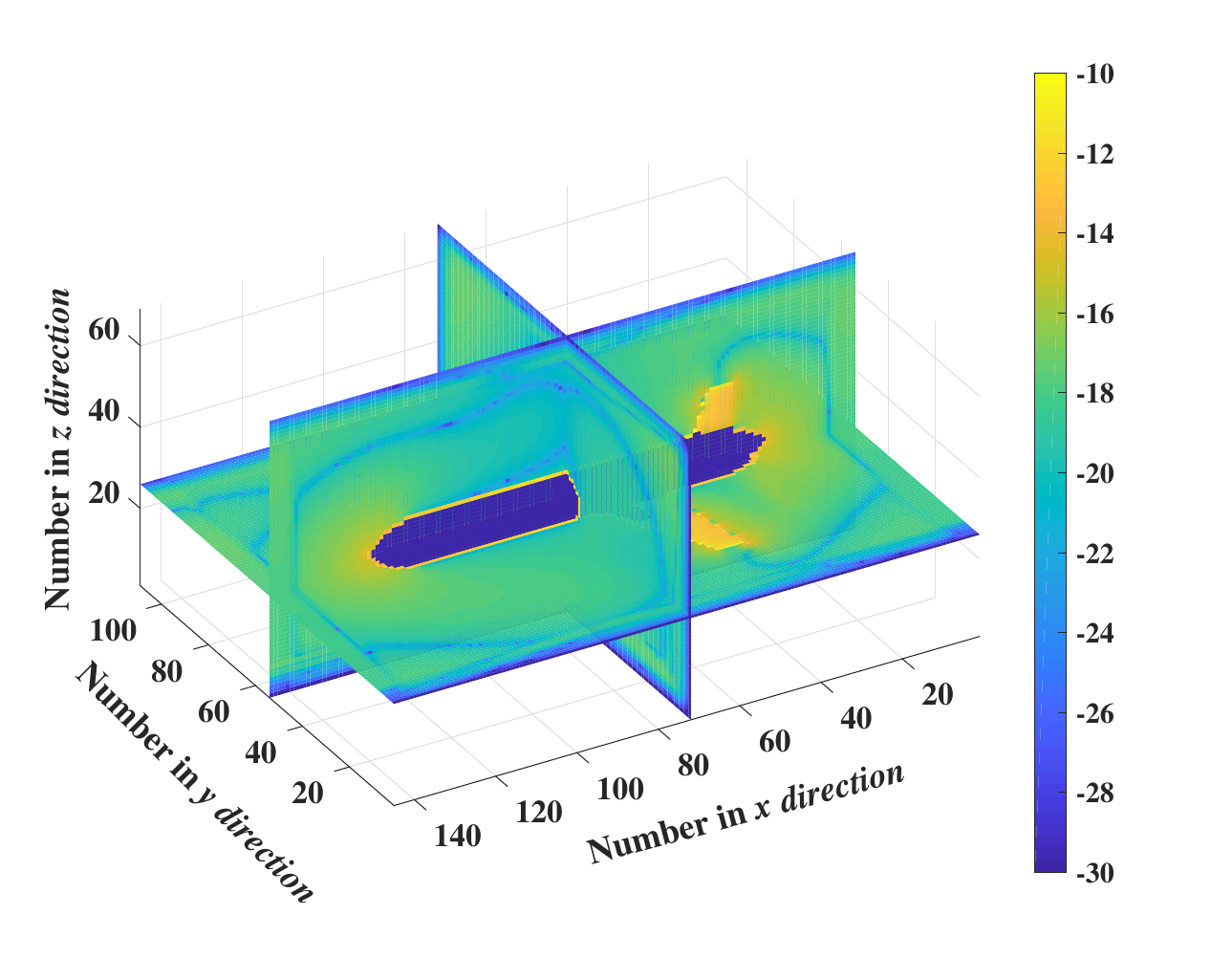}}
  \centerline{(c)}
 \end{minipage}
 \hfill
 \begin{minipage}[h]{0.23\linewidth}
  \centerline{\includegraphics[scale=0.26]{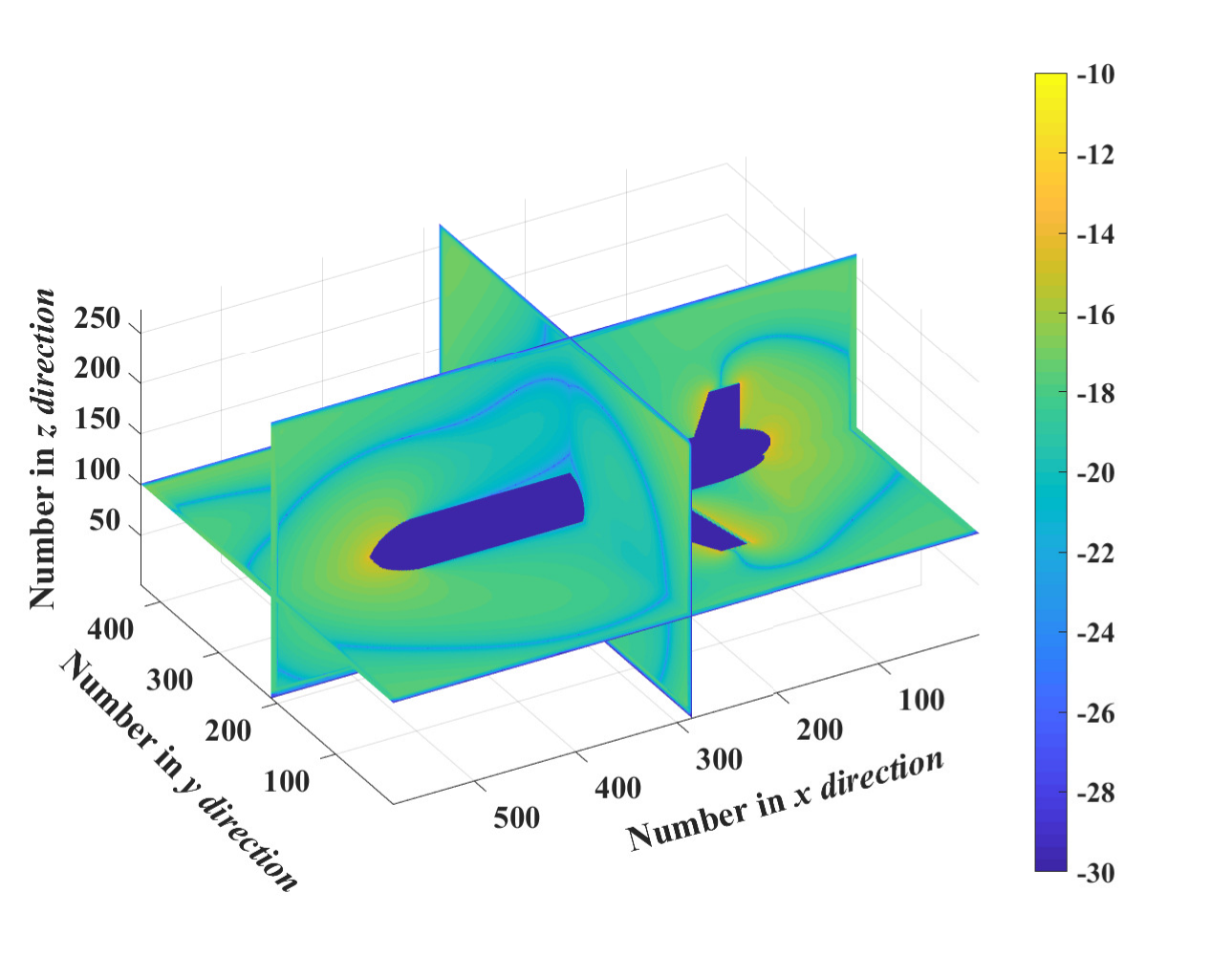}}
  \centerline{(d)}
 \end{minipage}
 \caption{The $log(\left|E_x\right|)$ at $66ns$: (a) the FDTD method with coarse meshes, (b) the LOD-FDTD method with coarse meshes, (c) the CLOD-FDTD method with coarse meshes, (d) the LOD-FDTD method with fine meshes.}
 \label{Fig.14}
\end{figure*}

$E_x$ at 66$ns$ is recorded when CFLN = 1, three slices parallel to the {\it xoy, yoz, xoz} plane are shown in Fig. \ref{Fig.14}. Figs. \ref{Fig.14}(a)-(c) show $log(\left|E_x\right|)$ obtained by the FDTD method, the LOD-FDTD method, and the CLOD-FDTD method with coarse meshes. Figs. \ref{Fig.14}(d) shows $log(\left|E_x\right|)$ obtained by the LOD-FDTD method with fine meshes. It can be found from Fig. \ref{Fig.14}(a)-(b) that results obtained by the FDTD and LOD-FDTD methods fit well with each other. However, comparing to Figs. \ref{Fig.14}(a)-(b), results obtained by the CLOD-FDTD method with coarse meshes have a different distribution of $E_x$ at the tail of the missile and the junction of field slices parallel to {\it xoy} and {\it yoz} plane. Meanwhile, results obtained by the LOD-FDTD method with fine meshes agree well with results obtained by the CLOD-FDTD method with coarse meshes in the exterior region.  

\begin{figure*}
 \begin{minipage}[h]{0.225\linewidth}
  \centerline{\includegraphics[scale=0.26]{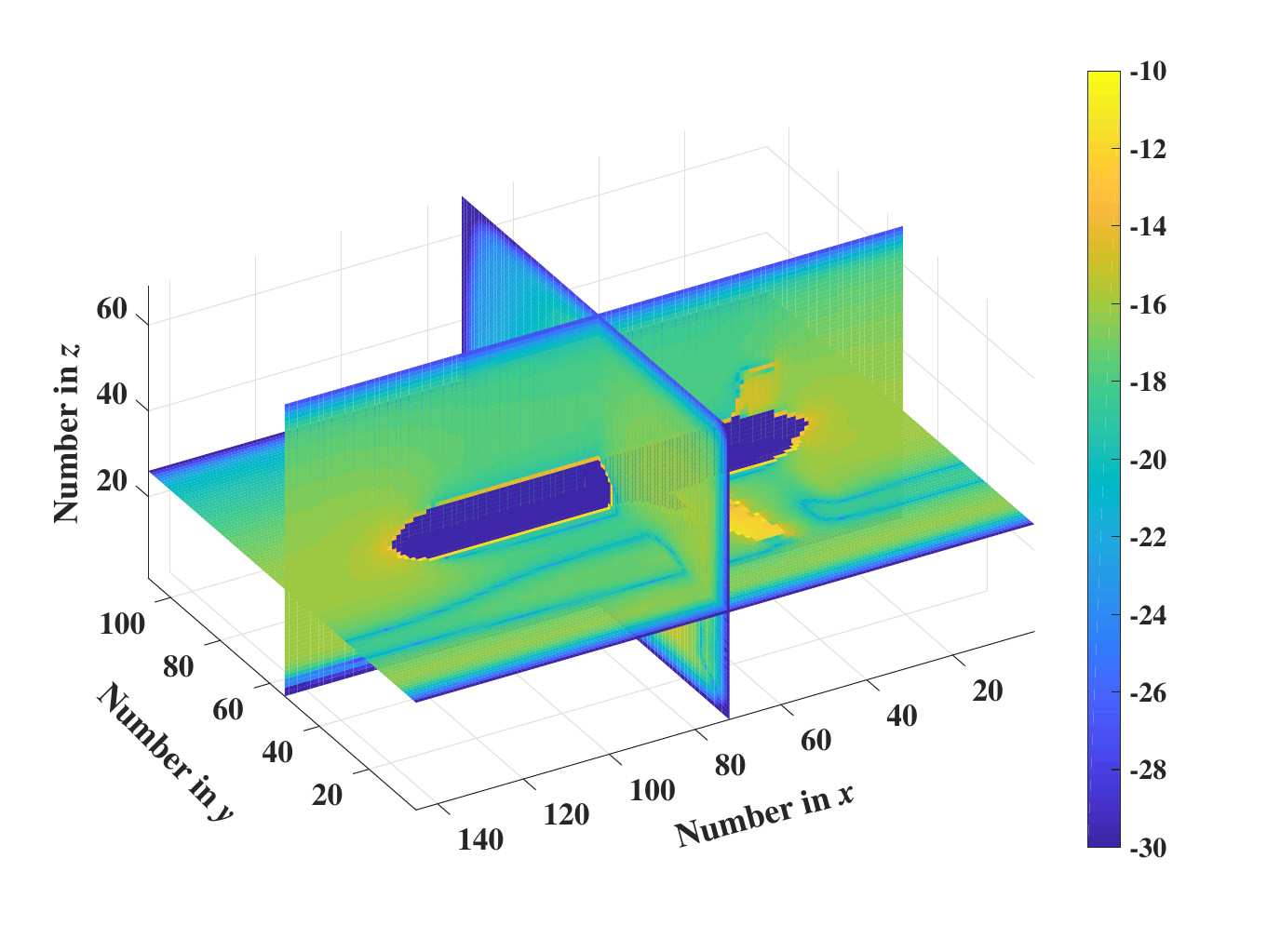}}
  \centerline{(a)}
 \end{minipage}
 \hfill
 \begin{minipage}[h]{0.225\linewidth}
  \centerline{\includegraphics[scale=0.26]{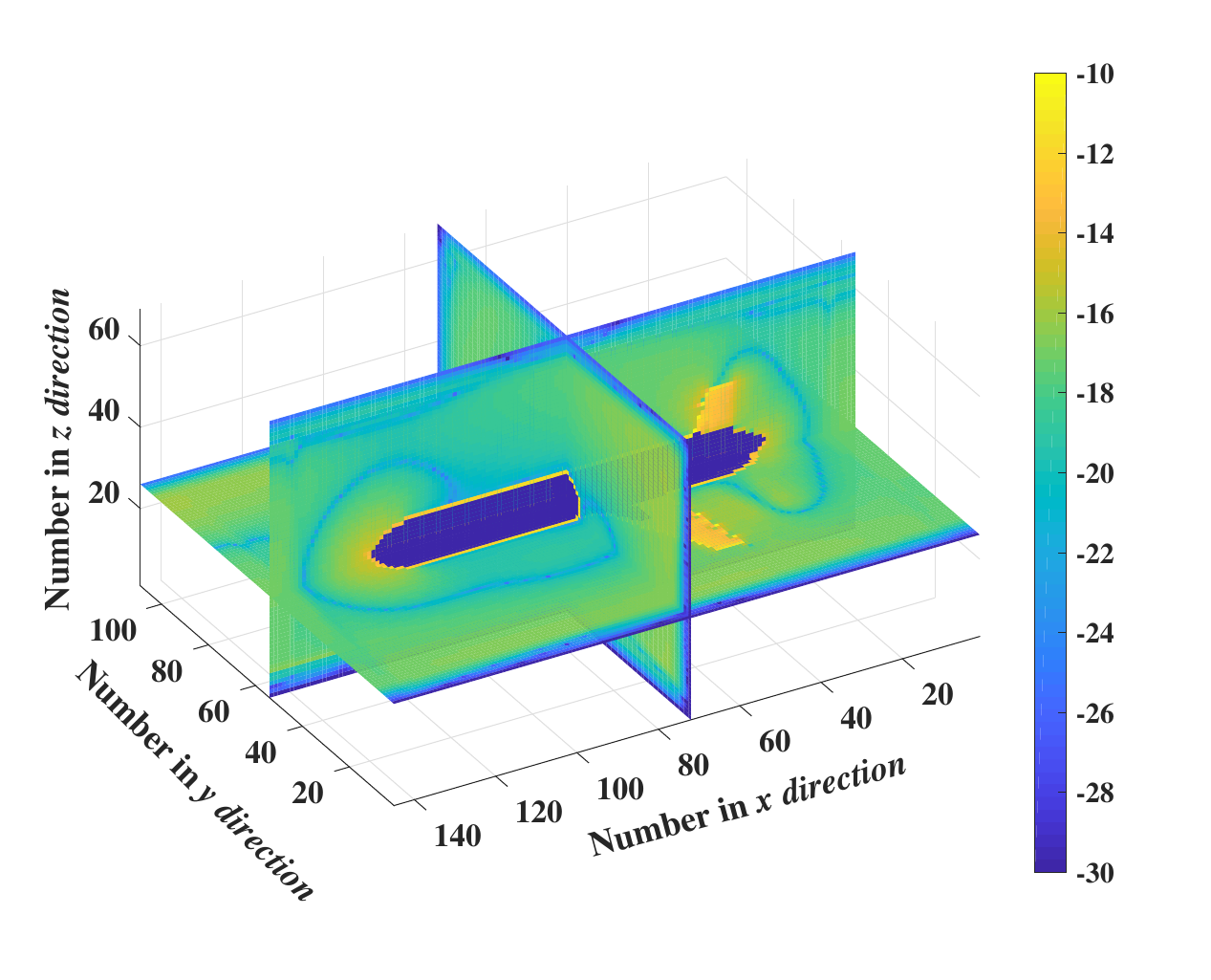}}
  \centerline{(b)}
 \end{minipage}
 \hfill
 \begin{minipage}[h]{0.225\linewidth}
  \centerline{\includegraphics[scale=0.26]{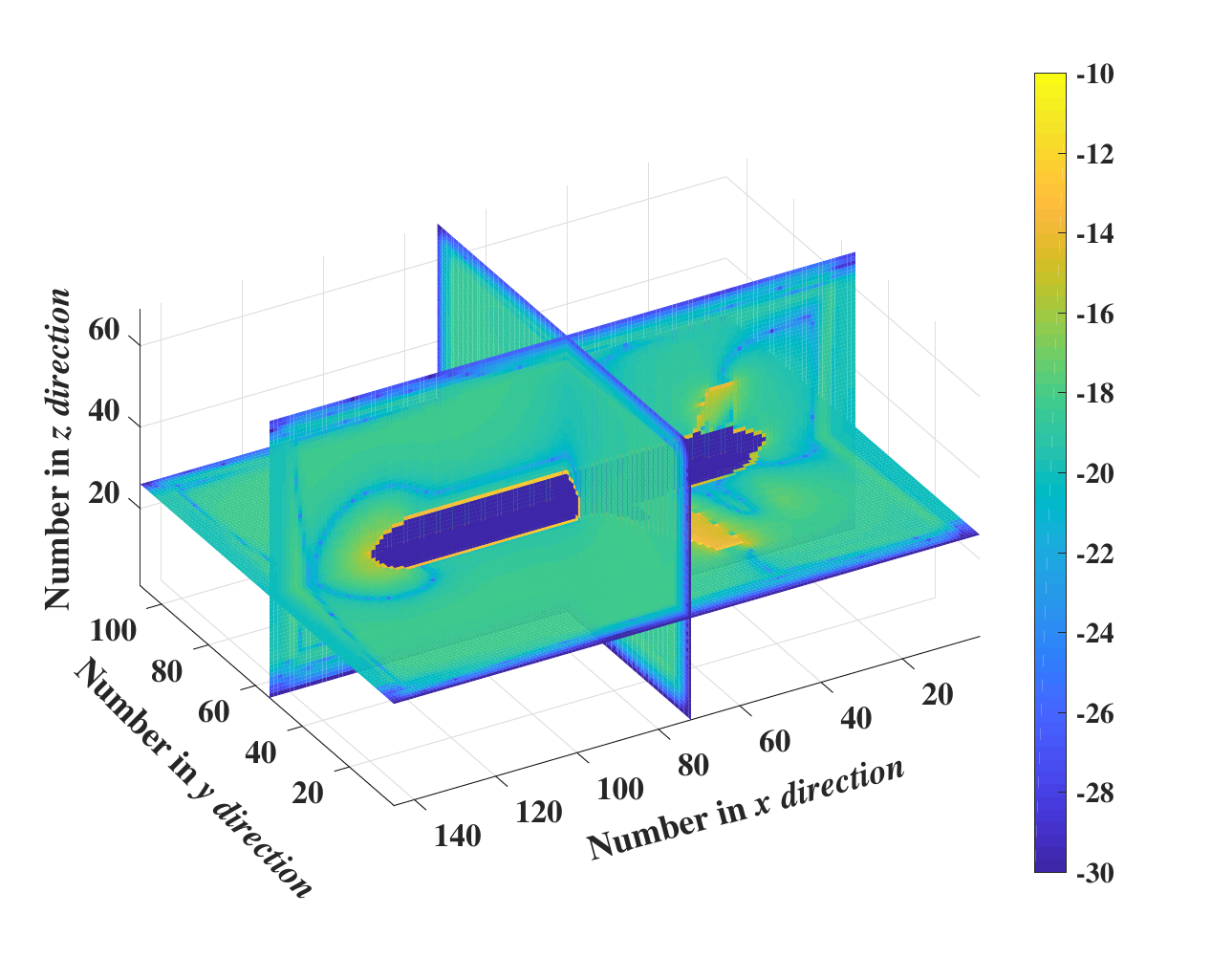}}
  \centerline{(c)}
 \end{minipage}
 \hfill
 \begin{minipage}[h]{0.225\linewidth}
  \centerline{\includegraphics[scale=0.26]{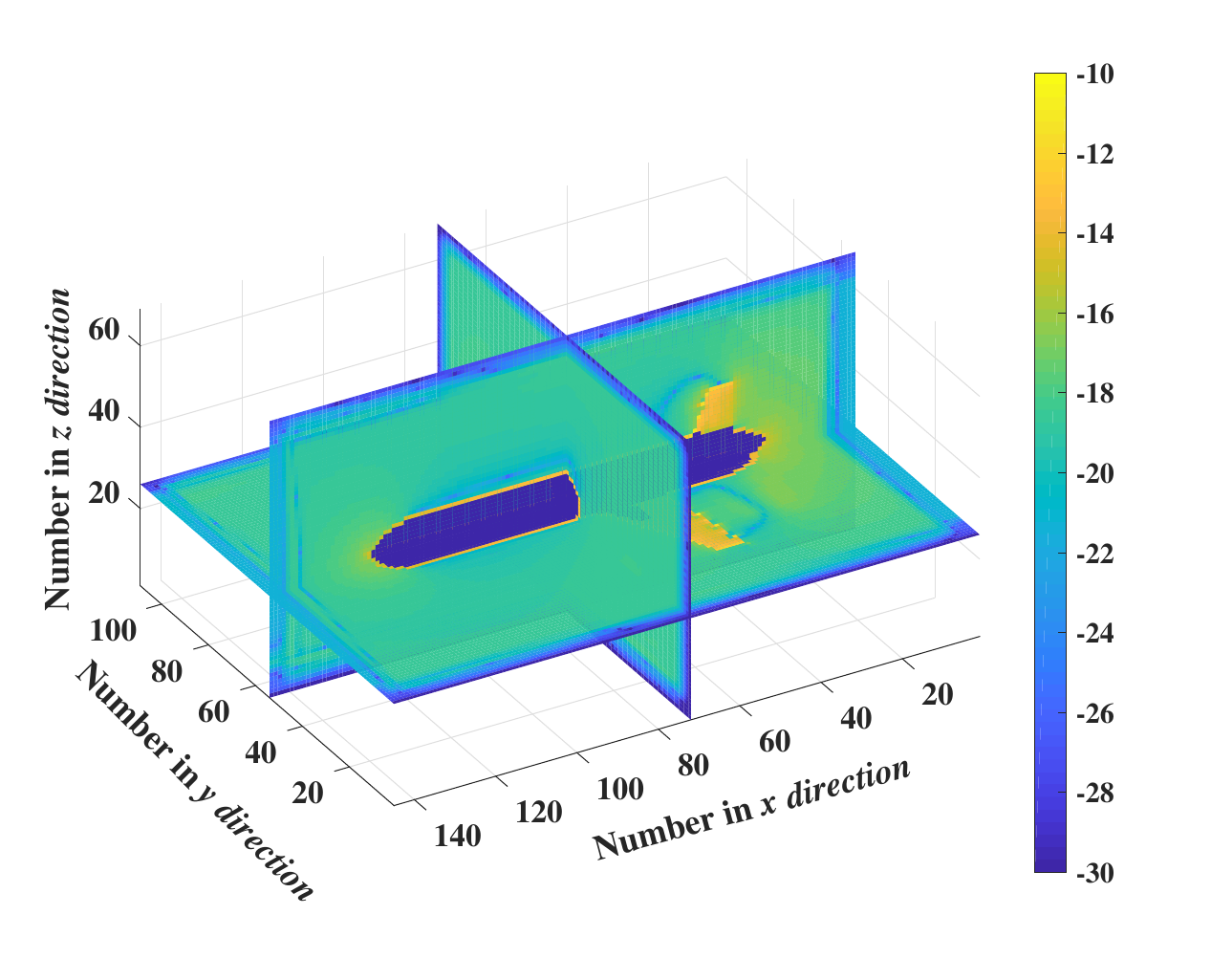}}
  \centerline{(d)}
 \end{minipage}
 \caption{The $log(\left|E_x\right|)$ obtained by the CLOD-FDTD method with coarse meshes: (a) at 30$ns$, (b) at 60$ns$, (c) at 90$ns$, (d) at 120$ns$.}
 \label{Fig.15}
\end{figure*}

\begin{figure*}
 \begin{minipage}[h]{0.225\linewidth}
  \centerline{\includegraphics[scale=0.26]{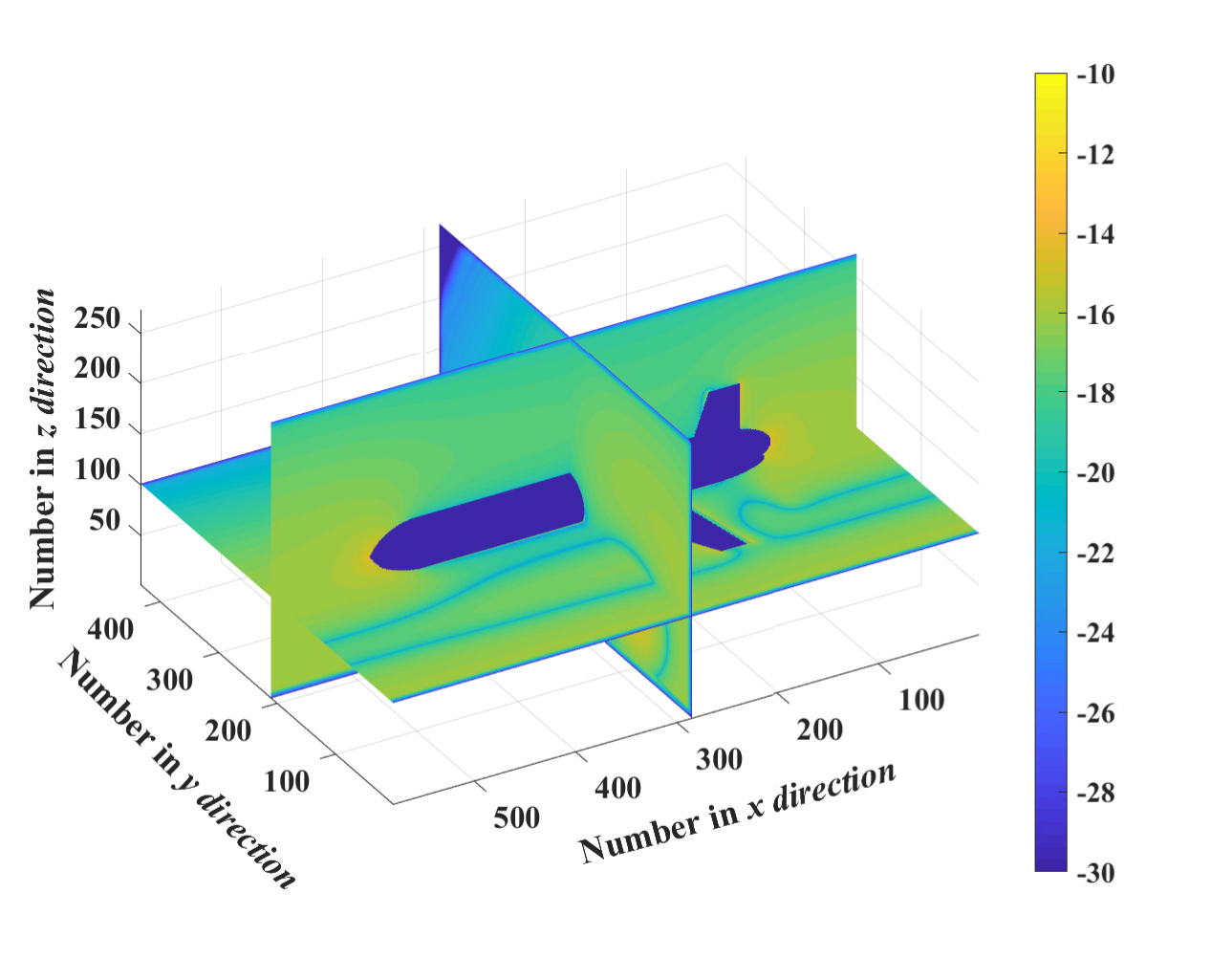}}
  \centerline{(a)}
 \end{minipage}
 \hfill
 \begin{minipage}[h]{0.225\linewidth}
  \centerline{\includegraphics[scale=0.26]{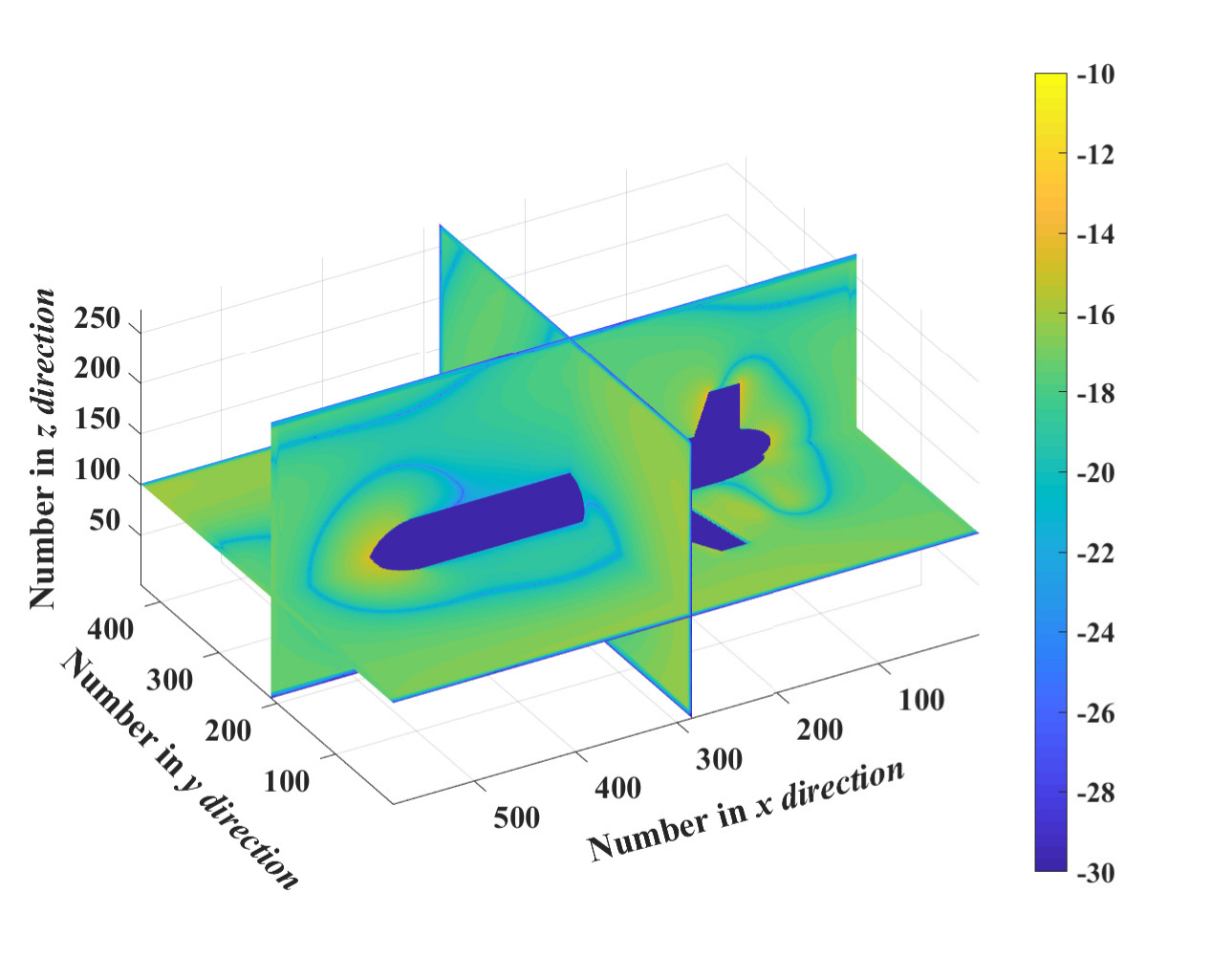}}
  \centerline{(b)}
 \end{minipage}
 \hfill
 \begin{minipage}[h]{0.225\linewidth}
  \centerline{\includegraphics[scale=0.26]{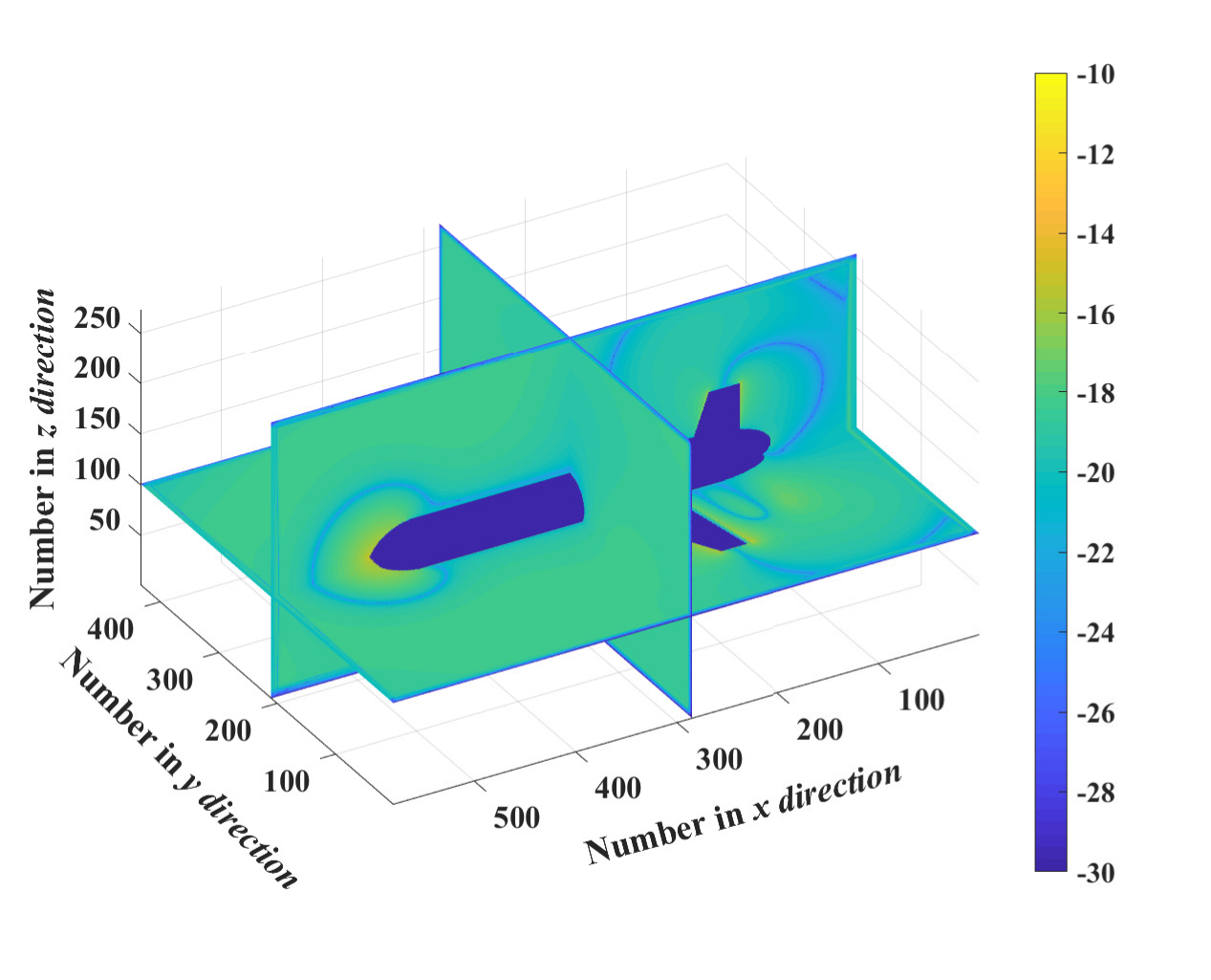}}
  \centerline{(c)}
 \end{minipage}
 \hfill
 \begin{minipage}[h]{0.225\linewidth}
  \centerline{\includegraphics[scale=0.26]{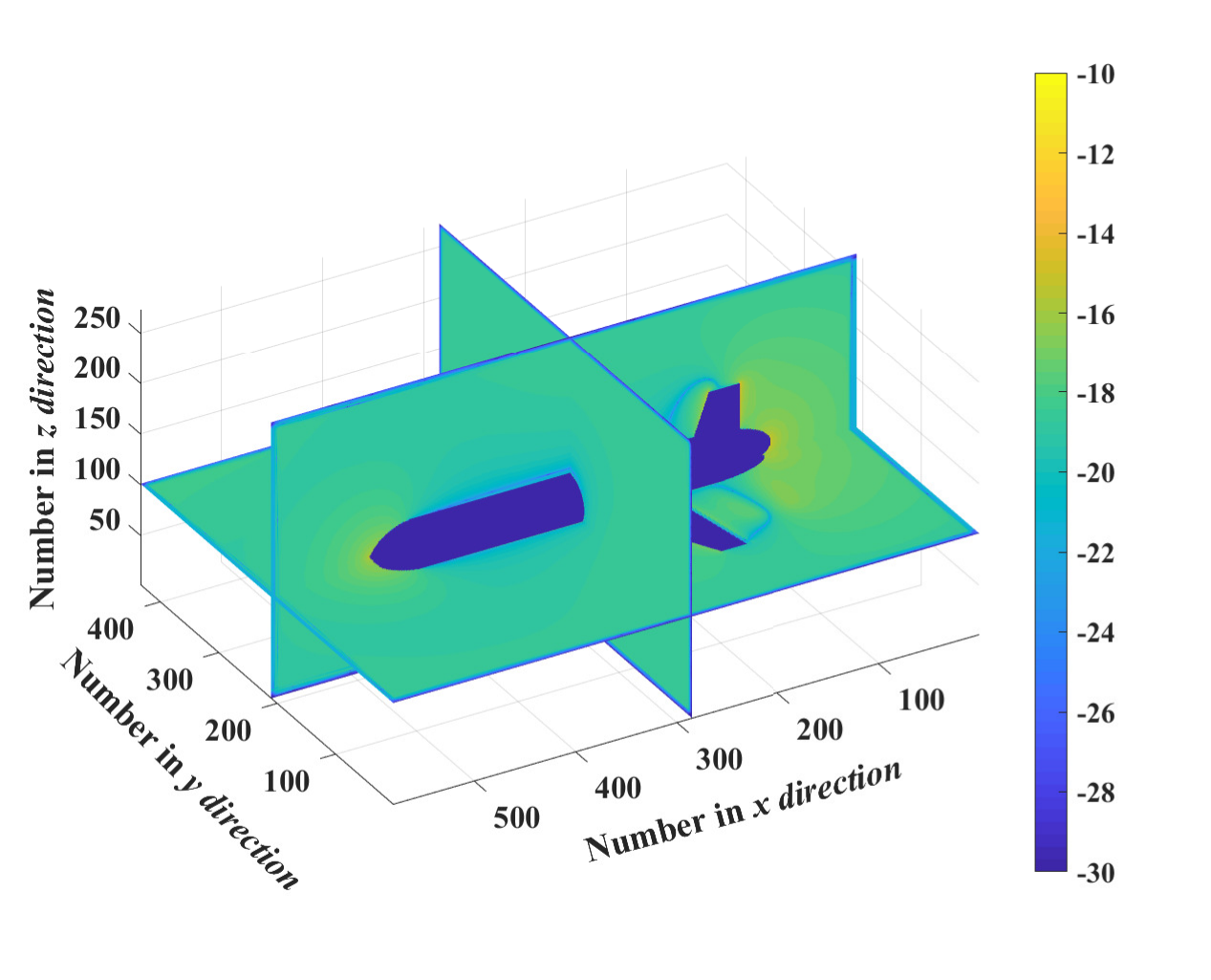}}
  \centerline{(d)}
 \end{minipage}
 \caption{The $log(\left|E_x\right|)$ obtained by the LOD-FDTD method with fine meshes: (a) at 30$ns$, (b) at 60$ns$, (c) at 90$ns$, (d) at 120$ns$.}
 \label{Fig.16}
\end{figure*}

Furthermore, $log(\left|E_x\right|)$ in the CLOD-FDTD method with coarse meshes and the LOD-FDTD method with fine meshes at 30$ns$, 60$ns$, 90$ns$, 120$ns$ are shown in Fig. \ref{Fig.15} and Fig. \ref{Fig.16}, relatively. Their results agree well with each other in the exterior region.

It should be noted that the results obtained by the CLOD-FDTD method in Figs. \ref{Fig.14}-\ref{Fig.15} has obvious difference with other results in the wings and tail. It is the conformal meshes that modify the magnetic field attached to the PEC surface, which makes $E_x$ near the PEC surface larger than those in the traditional FDTD method and the LOD-FDTD method. In fact, it can be seen from Fig. \ref{Fig.13}(b), the physical thickness of the wings is smaller than that of Yee's meshes in Fig. \ref{Fig.13}(a). It is obvious that electric fields at the meshes which are located at the bottom surface of wings are outside the PEC object. Therefore, those electric fields should not be zeros. However, the traditional FDTD method and the LOD-FDTD method enforce electric fields to be zeros at those positions due to staircase error. Since the conformal meshes in Fig. \ref{Fig.13}(b) are used in the proposed CLOD-FDTD method and partially filled cells are carefully considered, the error is significantly reduced in the CLOD-FDTD method. Therefore, the CLOD-FDTD method can obtain more accurate results than those from the LOD-FDTD method and the FDTD method with the same meshes.

\begin{figure}
\centerline{{\includegraphics[scale=0.38]{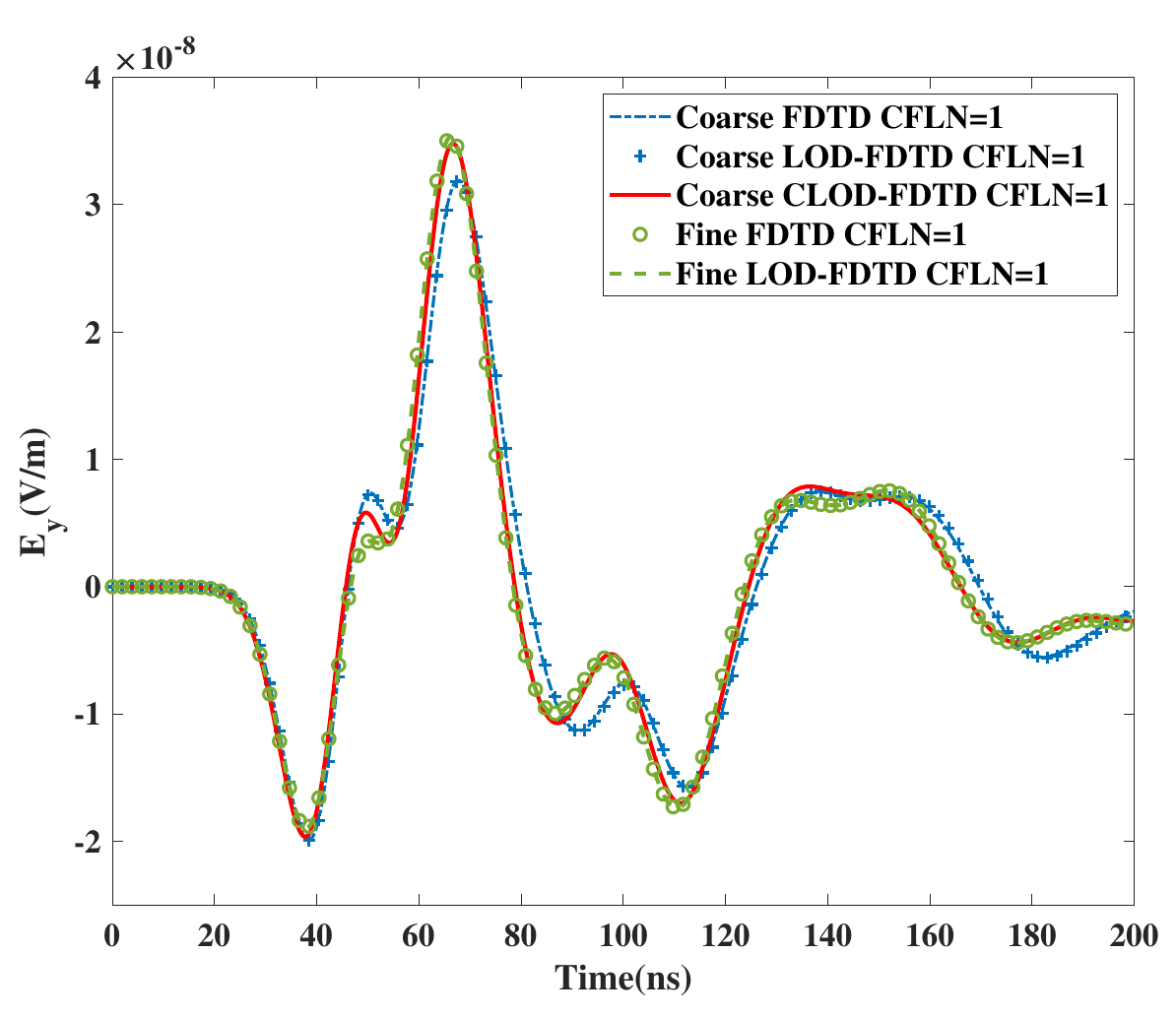}}}
\caption{$E_y$ obtained from the FDTD method, the LOD-FDTD method and the proposed CLOD-FDTD method from $0ns$ to $200ns$ with CFLN = 1.}
\label{Fig.17}
\end{figure}

Moreover, the electric field at the observation point was recorded to validate the effectiveness of the proposed CLOD-FDTD method. The observation point is set at (9, 3, 1.4)[$m$] in both fine and coarse mesh. The transient values of $E_y$ are shown in Fig. \ref{Fig.17}. It can be found that results obtained from the LOD-FDTD method, and the FDTD method show good agreement. Simultaneously, it can be found that results obtained from the CLOD-FDTD method with coarse meshes agree well with those from the LOD-FDTD method and the FDTD method with fine meshes. Therefore, the CLOD-FDTD method can obtain accurate results with coarse meshes. 

\begin{figure}
\centerline{{\includegraphics[scale=0.37]{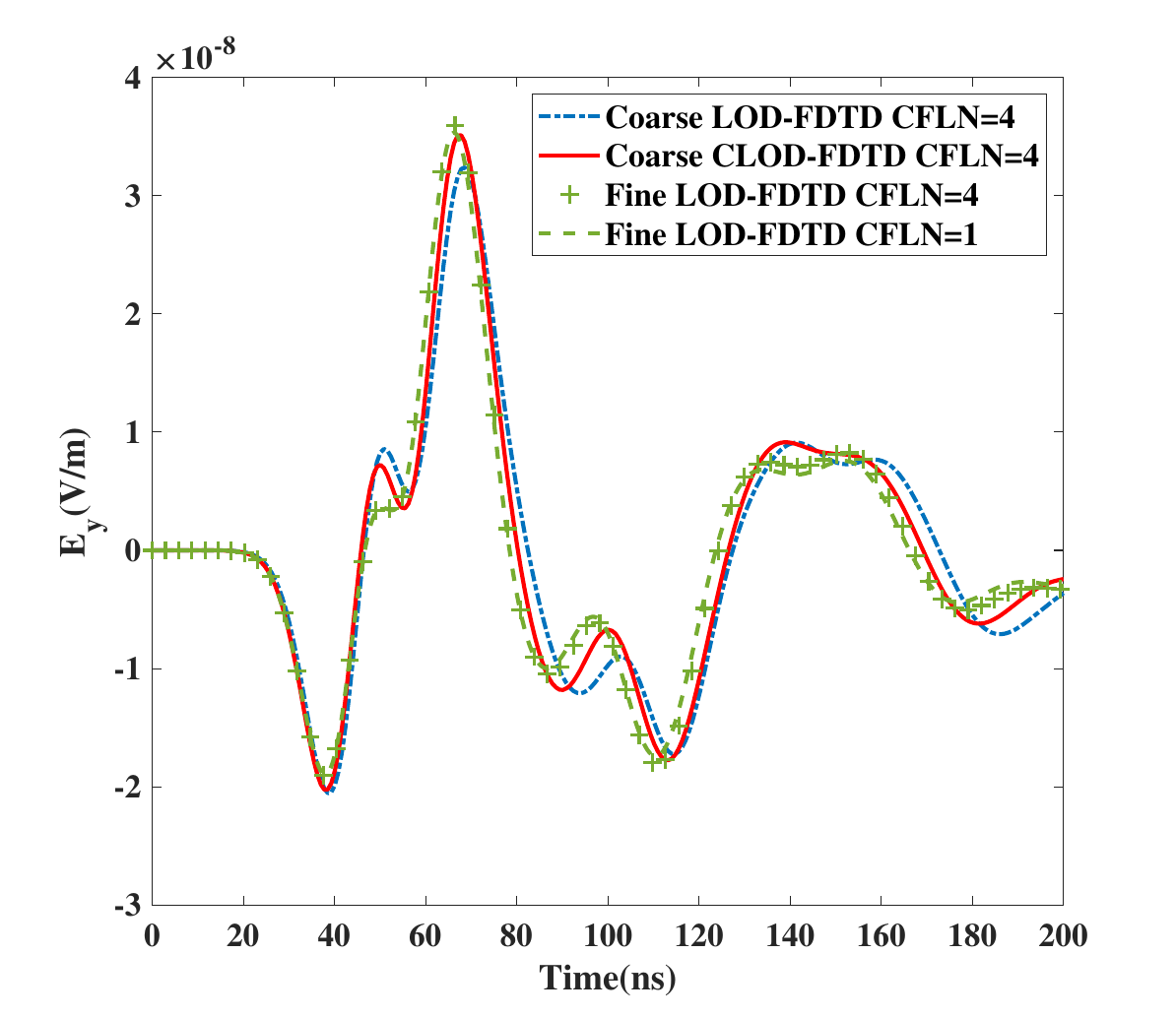}}}
\caption{$E_y$ obtained from the LOD-FDTD method and the proposed CLOD-FDTD method from $0ns$ to $200ns$ with CFLN = 1, 4.}
\label{Fig.18}
\end{figure}

$E_y$ obtained from the CLOD-FDTD method, and the LOD-FDTD method with different CFLN are shown in the Fig. \ref{Fig.18}. Fine meshes are used in the LOD-FDTD method with CFLN = 1, 4. Coarse meshes are used in the CLOD-FDTD method and the LOD-FDTD method with CFLN = 4. It can be seen that, the result in the CLOD-FDTD method agrees well with the results obtained from the LOD-FDTD method with fine meshes. Therefore, the proposed CLOD-FDTD method can indeed improve the accuracy with coarse meshes. 

Table \ref{T1} shows the overall number of cells and time cost for the LOD-FDTD method and the CLOD-FDTD method. It can be found from Table \ref{T1} that for the conventional LOD-FDTD method, the overall number of cells is 68,152,320 in fine meshes, which is 64 times that in coarse meshes. Besides, $\Delta t$ used in the simulations with fine meshes is only a quarter of time steps for coarse meshes with the same CFLN. Therefore, CPU time of the LOD-FDTD method with fine meshes is at least 256 times that for coarse meshes. However, due to the significant increase of CPU time in memory access for large-scale data set, the practical CPU time of the LOD-FDTD method with fine meshes is more than 360 times that for coarse meshes. By increasing CFLN to 4, the time cost for the LOD-FDTD method with coarse meshes can be reduced to 33,700.6$s$, which obtain a 5.2$\times$ efficiency gain. However, the overall time cost is still much longer than that with coarse meshes. 

On the other hand, the CLOD-FDTD method with coarse meshes takes 691.4$s$ and 173.3$s$ to complete the simulation when CFLN = 1, 4, relatively. Since additional modifications are required to be made in (\ref{E3})-(\ref{E8}) and (\ref{E9})-(\ref{E14}), it is slightly slower than the LOD-FDTD method with the same meshes. However, previous results have confirmed that the accuracy of the proposed CLOD-FDTD method with coarse meshes is almost the same as those obtained by the LOD-FDTD method with fine meshes, which implies that less CPU time can be used in the proposed CLOD-FDTD method without scarifying its accuracy by using coarse meshes. As shown in Table \ref{T1}, the speed-up ratio of the proposed CLOD-FDTD method with coarse meshes can be up to 255.3 and 1018.4 when CFLN = 1, 4, respectively.  

\begin{table}[h]
		\renewcommand\arraystretch{1.5}
		\centering
		\caption{The computational consumption of the LOD-FDTD method and the CLOD-FDTD method with different CFLNs and meshes}
		\label{T1}
		\resizebox{9cm}{!}{
		\begin{threeparttable}[b]
		\begin{tabular}{ c| c| c c c c c }
		\hline
  		\hline
			\multicolumn{2}{c }{\textbf{Method}}		 &\textbf{CFLN}		&${\bf{\Delta}t}$ [ns]	 &\textbf{No. of Cells}	&\textbf{Time Cost} [s]		&\textbf{Ratio$^*$}\cr
		\hline
		\hline
		\multirow{6}*{\textbf{LOD-FDTD}} &\multirow{3}*{Coarse}	 &1	&0.193		&1,064,880		&490.1		&360.1\\
										 &	&4	&0.770		&1,064,880		&122.0		&1446.6\\
										 &	&8	&1.54		&1,064,880		&63.5		&2779.3\\	
		\cline{2-7}
										 &\multirow{3}*{Fine}	&1	&0.048		&68,152,320		&176,489.0	&1.0\\
										 &	&4	&0.193		&68,152,320		&33,700.6		&5.2\\
										 &	&8	&0.385		&68,152,320		&20550.7		&8.5\\
		\hline
		\multirow{3}*{\textbf{CLOD-FDTD}} &\multirow{3}*{Coarse}	
									 	 &1	&0.193		&1,064,880		&691.4		&255.3\\
										 &	&4	&0.770		&1,064,880		&173.3		&1018.4\\ 
										 &	&8	&1.54		&1,064,880		&90.7		&1945.8\\ 
		\hline
		\multirow{2}*{\textbf{FDTD}} 
		&Coarse 	&1	&0.193		&1,064,880		&88.1		&2003.3\\
		\cline{2-7}
		&Fine 	&1	&0.048		&68,152,320		&16685.2		&10.58\\
		\hline
		\hline
\end{tabular}\textsf{}
%\tablefootnote{Ratio is defined as the ratio of time cost used in the LOD-FDTD method with fine grid to that in the correspond method.}
\begin{tablenotes}
\footnotesize
\item[*]Ratio is defined as the ratio of time cost used in the LOD-FDTD method with fine meshes to that of the corresponding method.
\end{tablenotes}
\end{threeparttable}
}
\end{table}

\begin{figure}[h]
	\centerline{{\includegraphics[scale=0.37]{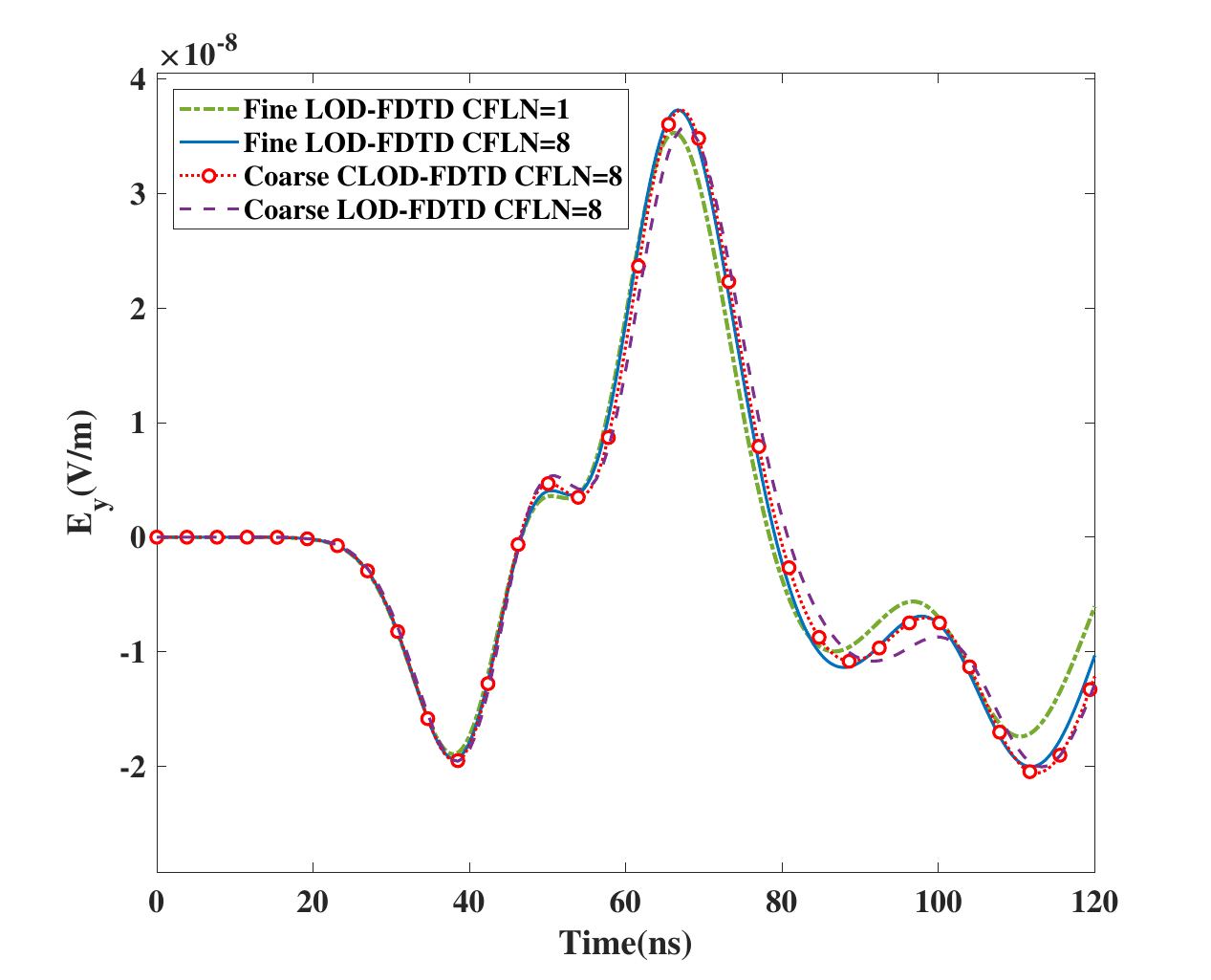}}}
	\caption{$E_y$ obtained from the LOD-FDTD method and the CLOD-FDTD method with CFLN = 1 and 8 from 0$ns$ to 120$ns$.}
	\label{Fig.missileCFLN8}
\end{figure}

We used coarse meshes with cell size as 0.05$m$ and fine meshes with cell size as 0.025$m$. The probe and the excitation are placed at (9, 3, 1.4)[$m$] and (9, 2.5, 1.4)[$m$], respectively. Results obtained by the LOD-FDTD and the CLOD-FDTD method with CFLNs = 1, 8 are shown in Fig. \ref{Fig.missileCFLN8}. It can be found that results with CFLN = 8 have slightly larger values and some delays because of the numerical dispersion error. However, results from the CLOD-FDTD method agree better with the reference results which obtained by the LOD-FDTD method with fine meshes.

Numerical results demonstrate that the CLOD-FDTD method is accurate in modeling curved surfaces with coarse meshes. Therefore, significant performance improvement in terms of CPU time and memory consumption can be obtained.

\section{CONCLUSION}
In summary, we have proposed a three-dimensional CLOD-FDTD method to accurately model the curved PEC objects through carefully considering partially filled cells. Unlike other existing CFDTD methods, which suffer from time step reduction to guarantee the stability when the PEC objects are involved, the proposed CLOD-FDTD method still preserves unconditional stability. Numerical results validate its accuracy and efficiency. Therefore, it can decrease staircase error without time step reduction, which implies that higher simulation efficiency can be obtained compared with the traditional LOD-FDTD method. It shows quite promising potential in solving the EMC problems.
% if have a single appendix:
%\appendix[Proof of the Zonklar Equations]
% or
%\appendix  % for no appendix heading
% do not use \section anymore after \appendix, only \section*
% is possibly needed

% use appendices with more than one appendix
% then use \section to start each appendix
% you must declare a \section before using any\partial
% \subsection or using \label (\appendices by itself
% starts a section numbered zero.)
%

%\appendices
%%\section{Proof of the First Zonklar Equation}
%Appendix one text goes here.

% you can choose not to have a title for an appendix
% if you want by leaving the argument blank
%\section{}
%Appendix two text goes here.

% use section* for acknowledgment

% Can use something like this to put references on a page
% by themselves when using endfloat and the captionsoff option.
\ifCLASSOPTIONcaptionsoff
  \newpage
\fi

% that's all folks
\end{document}